\documentclass[twocolumn,superscriptaddress,showpacs,nofootinbib]{revtex4}
\usepackage{bm,color}

\usepackage{graphicx}
\usepackage{amsmath}
\usepackage{amssymb}
\usepackage{amsfonts}
\usepackage{bm}
\usepackage{color}
\usepackage{lipsum}

\def\be{\begin{equation}}
\def\ee{\end{equation}}
\def\bea{\begin{eqnarray}}
\def\eea{\end{eqnarray}}

\begin{document}

\title{Cosmological evolution of a complex scalar field with repulsive \\
or attractive self-interaction}

% First author block:
\author{Abril Su\'arez}
\email{asuarez@upmh.edu.mx}
% \homepage{An author's web page; optional}
\affiliation{Laboratoire de Physique Th\'eorique, Universit\'e Paul
Sabatier, 118 route de Narbonne 31062 Toulouse, France}
\affiliation{Departamento de Aeron\'autica, Universidad Polit\'ecnica
Metropolitana de Hidalgo, Ex-Hacienda San Javier, Tolcayuca, Hgo. C.P.
43860, Mexico}
% You may list several affiliation, using separate commands for each:
%\affiliation{The third affiliation is shared by both co-authors}
\author{Pierre-Henri Chavanis}
\email{chavanis@irsamc.ups-tlse.fr}
\affiliation{Laboratoire de Physique Th\'eorique, Universit\'e Paul
Sabatier, 118 route de Narbonne 31062 Toulouse, France}
% For other authors please repeat the author block as needed
%\author{Second Author}
% Note how REVTeX 4 deals with identical affiliations
%\affiliation{The third affiliation is shared by both co-authors}

\begin{abstract}

We study the cosmological evolution of a complex scalar field with a
self-interaction potential $V(|\varphi|^2)$, possibly describing
self-gravitating Bose-Einstein condensates, using a fully general relativistic
treatment. We generalize the hydrodynamic representation of the
Klein-Gordon-Einstein equations in the weak field approximation
developed in our previous paper [Su\'arez and Chavanis, Phys. Rev. D 92, 023510
(2015)]. We establish the general equations
governing the evolution of a spatially homogeneous complex scalar field in an
expanding
background. We show how they can be simplified in the fast oscillation regime
(equivalent to the Thomas-Fermi, or semiclassical, approximation) and derive the
equation of state of the scalar field in parametric form for an arbitrary
potential $V(|\varphi|^2)$. We explicitly consider the case of a quartic
potential with repulsive or attractive self-interaction. For repulsive
self-interaction, the scalar field undergoes a stiff matter era followed by a
pressureless dark matter era in the weakly self-interacting regime and a stiff
matter era followed by a radiationlike era and a pressureless dark
matter era in the strongly self-interacting regime. For attractive
self-interaction, the scalar field undergoes an inflation era followed by a
stiff matter era and a pressureless dark matter era  in the weakly
self-interacting regime and an inflation era followed by a cosmic stringlike era
and a pressureless dark matter era in the strongly self-interacting
regime (the inflation era is suggested, not demonstrated). We also find a
peculiar branch on which the scalar field emerges
suddenly at a nonzero scale factor with a finite
energy density. At early times, it behaves as a gas of cosmic strings. At later
times,  it behaves as dark energy
with an almost constant energy density giving rise to a de Sitter evolution. 
This is due to spintessence. We
derive the effective cosmological constant produced by the scalar field. 
Throughout the paper, we analytically characterize the
transition scales of the scalar field and establish the domain of validity of
the fast oscillation regime. We analytically confirm and complement the
important results of 
Li, Rindler-Daller and Shapiro [Phys. Rev. D, 89, 083536 (2014)]. We
determine the phase
diagram of a scalar field with repulsive or attractive self-interaction. We show
that the transition between the weakly self-interacting regime and the strongly
self-interacting regime depends on how the scattering length of the bosons
compares with their effective Schwarzschild radius.  We also
constrain the parameters of the scalar field from astrophysical and cosmological
observations.
Numerical applications are made for ultralight bosons without
self-interaction (fuzzy dark matter), for bosons with repulsive
self-interaction, and for bosons with attractive self-interaction
(QCD axions and ultralight axions).

\end{abstract}

% Insert suggested PACS numbers (up to 4) in braces.
% The PACS (Physics and Astronomy Classification Scheme)
% can be accessed on the web at http://www.aip.org/pacs/

% Insert keywords (up to about 4) in braces; optional.
% \keywords{Up to four keywords}

\maketitle

% Here the text of your article begins
%--------------------------------------------------------------------------------------------------------------------------------------------------

\section{Introduction}

There is compelling observational evidence for the existence of dark matter
(DM) and dark energy (DE) in the Universe. The suggestion
that DM may constitute a large part of the Universe was
raised by Zwicky \cite{zwicky} in 1933. Using the virial theorem to infer the
average mass of galaxies within the Coma cluster, he obtained a 
much higher value than the mass of luminous material. He realized therefore
that some mass was ``missing'' to account for the observations. The existence
of DM has been confirmed by more precise observations of
rotation curves \cite{b1}, gravitational lensing \cite{massey}, and hot gas in
clusters \cite{b2}. On the other hand, DE is responsible for the ongoing
acceleration of the Universe revealed by the high redshift of type Ia supernovae
treated as standardized candles \cite{b6,b10,b11}. Recent observations of
baryonic acoustic oscillations 
provide another independent support to the DE hypothesis
\cite{b12}. In both cases (DM and DE) more indirect measurements come
from the Cosmic Microwave Background (CMB) and large scale structure
observations \cite{smoot,jarosik,planck}. 

The variations in the temperature of the thermal CMB radiation at $3$K
throughout the sky imply $\Omega_{k,0}\sim0$ and $\Omega_{\rm r,0}\sim10^{-4}$,
while the power spectrum of the spatial distributions of large scale structures
gives $\Omega_{\rm m,0}\sim 0.3$, where $\Omega_{k,0}$ is the effective
curvature of
spacetime, $\Omega_{\rm r,0}$ is the present energy density in the relativistic
CMB
radiation (photons) accompanied by the low mass neutrinos that almost
homogeneously fill the space, and $\Omega_{\rm m,0}$ is the current mean energy
density of nonrelativistic matter which mainly consists of baryons and
nonbaryonic DM. These observations give a value of $\Omega_{\Lambda,0}\sim0.7$
for the present DE density \cite{planck}.

One of the most fundamental problems in modern cosmology concerns the nature
of DM and DE. In the last decades, various DM and DE models have been studied.
The simplest model of DM consists in particles moving slowly compared to the
speed of light (they are cold) and interacting very weakly with ordinary
matter and electromagnetic radiation. These particles, known as weakly
interacting massive particles (WIMPS), behave as dust with an equation of
state (EOS)
parameter $w=P/\epsilon\simeq 0$ \cite{b1i,b1j,b1k}. They may correspond to
supersymmetric (SUSY) particles \cite{susy}. On the other hand, the
simplest manner
to explain the accelerated expansion of the Universe is to introduce
a cosmological constant $\Lambda$ in the Einstein equations \cite{b25}.  In that
case, the value of the energy density $\epsilon_{\Lambda}=\Lambda c^2/8\pi G$
stored in the cosmological constant represents the DE.

The standard  model of cosmological structure formation in the Universe is known
as the cold dark matter model with a cosmological constant ($\Lambda$CDM)
\cite{peebles1,b1b,b1c,b1a}. Cosmological observations at large
scales
support the $\Lambda$CDM model with a high precision.

However, this model has some problems at small (galactic) scales for the case
of DM \cite{kroupa,b1e,b1f,b1d}. In particular, it predicts that DM halos
should be cuspy \cite{nfw} while observations reveal that they have a flat 
core \cite{observations}. On the other hand, the $\Lambda$CDM model predicts
an over-abundance of small-scale structures (subhalos/satellites), much more
than what is observed around the Milky Way \cite{satellites}. These problems are
referred to as the ``cusp problem'' and  ``missing satellite problem''. The
expression ``small-scale crisis of CDM'' has been coined.

Furthermore, the value of the cosmological
constant $\Lambda$ assigned to DE has to face important fine tuning problems
\cite{b3,b1h,martin}. From the point of view of particle physics, the
cosmological
constant can be
interpreted naturally in terms of the vacuum energy density whose scale is of
the order of the Planck density  $\rho_P=5.16\times 10^{99}\, {\rm g}\, {\rm
m}^{-3}$. However, observationally, the cosmological constant is of
the order of the present value of the Hubble parameter squared, $\Lambda\sim
H_0^2=(2.18\times 10^{-18}\, {\rm s}^{-1})^2$,
which corresponds to a dark energy density
$\rho_\Lambda=\Lambda/8\pi G \sim 10^{-24}\, \mbox{g}\, \mbox{m}^{-3}$. The
Planck density and the cosmological density differ from
each other by $123$ orders of magnitude.
This leads to the so-called cosmological constant problem
\cite{b3,b1h,martin}.

Since the $\Lambda$CDM model poses problems, some efforts have been done in
trying to understand the nature of DM and DE from the framework of quantum field
theory. In particle physics and string theory, scalar fields (SF) arise in a
natural way as bosonic spin-$0$
particles described by the Klein-Gordon (KG) equation
\cite{kolb,zee}. Examples include the
Higgs particle, the inflaton, the dilaton field of superstring
theory, tachyons
etc. SFs also arise in the Kaluza-Klein and
Brans-Dicke theories \cite{b1m}. In cosmology, SFs were introduced to explain
the phase of inflation in the
primordial Universe \cite{linde}. SF models
have then been used in cosmology in
various contexts and they continue to play an important role as potential DM
and DE candidates.

For example, the source of DE can be attributed to a SF. 
A variety of SF models have been infered for this purpose (see for
example \cite{b5,ri,peebles1}). Quintessence \cite{quintessence,b14},
which is
the simplest case, is described by an ordinary SF  minimally coupled
to gravity. It generally has a density and EOS parameter $w(t)$ that vary with
time, hence making it dynamic. By contrast,
a cosmological
constant is static, with a fixed energy density and $w=-1$.
Phantom fields \cite{b16,bigrip,b17,b18}
are associated to a negative kinetic term. This strange property leads to an EOS
parameter $w\le -1$ implying that the energy density increases as the Universe
expands, possibly leading to a big rip. It has also
been suggested that, in
a class of string theories, tachyonic SF \cite{b21} can condense and have
cosmological
applications. Tachyons have an interesting EOS whose parameter smoothly
interpolates between $-1$ and $0$, thus behaving as DE and pressureless DM. SF
models describing DE usually feature masses of
the order of the current Hubble scale ($m\sim H_0\hbar/c^2\sim 10^{-33}\, {\rm
eV}/c^2$)
\cite{b54,b53}.

Concerning DM, it has been proposed that DM halos can be made of a SF
described by the Klein-Gordon-Einstein (KGE) equations (see,
e.g., \cite{b26,b27,b28,marshrev} for reviews and \cite{nature} for high
resolution numerical simulations showing the viability of this scenario). In
general, SFDM
models suppose that DM is a real or complex SF
minimally coupled to gravity. This SF can be self-interacting but it does
not interact with the other particles and fields, except
gravitationally. SF that interact only with gravity could be gravitationally
produced by inflation \cite{ford}. The SF may represent the wave function of
the bosons
having formed a 
Bose-Einstein condensate (BEC). The KGE equations describe a relativistic
SF/BEC. General relativity is necessary to describe compact SF objects such as
boson
stars \cite{kaup,rb,colpi} and neutron stars with a superfluid core
\cite{page,b45}. It is also necessary
in
cosmology to describe the phase of inflation and the evolution of the
early Universe \cite{linde}. However, in the context of DM halos, Newtonian
gravity is
sufficient. The evolution of a nonrelativistic SF/BEC is described by
the Gross-Pitaevskii-Poisson (GPP) equations. There
are several models of SFDM, e.g. noninteracting (fuzzy) DM \cite{b29},
self-interacting DM \cite{ss}, or
axionic DM
\cite{b30,b31,dine,turneraxion,b32}.\footnote{Axions can
be produced in the early Universe through two mechanisms. At the quantum
chromodynamics (QCD) phase transition where a BEC of axions forms
and these very cold particles behave as CDM; and through the decay
of strings formed at the
Peccei-Quinn phase transition \cite{davis1,davis2}. Unless inflation occurs
after the Peccei-Quinn phase transition, strings are thought to be the
dominant
mechanism for axion production \cite{hagmann}. Recent analysis confirms that
strings are likely to be the dominant source of axions, even though strings will
not produce an interesting level of density fluctuations as their predicted mass
per unit length is far too small to be cosmologically interesting
\cite{battye}.} Most of these models are based on the
assumption that DM
is made of extremely light scalar particles with masses between $10^{-23}\, {\rm
eV}/c^2\leq m\leq 10^{-2}\,  {\rm eV}/c^2$. Within this mass scale, SFDM
displays
a wave (quantum) behavior at galactic scales that could solve many of the
problems of the $\Lambda$CDM model. Indeed, the wave properties
of bosonic DM may stabilize the system against gravitational collapse, providing
halo cores and sharply suppressing small-scale linear power. This may solve
the cusp problem and the
missing satellite problem. Therefore, the main
virtues of the SF/BEC model is that it can reproduce
the cosmological evolution of the Universe for the background and behave as
CDM at large scales where its wave nature is invisible,
while at the same time it solves the problems
of the CDM model at small scales where its wave nature manifests itself.

In quantum field theory, ultralight SFs seem unnatural but renormalization
effects tend to drive these scalar masses up to the scale of a new physics.
Given the present observational status of cosmology, and despite all the
efforts that have been made, it is fair to say that the nature of DM
and DE remains a mystery. As a result, the SF scenario is an interesting
suggestion that deserves to be studied in more detail.

Instead of working directly in terms of field variables, a fluid approach can be
adopted. In the nonrelativistic case, this hydrodynamic approach was introduced
by Madelung \cite{b37} who showed that the Schr\"odinger equation is
equivalent to the
Euler equations for an irrotational fluid with an additional quantum potential
arising from the finite value of $\hbar$ and accounting for
Heisenberg's uncertainty principle. This approach has been
generalized to the GPP equations in the context of DM halos by
\cite{bohmer,b34,rindler} among others. In the relativistic case,
de Broglie
\cite{broglie1927a,broglie1927b,broglie1927c} in his
so-called pilot wave theory, showed that the KG equations are equivalent to
hydrodynamic equations including a covariant quantum
potential. This
approach has been generalized to the Klein-Gordon-Poisson (KGP) and KGE
equations in the context
of DM halos by \cite{b38,b39,b40,b41,chavmatos}.\footnote{The pilot wave theory
of
de Broglie \cite{broglie1927a,broglie1927b,broglie1927c}  is the
relativistic version of
Madelung's hydrodynamics \cite{b37}. The works of de Broglie and Madelung were
developed independently.
See the Introduction of \cite{chavmatos} for a short
historic of the early development of quantum mechanics.} In this hydrodynamic
representation, DM halos
result from the balance between the gravitational attraction and the quantum
pressure arising from the Heisenberg uncertainty principle or from the
self-interaction of the bosons. At small scales, pressure effects are important
and can prevent the formation of singularities and solve the cusp problem and
the missing satellite problem. At large scales, pressure effects are
generally negligible (except in the early Universe) and one recovers the 
$\Lambda$CDM model.

The formation of large-scale structures is an important topic of cosmology.
This problem was first considered by Jeans  \cite{jeans}  (before the discovery
of the expansion of the Universe) who studied the instability of an infinite
homogeneous self-gravitating classical collisional gas (see \cite{chavjeans} for
a review). This
study has been generalized in the context of SF theory.
The Jeans instability of an infinite
homogeneous  self-gravitating system in a static background was studied by
\cite{khlopov} for a relativistic SF described by the generalized KGP equations,
using the field representation. The same
problem was studied in \cite{b34,b35} for a nonrelativistic SF described
by the GPP equations in the context of Newtonian cosmology, and in
\cite{b40,b41}
for a relativistic SF described by the KGE equations, using the hydrodynamic
representation.

The growth of perturbations of a relativistic real SF in an
expanding Universe was considered in \cite{mul,jcap} using the field
representation. The same problem was
addressed in \cite{b35,b40} for a complex SF using the hydrodynamic
representation. Analytical results  were obtained in the
(nonrelativistic) matter era where the background Universe has an Einstein-de
Sitter (EdS) evolution \cite{b35,b40}. The matter era is valid at sufficiently
late times, after the radiation-matter equality. At earlier times, the SF
affects the background evolution of the
Universe so we can no more assume that the scale factor follows the EdS
solution.

The classical evolution of a  real SF described by the
KGE equations with a potential of the form $V(\varphi)=a\varphi^{n}$ in an
isotropic and homogeneous cosmology was first investigated by Turner
\cite{turner} (see also the subsequent works of \cite{ford,greene,pv}). He
showed that the SF experiences damped oscillations but that, in average, it is
equivalent to a perfect fluid with an EOS $P=[(n-2)/(n+2)]\epsilon$ (this
result is valid if we neglect particle creation due to the time variation of
$\varphi$). For $n=2$ the SF behaves as pressureless matter and for $n=4$
it behaves as radiation. Turner also mentioned the possibility of a stiff EOS. 
The cosmological evolution of a spatially homogeneous real self-interacting SF
with a repulsive $\varphi^4$ potential described by the KGE equations competing
with
baryonic matter, radiation and
dark energy was considered by \cite{mul}. In this work, it is found that
a real self-interacting  SF displays fast oscillations and that, on the
mean, it undergoes a radiationlike era followed by a matterlike era. In the
noninteracting case, the SF undergoes only a matterlike era \cite{b1p}. In any
case, at sufficiently late times, the SF reproduces the cosmological predictions
of the standard $\Lambda$CDM model.

The cosmological evolution a complex self-interacting SF 
representing BECDM has  been considered by \cite{harkocosmo,b35} who solved
the (relativistic) Friedmann equations with the EOS of the BEC derived from the
(nonrelativistic) GP equation
after identifying $\rho_m c^2$, where $\rho_m$ is the rest-mass density, with
the energy density $\epsilon$. However, as clarified in \cite{stiff}, this
approach is not valid in the early Universe as it combines relativistic and
nonrelativistic equations. These studies may still have 
interest in cosmology in a different context, as discussed
in \cite{cosmopoly1,cosmopoly2}.

The exact relativistic cosmological evolution of a complex
self-interacting
SF/BEC described by the KGE equations with a repulsive $|\varphi|^4$ potential
has been considered by Li {\it et al.}
\cite{b36} (see also the previous
works of \cite{sj,js,spintessence,arbeycosmo}). In
this work, the evolution of the homogeneous background is studied. It is shown
that the SF undergoes three successive phases: a stiff matter era, followed by
a radiationlike era (that only exists for self-interacting SFs), and finally a
matterlike era similar to the one appearing in the CDM model. Another
cosmological model displaying a
primordial stiff matter era has been developed
in \cite{stiff}. Interestingly, it leads to a completely
analytical cosmological solution generalizing the EdS model and the
(anti)-$\Lambda$CDM
model.

In general, the SF oscillates in time and it is
not clear how these oscillations can be measured in practice because there is no
direct access to field variables such as $\varphi$. As a result, the
hydrodynamic representation of
the SF may be more physical than the KG equation itself because it is easier to
measure hydrodynamic variables such as the energy density $\epsilon$, the
rest-mass density $\rho_m$, and the pressure $P$. In our previous paper
\cite{b40}, we showed that the three phases of a relativistic SF with a
repulsive  $|\varphi|^4$ potential (stiff matter, radiation and pressureless
matter) could be obtained from the hydrodynamic approach in complete agreement
with the field theoretic approach of Li {\it et al.} \cite{b36}.

In the present paper, we complete and generalize our study in different
directions: we formulate the problem for an arbitrary SF potential 
$V(|\varphi|^2)$, not just for a $|\varphi|^4$ potential; we solve the
equations in the fast oscillation regime and obtain several analytical
results in different asymptotic limits that complement the work of Li {\it et
al.} \cite{b36}; we consider repulsive and
attractive self-interaction and show that the later can lead to very
peculiar results. The case of attractive self-interaction is of considerable
interest since
axions, that have been proposed as a serious DM candidate, usually have an
attractive self-interaction. The case of attractive self-interaction
has been studied previously in \cite{b34,cd,b35,b40,bectcoll}. It is shown
in \cite{b35,b40} that an attractive self-interaction can accelerate the growth
of structures is cosmology. On the other hand, it is shown in
\cite{b34,cd,bectcoll} that
stable DM halos with an attractive self-interaction can exist only below a
maximum mass that severely constrains the parameters of the SF.

The paper is organized as follows. In Sec. \ref{sec_tb}, we
introduce the KG and Friedmann equations describing the cosmological evolution
of a spatially homogeneous complex\footnote{Complex
SFs are potentially more relevant
than real SFs because they can form stable DM halos while DM halos made of real
SFs are either dynamically unstable or oscillating. If DM halos were stable
``oscillatons'' \cite{oscillaton}, their oscillations would probably have been
detected. On the other hand, bosons described by a complex SF 
with a global $U(1)$ symmetry associated with a conserved charge (Noether
theorem) can form BECs even in the early Universe while this is more difficult
for boson described by a real SF (like the QCD axion).} SF with an
arbitrary self-interaction
potential $V(|\varphi|^2)$ in an expanding background and
provide their hydrodynamic representation. We show that these
hydrodynamic equations can be simplified
in the fast oscillation
regime equivalent to the Thomas-Fermi (TF), or semiclassical, approximation
where
the quantum potential can be neglected. We derive the EOS of the
SF in parametric form for an
arbitrary potential $V(|\varphi|^2)$. In Sec. \ref{sec_pos}, we consider the
cosmological evolution of a spatially homogeneous SF with a
repulsive quartic self-interaction. In agreement with previous works
\cite{b36,b40},
we show that the SF undergoes a stiff matter era ($w=1$) in the slow
oscillation regime, followed by a radiationlike era ($w=1/3$)  and a
pressureless dark matter era ($w\simeq 0$)  in the fast oscillation regime. We
analytically determine the transition scales between these different periods and
show that the radiationlike era can only exist for sufficiently large values of
the self-interaction parameter. More precisely, the transition between the
weakly self-interacting and strongly self-interacting regimes depends on how the
scattering length of the bosons $a_s$ compares with their effective
Schwarzschild radius
$r_S=2Gm/c^2$. We determine the phase diagram of a SF with repulsive
self-interaction. We
also analytically recover the bounds on the ratio $a_s/m^3$
obtained by
Li {\it et al.} \cite{b36} by requiring that the SF must be
nonrelativistic at the epoch of matter-radiation equality 
and by using constraints from the big bang nucleosynthesis (BBN). In Sec. 
\ref{sec_neg}, we consider the evolution of a spatially homogeneous SF with an
attractive quartic self-interaction. In the fast oscillation regime, the
SF emerges at a nonzero scale factor with a finite energy
density. At early time, it behaves as a gas of cosmic strings
($w=-1/3$). At later time, two evolutions are possible. On the normal branch,
the SF behaves as pressureless DM ($w\simeq 0$). On the
peculiar branch, it behaves as DE ($w=-1$) with an almost constant
energy density giving rise to a de Sitter evolution. We derive the effective
cosmological constant produced by the SF. We establish the domain
of validity of the fast oscillation regime. We argue that, in the very early
Universe, a complex SF with an attractive self-interaction undergoes an
inflation era. If the self-interaction constant
is sufficiently small, the inflation era is followed by a stiff matter era. 
We determine the phase diagram of a SF with attractive self-interaction.
We also set constraints on the parameters
of the SF using cosmological observations. Numerical applications are made for
standard (QCD) axions and ultralight axions. This is indicative because QCD
axions are real SFs while certain of our results are only valid for complex
SFs. In Sec. \ref{sec_tp}, we study the
evolution of the SF in the total potential $V_{\rm tot}(|\varphi|^2)$
incorporating the rest-mass energy. A SF with
repulsive self-interaction descends the potential.  A SF with attractive
self-interaction descends the potential on the normal branch and ascends the
potential on the peculiar branch. This is possible because of the effect of a
centrifugal force that is specific to a complex SF. This is called
spintessence \cite{spintessence}. The concluding Sec. \ref{sec_con}
summarizes the main
results of
our study and regroups the numerical applications of astrophysical relevance. 
The Appendices contain additional material that is needed to interpret our
results.

\section{Spatially homogeneous complex SF}
\label{sec_tb}

In our previous paper \cite{b40}, we have derived a hydrodynamic representation
of
the KGE equations in an expanding background in the weak field approximation. We
considered a complex SF with a quartic self-interaction potential. This study
was extended to the case of an arbitrary SF potential of the form
$V(|\varphi|^2)$ in \cite{b41,chavmatos}. In this section, we consider the case
of a spatially homogeneous complex SF. For the clarity and the 
simplicity of the presentation, we assume that the Universe is only composed of
a SF, although it would be straightforard to include in the formalism other
components such as normal radiation, baryonic matter, and dark energy
(e.g., a cosmological constant).

\subsection{The KG equation for a spatially homogeneous complex SF}

The cosmological evolution of a spatially homogeneous complex SF
$\varphi(t)$ with a self-interaction potential $V(|\varphi|^2)$ in a
Friedmann-Lema\^itre-Robertson-Walker (FLRW) universe is described by
the KG equation
\begin{eqnarray}
\frac{1}{c^2}\frac{d^2\varphi}{dt^2}+\frac{3H}{c^2}\frac{d\varphi}{dt}+\frac
{m^2
c^2}{\hbar^2}\varphi
+2\frac{dV}{d|\varphi|^2}\varphi=0,
\label{h1}
\end{eqnarray}
where $H=\dot a/a$ is the Hubble parameter and $a(t)$ is the scale factor. The
second term in Eq. (\ref{h1}) is the Hubble drag. The
rest-mass term (third term) can be written as $\varphi/\lambda_C^2$ where
$\lambda_C=\hbar/mc$ is the
Compton wavelength of the bosons.

The energy density
$\epsilon(t)$ and the pressure $P(t)$ of the SF are given by
\begin{equation}
\epsilon=\frac{1}{2c^2}\left |\frac{d\varphi}{d
t}\right|^2+\frac{m^2c^2}{2\hbar^2}|\varphi|^2+V(|\varphi|^2),
\label{h2}
\end{equation}
\begin{equation}
P=\frac{1}{2c^2}\left |\frac{d\varphi}{d
t}\right|^2-\frac{m^2c^2}{2\hbar^2}|\varphi|^2-V(|\varphi|^2).
\label{h3}
\end{equation}
The EOS parameter is defined by $w=P/\epsilon$.

\subsection{The Friedmann equations}
\label{sec_frid}

From Eqs. (\ref{h1})-(\ref{h3}), we can obtain the energy equation
\begin{equation}
\frac{d\epsilon}{dt}+3H(\epsilon+P)=0.
\label{h4}
\end{equation}
This equation can also be directly obtained from the Einstein field equations
and constitutes the first Friedmann equation \cite{b20e}. From this equation we
deduce that, as the Universe expands, the energy density decreases when
$w>-1$, increases when $w<-1$, and remains constant when $w=-1$. In the second
case, the Universe is ``phantom'' \cite{b16}. The second
Friedmann equation, obtained from the Einstein field equations,
writes
\begin{eqnarray}
H^2=\frac{8\pi G}{3c^2}\epsilon.
\label{h5}
\end{eqnarray}
We have assumed that the Universe is flat in agreement with the observations
of the CMB. From Eqs. (\ref{h4}) and (\ref{h5}), we easily obtain the
acceleration equation
\begin{eqnarray}
\frac{\ddot a}{a}=-\frac{4\pi G}{3c^2}(\epsilon+3P)
\label{h6}
\end{eqnarray}
which constitutes the third Friedmann equation. From this equation, we deduce
that the
expansion of the Universe is
decelerating when $w>-1/3$ and accelerating when  $w<-1/3$. The intermediate
case, in which the scale factor increases linearly with time, corresponds to
$w=-1/3$.

\subsection{Hydrodynamic representation of a spatially homogeneous complex SF}

Instead of working with the SF $\varphi(t)$, we will use hydrodynamic variables
like those considered in our previous works  \cite{b40,b41,chavmatos}. We define
the pseudo rest-mass density by
\begin{eqnarray}
\rho=\frac{m^2}{\hbar^2}|\varphi|^2.
\label{kge7}
\end{eqnarray}
We stress that it is only in the nonrelativistic limit $c\rightarrow +\infty$
that $\rho$
has the interpretation of a rest-mass density. In the relativistic regime,
$\rho$ does not have a clear physical interpretation but it can
always be defined as a convenient notation \cite{b40,b41,chavmatos}.
We write the SF in the de Broglie form
\begin{eqnarray}
\varphi(t)=\frac{\hbar}{m}\sqrt{\rho(t)}e^{i
S_{\rm tot}(t)/\hbar},
\label{ic1b}
\end{eqnarray}
where $\rho$ is the pseudo rest-mass density
and $S_{\rm tot}=(1/2)i\hbar\ln(\varphi^*/\varphi)$  is the real
action. The total energy  of the SF (including its rest mass $mc^2$ energy) is
\begin{eqnarray}
E_{\rm tot}(t)= -\frac{d S_{\rm tot}}{d t}.
\label{ic3z}
\end{eqnarray}

Substituting Eq. (\ref{ic1b}) into the KG equation (\ref{h1}) and separating
real and imaginary
parts, we get
\begin{eqnarray}
\frac{1}{\rho}\frac{d\rho}{dt}+\frac{3}{a}\frac{da}{dt}+\frac{1}{E_{\rm
tot}} \frac{dE_{\rm tot}}{dt}=0,
\label{b4}
\end{eqnarray}
\begin{eqnarray}
E_{\rm tot}^2=\hbar^2\frac{\frac{d^2\sqrt{\rho}}{d
t^2}}{\sqrt{\rho}}+3H\hbar^2\frac{\frac{d\sqrt{\rho}}{dt}}{\sqrt{
\rho}}
+m^2c^4+2m^2c^2 V'(\rho).
\label{b2}
\end{eqnarray}
On the other hand, from Eqs. (\ref{h2}) and (\ref{ic1b}), we find that the
Friedmann equation (\ref{h5}) takes the form
\begin{eqnarray}
\frac{3H^2}{8\pi G}=\frac{\hbar^2}{8m^2c^4}\frac{1}{\rho}\left
(\frac{d\rho}{dt}\right
)^2+\frac{\rho E_{\rm tot}^2}{2m^2c^4}+\frac{1}{2}\rho+\frac{1}{c^2}V(\rho).
\label{b3}
\end{eqnarray}
Equations (\ref{b4})-(\ref{b3}) can also be obtained from the
general hydrodynamic
equations derived in  \cite{b40,b41,chavmatos} by considering the particular
case of a
spatially homogeneous SF
($\rho(\vec x,t)=\rho(t)$, $\vec{v}({\vec x},t)=\vec{0}$, $\Phi({\vec
x},t)=0$, and $S({\vec x},t)=S(t)$). In that case, Eq. (\ref{b4}) is deduced
from the continuity equation, Eq. (\ref{b2}) from the quantum Bernoulli or
Hamilton-Jacobi equation, and Eq. (\ref{b3}) from the Einstein equations.
In this connection, we note that the first two terms (the terms
proportional to $\hbar^2$)  in the r.h.s.
of Eq. (\ref{b2}) correspond to the relativistic de
Broglie quantum
potential
\begin{eqnarray}
Q_{\rm dB}=\frac{\hbar^2}{2m}\frac{\square\sqrt{\rho}}{\sqrt{\rho}}
\end{eqnarray}
for a spatially homogeneous SF. We stress that the
hydrodynamic equations (\ref{b4})-(\ref{b3}) are
equivalent to the KGE equations (\ref{h1}), (\ref{h2}) and (\ref{h5}).
Finally, we note that the hydrodynamic equations
(\ref{b4})-(\ref{b3})
with the terms in $\hbar$ neglected provide a TF, or semiclassical,
description of
relativistic SFs.

The  continuity equation (\ref{b4}) can be rewritten as a conservation law
\begin{eqnarray}
\frac{d}{dt}(E_{\rm tot}\, \rho
a^3)=0. 
\label{b5}
\end{eqnarray}
Therefore, the total energy of the SF is exactly given by
\begin{eqnarray}
\frac{E_{\rm tot}}{mc^2}=\frac{Qm}{\rho a^3},
\label{b6}
\end{eqnarray}
where $Q$ is a constant which represents the
conserved charge of the complex SF
\cite{arbeycosmo,gh,b36,b40}.\footnote{The
conserved charge  (normalized by the
elementary charge $e$) of a complex SF is given by $Q=\frac{1}{e c}\int
J_e^0\sqrt{-g}\, d^3x$, where 
$(J_e)_{\mu}=\frac{ie}{2\hbar}(\varphi^*\partial_{\mu}\varphi-\varphi\partial_{
\mu}
\varphi^*)$ is the quadricurrent of charge of the SF (see, e.g.,
\cite{chavmatos} for
details). The charge density is $\rho_e=(J_e)_0/c$. Using Eqs.
(\ref{ic1b}) and (\ref{ic3z}), we find that  $\rho_e=e\rho
E_{\rm tot}/m^2c^2$. The
conserved charge of a
spatially homogeneous SF in an expanding Universe is 
$Q=\rho_e a^3/e=\rho E_{\rm tot}a^3/m^2c^2$, corresponding to Eq. (\ref{b6}).} 

In the hydrodynamic representation, the energy density and the
pressure of a homogeneous SF can be expressed as
\begin{eqnarray}
\epsilon=\frac{\hbar^2}{8m^2c^2}\frac{1}{\rho}\left
(\frac{d\rho}{dt}\right
)^2+\frac{\rho E_{\rm tot}^2}{2m^2c^2}+\frac{1}{2}\rho c^2+V(\rho),
\label{b7}
\end{eqnarray}
\begin{eqnarray}
P=\frac{\hbar^2}{8m^2c^2}\frac{1}{\rho}\left
(\frac{d\rho}{dt}\right
)^2+\frac{\rho E_{\rm tot}^2}{2m^2c^2}-\frac{1}{2}\rho c^2-V(\rho).
\label{b8}
\end{eqnarray}

\subsection{Cosmological evolution of a spatially homogeneous complex SF in the
fast
oscillation regime}
\label{sec_rm}

The exact equations (\ref{b4})-(\ref{b3}) are
complicated. In the case of a
quartic potential with a positive scattering length, Li {\it et al.} \cite{b36}
have identified two regimes in which these equations can be simplified. When the
oscillations of the SF are slower than the Hubble expansion ($\omega\ll H$), the
SF is equivalent to a stiff fluid with
an EOS $P=\epsilon$. This approximation is valid in the early
Universe. At later times, when the oscillations of the
SF are faster than the Hubble expansion ($\omega\gg H$), it is possible to
average over the fast oscillations in order to obtain a simpler dynamics. The
resulting equations can be obtained either from the field theoretic approach
\cite{b36} or from the hydrodynamic approach \cite{b40}. We note that the
equations obtained in the fast oscillation regime specifically depend on the
form of the SF potential. In this section, we generalize
these results to the case of an arbitrary SF potential $V(|\varphi|^2)$. We use
the hydrodynamic approach. The  field theoretic approach is exposed in Appendix
\ref{sec_ext}.

The simplified equations valid in the fast oscillation regime can be
obtained from  Eqs. (\ref{b2}) and (\ref{b3}) by neglecting the terms involving
a time derivative. Interestingly, this is equivalent to neglecting the
terms in $\hbar$. Therefore, the fast oscillation regime is equivalent
to the TF, or semiclassical, approximation where the quantum
potential (arising from Heisenberg's uncertainty principle) is
neglected. In that case, we obtain
\begin{eqnarray}
E_{\rm tot}^2=m^2c^4+2m^2c^2 V'(\rho),
\label{b9}
\end{eqnarray}
\begin{eqnarray}
\frac{3H^2}{8\pi G}=\frac{\rho E_{\rm
tot}^2}{2m^2c^4}+\frac{1}{2}\rho+\frac{1}{c^2}V(\rho).
\label{b10}
\end{eqnarray}
Keeping only
the solution of Eq. (\ref{b9}) that leads to a positive total energy (the
solution with a negative
total energy corresponds to antibosons), we
get
\begin{eqnarray}
E_{\rm tot}=mc^2\sqrt{1+\frac{2}{c^2} V'(\rho)}. 
\label{b12}
\end{eqnarray}
Combining Eqs. (\ref{b6})  and (\ref{b12}), we obtain
\begin{eqnarray}
\rho \sqrt{1+\frac{2}{c^2}V'(\rho)}=\frac{Qm}{a^3}.
\label{b13}
\end{eqnarray}
This equation determines the pseudo rest-mass density $\rho$ as a function of
the scale factor $a$. Substituting Eq. (\ref{b9}) into Eq. (\ref{b10}), we find
\begin{eqnarray}
\frac{3H^2}{8\pi G}=\rho+\frac{1}{c^2}\left\lbrack V(\rho)+\rho
V'(\rho)\right\rbrack.
\label{b14}
\end{eqnarray}
Equations (\ref{b13}) and (\ref{b14}) determine the evolution of the scale
factor $a(t)$ of the Universe
induced by a spatially  homogeneous SF in the regime where its oscillations
are faster than the Hubble expansion. The energy $E_{\rm tot}$ of the SF is then
given by Eq. (\ref{b12}).

It is not convenient to solve the differential equation (\ref{b14}) for the
scale factor $a$ because we would need to inverse Eq. (\ref{b13}) in order to
express
$\rho$ as a function of $a$ in the r.h.s. of Eq. (\ref{b14}). Instead, it is
more convenient to view $a$ as a function of $\rho$, given by Eq.
(\ref{b13}), and
transform Eq. (\ref{b14}) into a differential equation for $\rho$. Taking the
logarithmic derivative of Eq. (\ref{b13}), we get
\begin{eqnarray}
\frac{\dot a}{a}=-\frac{1}{3}\frac{\dot \rho}{\rho}\left \lbrack
1+\frac{\rho V''(\rho)}{c^2+2V'(\rho)}\right\rbrack.
\label{b15}
\end{eqnarray}
Substituting this expression into Eq. (\ref{b14}), we obtain the differential
equation
\begin{eqnarray}
\frac{c^2}{24\pi G}\left
(\frac{\dot\rho}{\rho}\right )^2=\frac{\rho
c^2+V(\rho)+\rho V'(\rho)}{\left\lbrack 1+\frac{\rho
V''(\rho)}{c^2+2V'(\rho)}\right\rbrack^2}.
\label{b16}
\end{eqnarray}
For a given SF potential $V(\rho)$, this equation can be solved easily as it
is just a first order differential equation for $\rho$. The temporal evolution
of the scale factor $a$ is then obtained by plugging the solution of Eq.
(\ref{b16}) into Eq. (\ref{b13}).

In the fast oscillation regime, the energy density and the pressure are given by
\begin{eqnarray}
\epsilon=\frac{\rho E_{\rm
tot}^2}{2m^2c^2}+\frac{1}{2}\rho c^2+V(\rho),
\label{b17}
\end{eqnarray}
\begin{eqnarray}
P=\frac{\rho E_{\rm
tot}^2}{2m^2c^2}-\frac{1}{2}\rho c^2-V(\rho).
\label{b18}
\end{eqnarray}
Using Eq. (\ref{b9}), we get
\begin{eqnarray}
\epsilon=\rho c^2+V(\rho)+\rho V'(\rho), 
\label{b19}
\end{eqnarray}
\begin{eqnarray}
P=\rho
V'(\rho)-V(\rho).
\label{b20}
\end{eqnarray}
The pseudo velocity of sound is 
\begin{eqnarray}
c_s^2=P'(\rho)=\rho V''(\rho).
\label{kge29}
\end{eqnarray}
We note that the pressure $P(t)$ of a spatially homogeneous SF in the
fast oscillation regime coincides with the pseudo pressure $p({\vec x},t)=p(t)$
that arises in the Euler equation obtained in the hydrodynamic representation of
a complex SF \cite{b40,b41,chavmatos}, i.e. $P(t)=p(t)$ (compare Eq. (\ref{b20})
with Eq. (38) of
\cite{b41}). This extends to
an arbitrary SF potential $V(|\varphi|^2)$ the
result obtained in \cite{b40} for a quartic potential (we note that this
equivalence is not true for a spatially inhomogeneous SF and for a
homogeneous SF outside of the fast oscillation regime).

On the other
hand, Eqs. 
(\ref{b19}) and (\ref{b20}) define the EOS $P(\epsilon)$ of the
SF in parametric form for an arbitrary potential. The EOS
parameter can be written as
\begin{eqnarray}
w=\frac{P}{\epsilon}=\frac{\rho
V'(\rho)-V(\rho)}{\rho c^2+V(\rho)+\rho V'(\rho)}.
\label{b22}
\end{eqnarray}
The Universe is accelerating ($w<-1/3$) when $4\rho
V'(\rho)-2V(\rho)<-\rho c^2$. Introducing the total potential $V_{\rm
tot}(\rho)=V(\rho)+\rho c^2/2$ (see Sec. \ref{sec_tp}), this condition can be
rewritten as $2\rho V_{\rm tot}'(\rho)<V_{\rm
tot}(\rho)$. The Universe is phantom ($w<-1$) when $2
V'(\rho)/c^2<-1$ or, equivalently, when $V_{\rm tot}'(\rho)<0$. However, this
condition is never realized in the fast oscillation regime because of the 
constraint imposed by Eq. (\ref{b12}).

For a given EOS $P(\epsilon)$, we can obtain the
potential $V(\rho)$ as follows (inverse problem \cite{bilic}). Eqs. (\ref{b19})
and (\ref{b20}) can be rewritten as $\epsilon=V_{\rm tot}(\rho)+\rho V_{\rm
tot}'(\rho)$ and $P=\rho V_{\rm tot}'(\rho)-V_{\rm tot}(\rho)$ leading to
$\epsilon-P=2V_{\rm tot}(\rho)$ and $\epsilon+P=2\rho V_{\rm tot}'(\rho)$. From
these equations, we obtain
\begin{eqnarray}
\int\frac{1-P'(\epsilon)}{\epsilon+P(\epsilon)}\, d\epsilon=\ln\rho,\qquad
V_{\rm tot}(\rho)=\frac{1}{2}[\epsilon-P(\epsilon)].
\label{bi1}
\end{eqnarray}
The first equation determines the relationship between $\rho$ and $\epsilon$.
The second relation then determines the total potential $V_{\rm tot}(\rho)$.

{\it Remark:} From
Eqs. (\ref{b19})
and (\ref{b20}), we can obtain the EOS $P(\epsilon)$. Solving the energy
equation (\ref{h4}) with this EOS, we can obtain $\epsilon(a)$. The relation
$\epsilon(a)$ can also be obtained from Eqs. (\ref{b13}) and (\ref{b19}). We can
easily check that the relations are the same. Indeed, from Eqs. (\ref{h4}),
(\ref{b19}) and (\ref{b20}) we obtain the differential equation
\begin{equation}
\left\lbrack c^2+2V'(\rho)+\rho V''(\rho)\right \rbrack
\frac{d\rho}{da}+\frac{3}{a}\left\lbrack \rho c^2+2\rho V'(\rho)\right
\rbrack=0.
\label{che1}
\end{equation}
This differential equation is equivalent to Eq. (\ref{b13}). This can be
 seen easily by taking the logarithmic derivative of Eq. (\ref{b13}) which leads
to Eq.
(\ref{che1}). This shows the
consistency of our approximations.

\subsection{The nonrelativistic limit}
\label{sec_nr}

In order to take the nonrelativistic limit $c\rightarrow +\infty$ of the
previous equations, we need to subtract the contribution of the rest mass
energy $mc^2$ of the SF. To that purpose, we make the Klein transformation
\begin{equation}
\varphi(t)=\frac{\hbar}{m}e^{-imc^2t/\hbar}\psi(t),
\label{nr1}
\end{equation}
where $\psi$ is the wave function such that $\rho=|\psi|^2$.
Substituting Eq. (\ref{nr1}) into Eq. (\ref{h1}) and taking the limit
$c\rightarrow +\infty$, we obtain the GP equation
 \begin{eqnarray}
i\hbar \frac{d\psi}{dt}+\frac{3}{2}i\hbar
H\psi=m\frac{dV}{d|\psi|^2}\psi
\label{nr2}
\end{eqnarray}
for a nonrelativistic spatially homogeneous SF.  On the other hand, in the
nonrelativistic limit, Eqs. (\ref{h2}) and (\ref{h3}) become
 \begin{eqnarray}
\epsilon\sim \rho c^2,\qquad 
P/c^2\rightarrow 0.
\label{nr3}
\end{eqnarray}
As explained previously, it is convenient to work in terms of hydrodynamic
variables. We write the wave
function under the Madelung form
\begin{equation}
\psi(t)=\sqrt{\rho(t)}e^{iS(t)/\hbar}
\label{nr4}
\end{equation}
and introduce the energy
\begin{equation}
E(t)=-\frac{dS}{dt}.
\label{nr5}
\end{equation}
Substituting Eq. (\ref{nr4}) into the GP equation (\ref{nr2}) and
separating
real and imaginary
parts, we get
\begin{eqnarray}
\frac{1}{\rho}\frac{d\rho}{dt}+\frac{3}{a}\frac{da}{dt}=0,
\label{nr6}
\end{eqnarray}
\begin{eqnarray}
E=m V'(\rho).
\label{nr7}
\end{eqnarray}
On the other hand, using Eq. (\ref{nr3}), we find that the
Friedmann equation (\ref{h5}) takes the form
\begin{eqnarray}
\frac{3H^2}{8\pi G}=\rho.
\label{nr8}
\end{eqnarray}
We also note that the energy equation (\ref{h4}) reduces to Eq. (\ref{nr6}). It
can be integrated into $\rho\propto 1/a^3$ which, together with Eq. (\ref{nr8}),
leads to the EdS solution. Equations (\ref{nr6})-(\ref{nr8}) can
also be
obtained from the general hydrodynamic
equations derived in  \cite{b40,b41,chavmatos} by considering the particular
case of a
spatially homogeneous SF in
the nonrelativistic limit $c\rightarrow +\infty$. 

Finally, comparing Eqs. (\ref{ic1b}), (\ref{nr1}) and (\ref{nr4}) we find that
$S_{\rm tot}=S-mc^2 t$ and  $E_{\rm tot}=E+mc^2$. Substituting this
decomposition into Eqs. (\ref{b4})-(\ref{b3})  and taking the limit
$c\rightarrow +\infty$ we recover  Eqs. (\ref{nr6})-(\ref{nr8}). We also find
that Eq. (\ref{b6}) reduces to 
\begin{eqnarray}
\rho=\frac{Qm}{a^3}.
\end{eqnarray}

{\it Remarks:} the hydrodynamic equations (\ref{b4})-(\ref{b3}) and
(\ref{nr6})-(\ref{nr8}) do not involve viscous
terms because they are
equivalent to the KG and GP equations. As a result, they describe a
superfluid.  We note that Eq. (\ref{b2}) for $S_{\rm tot}(t)$ or $E_{\rm
tot}(t)$ is necessary in the relativistic case in order to have a closed system
of equations (since $E_{\rm tot}$ appears explicitly in Eqs.
(\ref{b4}) and (\ref{b3})) while Eq. (\ref{nr7})
for $S(t)$ or $E(t)$   is
not strictly necessary in the nonrelativistic case (since $E$ does not appear in
Eq . (\ref{nr6}) and (\ref{nr8})).

\subsection{The quartic potential}
\label{sec_qua}

In the case where the SF describes a BEC at zero temperature, the
self-interaction potential can be written as
\begin{equation}
V(|\varphi|^2)=\frac{2\pi a_sm}{\hbar^2}|\varphi|^4,
\label{kge3}
\end{equation}
where $m$ is the mass of the bosons and $a_s$ is their scattering length (see
Appendix \ref{sec_sic} for other expressions of the self-interaction constant).
A repulsive self-interaction corresponds to $a_s>0$ and an attractive
self-interaction
corresponds to $a_s<0$. In the first case, $a_s$ may be interpreted as the
``effective radius'' of the bosons if we make an analogy with a classical hard
spheres gas.

In terms of the pseudo rest-mass density $\rho$ and wave function $\psi$, the
quartic potential (\ref{kge3}) can be rewritten as
\begin{eqnarray}
V(\rho)=\frac{2\pi
a_s\hbar^2}{m^3}\rho^2,   \qquad V(|\psi|^2)=\frac{2\pi
a_s\hbar^2}{m^3}|\psi|^4.
\label{kge30}
\end{eqnarray}
From to Eqs. (\ref{b20}) and (\ref{kge29}), we obtain
\begin{eqnarray}
P(\rho)=\frac{2\pi a_s\hbar^2}{m^3}\rho^2,\qquad 
c_s^2=\frac{4\pi a_s\hbar^2}{m^3}\rho.
\label{kge31}
\end{eqnarray}
The pressure law $P(\rho)$ corresponds to a polytropic EOS of index $\gamma=2$
(quadratic).

\section{The case of a quartic potential with a positive scattering
length}
\label{sec_pos}

From now on, we restrict ourselves to a SF with a quartic potential given by
Eq. (\ref{kge3}). We focus on the evolution of a homogeneous SF in the regime
where its oscillations are faster than the Hubble expansion. We first consider
the case of a SF with a positive scattering length $a_s\ge 0$ corresponding to a
repulsive self-interaction (the noninteracting case corresponds to $a_s=0$).
This is the most studied case in the literature. A very nice study has been done
by Li {\it et al.} \cite{b36}. Here, we complement their study and provide more
explicit analytical results.

\subsection{The basic equations}
\label{sec_posba}

The equations of the problem are 
\begin{eqnarray}
\rho
\sqrt{1+\frac{8\pi
a_s\hbar^2}{m^3c^2}\rho}=\frac{Qm}{a^3},
\label{posba1}
\end{eqnarray}
\begin{eqnarray}
\frac{3H^2}{8\pi
G}=\rho\left (1+\frac{6\pi
a_s\hbar^2}{m^3c^2}\rho\right ),
\label{posba2}
\end{eqnarray}
\begin{eqnarray}
\epsilon=\rho c^2\left (1+\frac{6\pi
a_s\hbar^2}{m^3c^2}\rho\right ),
\label{posba3}
\end{eqnarray}
\begin{eqnarray}
P=\frac{2\pi a_s\hbar^2}{m^3}\rho^2,
\label{posba4}
\end{eqnarray}
\begin{eqnarray}
w=\frac{\frac{2\pi a_s\hbar^2}{m^3c^2}\rho}{1+\frac{6\pi
a_s\hbar^2}{m^3c^2}\rho},
\label{posba4b}
\end{eqnarray}
\begin{eqnarray}
E_{\rm tot}=mc^2\sqrt{1+\frac{8\pi
a_s\hbar^2}{m^3c^2}\rho}.
\label{posba5}
\end{eqnarray}
Equation (\ref{posba3}) gives the relation between the energy density
$\epsilon$ and the pseudo rest-mass density $\rho$. This is a second degree
equation for $\rho$. The only physically acceptable solution (the one that is
positive) is
\begin{eqnarray}
\rho=\frac{m^3c^2}{12\pi a_s\hbar^2}\left (\sqrt{1+\frac{24\pi
a_s\hbar^2}{m^3c^4}\epsilon}-1\right ).
\label{posba6}
\end{eqnarray}
Combining Eqs. (\ref{posba4}) and (\ref{posba6}), we obtain the
EOS \cite{mul,pv,b36}:
\begin{eqnarray}
P=\frac{m^3c^4}{72\pi a_s\hbar^2}\left (\sqrt{1+\frac{24\pi
a_s\hbar^2}{m^3c^4}\epsilon}-1\right )^2.
\label{posba7}
\end{eqnarray}
It coincides with the EOS obtained by Colpi {\it et al.}
\cite{colpi} in the context of boson stars (see also
\cite{b45}). For a noninteracting SF ($a_s=0$), Eq.
(\ref{posba7}) reduces to $P=0$ meaning that a noninteracting SF behaves as
pressureless matter.

\subsection{The evolution of the parameters with the scale factor $a$}
\label{sec_posea}

The evolution of the pseudo rest-mass density $\rho$ with the scale factor
$a$ is plotted in Fig. \ref{arhopos} (in the Figures, unless
otherwise specified, we
use the dimensionless parameters defined in Appendix \ref{sec_dv}). It starts
from
$+\infty$ at $a=0$ and decreases to $0$ as $a\rightarrow +\infty$. For
$a\rightarrow 0$:
\begin{eqnarray}
\rho\sim \left (\frac{Q^2m^5c^2}{8\pi a_s\hbar^2}\right
)^{1/3}\frac{1}{a^2}.
\label{posea1}
\end{eqnarray}
For $a\rightarrow +\infty$:
\begin{eqnarray}
\rho\sim \frac{Qm}{a^3}. 
\label{posea2}
\end{eqnarray}

\begin{figure}[h]
\scalebox{0.33}{\includegraphics{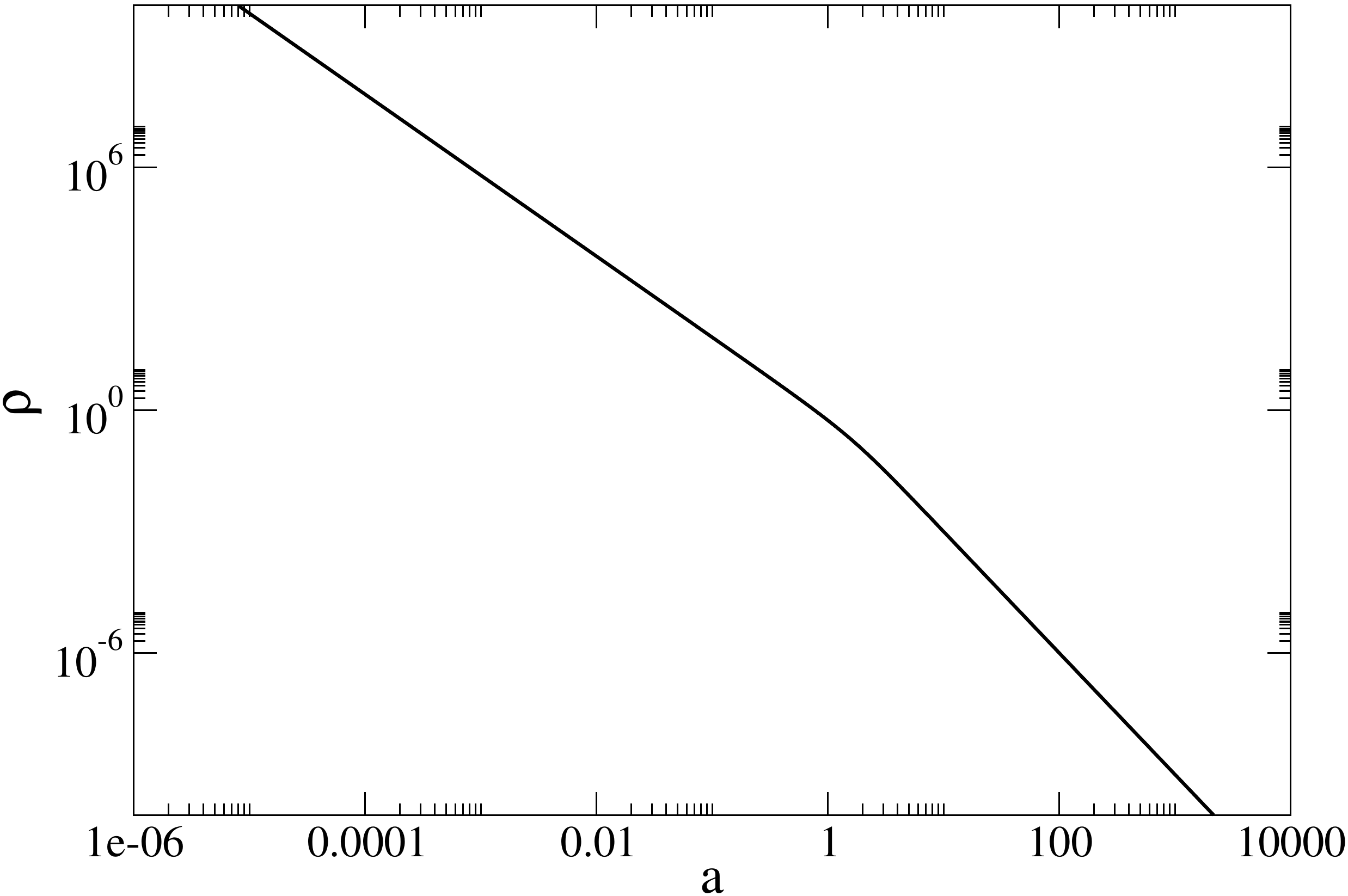}} 
\caption{Pseudo rest-mass density $\rho$ as a function of the scale factor $a$.}
\label{arhopos}
\end{figure}

\begin{figure}[h]
\scalebox{0.33}{\includegraphics{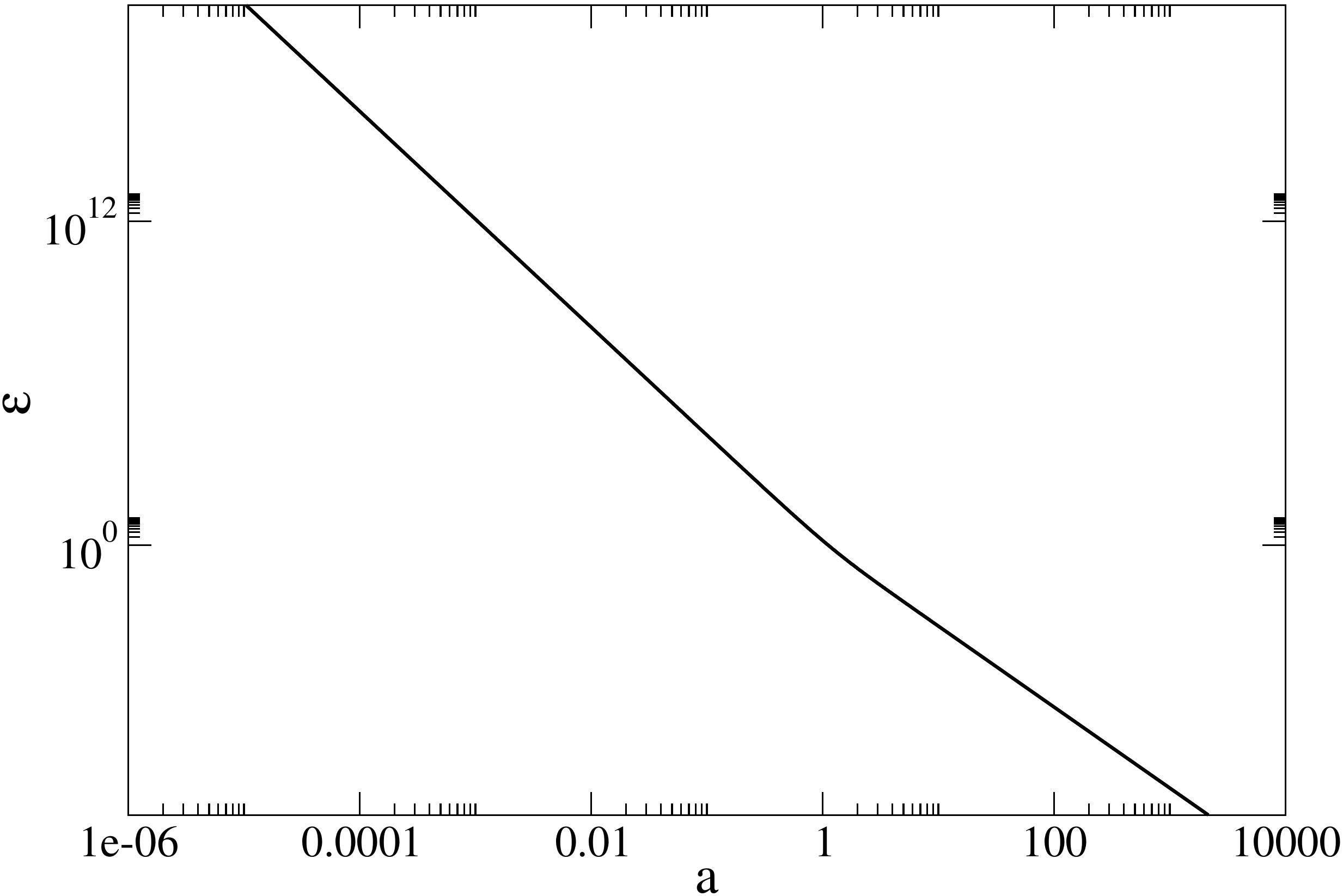}} 
\caption{Energy density $\epsilon$ as a function of the scale factor $a$.}
\label{aepspos}
\end{figure}

The evolution of the energy density $\epsilon$
with the scale factor $a$ is
plotted in Fig. \ref{aepspos}. It starts from
$+\infty$ at $a=0$ and decreases to $0$ as $a\rightarrow
+\infty$.  For
$a\rightarrow 0$:
\begin{eqnarray}
\epsilon\sim \frac{6\pi a_s
\hbar^2}{m^3}\rho^2\sim \frac{3}{2}\left ({Q^4 \pi
ma_s\hbar^2c^4}\right
)^{1/3}\frac{1}{a^4}.
\label{posea3}
\end{eqnarray}
For $a\rightarrow +\infty$:
\begin{eqnarray}
\epsilon\sim \rho c^2\sim \frac{Qmc^2}{a^3}.
\label{posea4}
\end{eqnarray}

\begin{figure}[h]
\scalebox{0.33}{\includegraphics{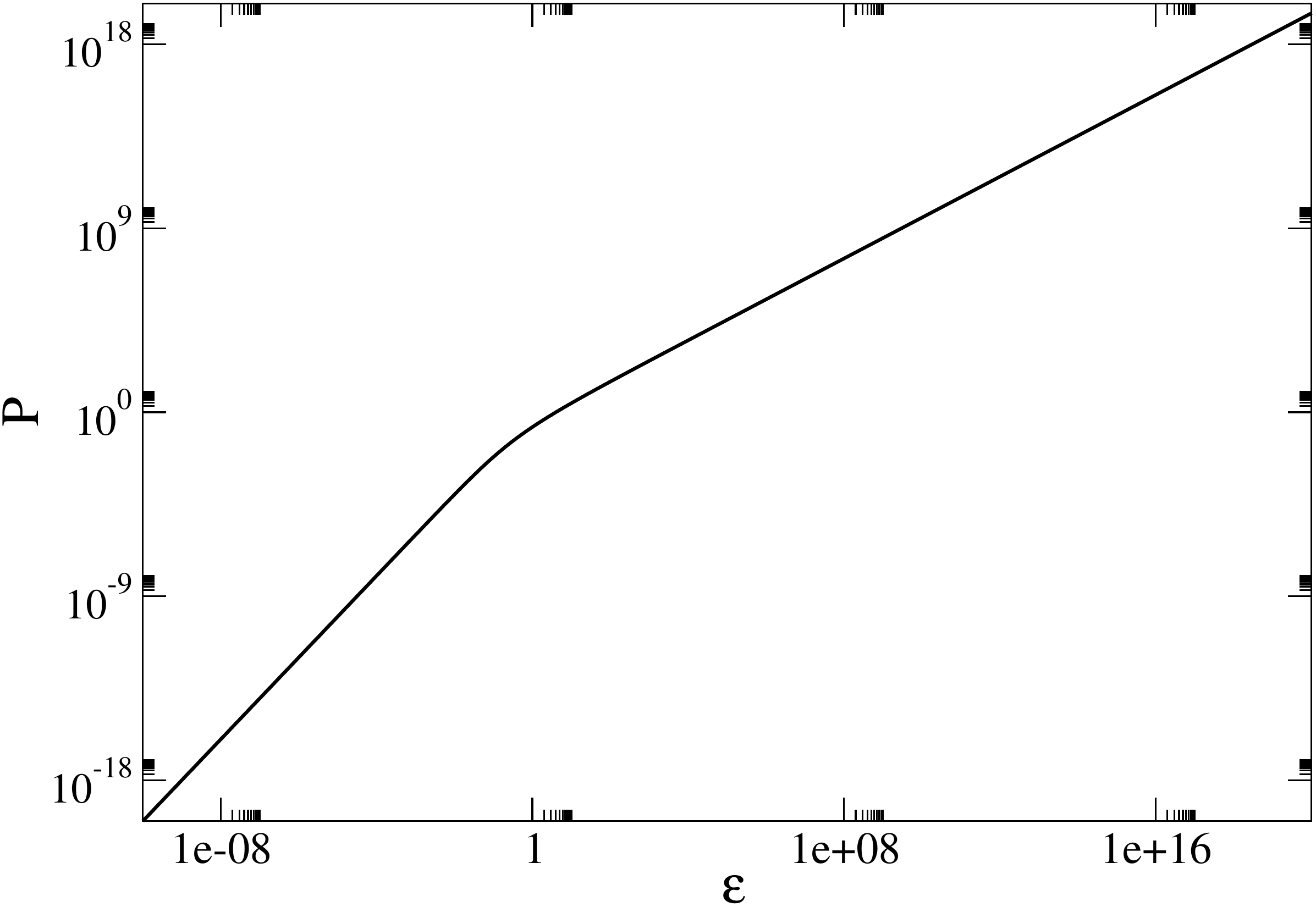}} 
\caption{Pressure $P$ as a function of the energy density
$\epsilon$.}
\label{epsPpos}
\end{figure}

The pressure is always positive. It starts from 
$+\infty$ at $a=0$ and decreases to $0$ as $a\rightarrow +\infty$. For
$a\rightarrow 0$:
\begin{eqnarray}
 P\sim \frac{1}{3}\epsilon\sim \frac{1}{2}\left ({Q^4 \pi
ma_s\hbar^2c^4}\right
)^{1/3}\frac{1}{a^4}.
\label{posea5}
\end{eqnarray}
For $a\rightarrow +\infty$:
\begin{eqnarray}
 P\sim \frac{2\pi a_s
\hbar^2}{m^3c^4}\epsilon^2 \sim \frac{2\pi a_s\hbar^2Q^2}{ma^6}\simeq 0.
\label{posea6}
\end{eqnarray}
The relationship
between
the
pressure and the energy density is
plotted in Fig. \ref{epsPpos}.

\begin{figure}[h]
\scalebox{0.33}{\includegraphics{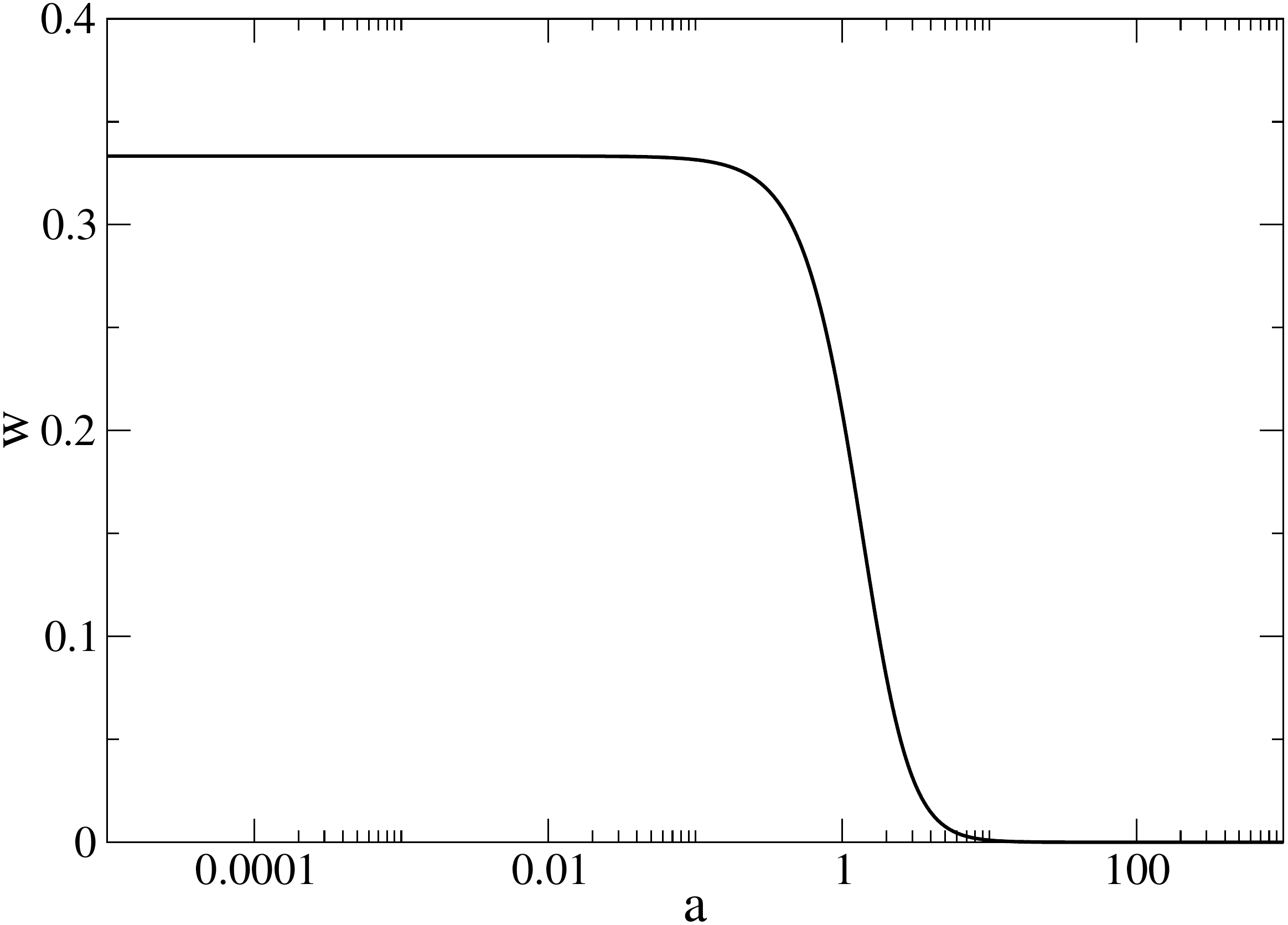}} 
\caption{EOS parameter $w$ as a function of the scale factor $a$.}
\label{awpos}
\end{figure}

The evolution of the EOS parameter $w=P/\epsilon$ with the scale
factor $a$ is plotted in Fig. \ref{awpos}. It  starts from
\begin{equation}
w_{i}=\frac{1}{3}
\label{posea7}
\end{equation}
when $a=0$ and
decreases to $0$ as $a\rightarrow +\infty$. For
$a\rightarrow 0$:
\begin{eqnarray}
w\simeq \frac{1}{3}-\frac{1}{6}\left (\frac{m^2c^2}{\pi a_s\hbar^2 Q}\right
)^{2/3}a^2.
\label{posea8}
\end{eqnarray}
For $a\rightarrow +\infty$:
\begin{eqnarray}
w\sim \frac{2\pi a_s
\hbar^2Q}{m^2c^2a^3}.
\label{posea9}
\end{eqnarray}

The total energy $E_{\rm tot}$ starts from $+\infty$ at
$a=0$ and
decreases up to $mc^2$ as $a\rightarrow +\infty$.  For
$a\rightarrow 0$: 
\begin{eqnarray}
 \frac{E_{\rm tot}}{mc^2}\sim
\left
(\frac{8\pi a_s\hbar^2 Q}{m^2 c^2}\right
)^{1/3}\frac{1}{a}.
\label{posea10}
\end{eqnarray}
For $a\rightarrow +\infty$:
\begin{eqnarray}
 \frac{E_{\rm tot}}{mc^2}\simeq 1+ \frac{4\pi a_s \hbar^2
Q}{m^2c^2a^3}.
\label{posea11}
\end{eqnarray}

\subsection{The temporal evolution of the parameters}
\label{sec_post}

\begin{figure}[h]
\scalebox{0.33}{\includegraphics{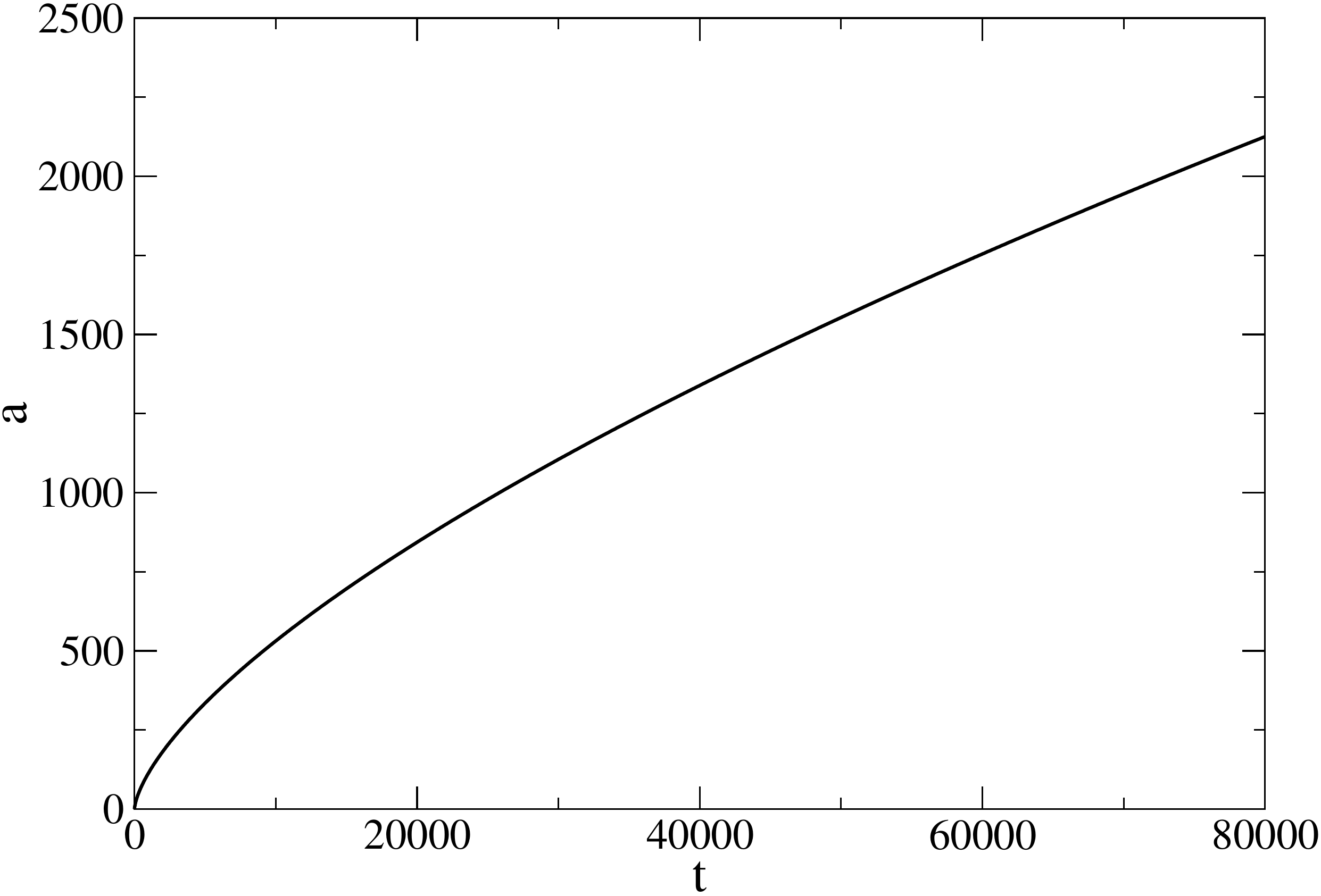}} 
\caption{Temporal evolution of the scale factor $a$.}
\label{tapos}
\end{figure}

In this section, we determine the temporal evolution of the parameters
assuming that the Universe contains only the SF. For a quartic potential with
$a_s\ge 0$, the differential equation (\ref{b16})
becomes
\begin{eqnarray}
\left (\frac{d\rho}{dt}\right )^2=24\pi G\rho^3\frac{\left (1+\frac{6\pi
a_s\hbar^2}{m^3c^2}\rho\right )\left (1+\frac{8\pi
a_s\hbar^2}{m^3c^2}\rho\right )^2}{\left (1+\frac{12\pi
a_s\hbar^2}{m^3c^2}\rho\right )^2}.\nonumber\\
\label{post1}
\end{eqnarray}
The solution of this differential equation which satisfies the
condition that $\rho\rightarrow +\infty$ as $t\rightarrow 0$  is
\begin{eqnarray}
\int_{\frac{2\pi
a_s\hbar^2}{m^3c^2}\rho}^{+\infty}\frac{(1+6x)\,
dx}{x^{3/2}(1+3x)^{1/2}(1+4x)}=\left (\frac{12 G m^3c^2}{a_s\hbar^2}\right
)^{1/2}t.\nonumber\\
\label{post2}
\end{eqnarray}
The integral can be computed analytically:
\begin{eqnarray}
\int\frac{(1+6x)\,
dx}{x^{3/2}(1+3x)^{1/2}(1+4x)}\nonumber\\
=4\tan^{-1}\left
(\sqrt{\frac{x}{1+3x}}\right )-2\sqrt{\frac{1+3x}{x}}.
\label{post3}
\end{eqnarray}
From these equations, we can obtain the temporal evolution of the
pseudo rest-mass density $\rho(t)$. Then, using Eqs.
(\ref{posba1})-(\ref{posba5}), we can obtain the temporal evolution of the
all the parameters. The temporal evolution of the scale factor $a$ is plotted in
Fig. \ref{tapos}. It starts
from $a=0$ at $t=0$ and increases to $+\infty$ as $t\rightarrow +\infty$. We do
not show the other curves because they can be easily deduced from Figs.
\ref{arhopos}, \ref{aepspos} and  \ref{awpos} 
since $a$ is a monotonic function of time. However, we provide below the
asymptotic behaviors of all the parameters.

For $t\rightarrow 0$:
\begin{eqnarray}
a\sim 2\left (\frac{\pi^4 G^3 Q^4 m a_s \hbar^2}{c^2}\right )^{1/12}t^{1/2},
\label{post5}
\end{eqnarray}
\begin{eqnarray}
\rho\sim \frac{m^{3/2}c}{8\pi\hbar a_s^{1/2}G^{1/2}t},
\label{post7}
\end{eqnarray}
\begin{eqnarray}
\epsilon\sim \frac{3c^2}{32\pi G t^2}, 
\label{post9}
\end{eqnarray}
\begin{eqnarray}
P\sim \frac{c^2}{32\pi G t^2}, 
\label{post11}
\end{eqnarray}
\begin{eqnarray}
w\simeq \frac{1}{3}-\frac{2m^{3/2}G^{1/2}c}{3\hbar a_s^{1/2}}t,
\label{post13}
\end{eqnarray}
\begin{eqnarray}
\frac{E_{\rm tot}}{mc^2}\sim \left (\frac{a_s \hbar^2}{m^3 G c^2}\right
)^{1/4}\frac{1}{t^{1/2}}.
\label{post15}
\end{eqnarray}

For $t\rightarrow +\infty$:
\begin{eqnarray}
 a\sim  (6\pi
GQmt^2)^{1/3},
\label{post6}
\end{eqnarray}
\begin{eqnarray}
\rho\sim \frac{1}{6\pi G t^2},
\label{post8}
\end{eqnarray}
\begin{eqnarray}
\epsilon\sim \frac{c^2}{6\pi G t^2},
\label{post10}
\end{eqnarray}
\begin{eqnarray}
P\sim \frac{a_s\hbar^2}{18\pi G^2 m^3 t^4},
\label{post12}
\end{eqnarray}
\begin{eqnarray}
w\sim \frac{a_s\hbar^2}{3Gm^3c^2t^2},
\label{post14}
\end{eqnarray}
\begin{eqnarray}
\frac{E_{\rm tot}}{mc^2}\simeq 1+\frac{2a_s \hbar^2}{3m^3 G c^2 t^2}.
\label{post16}
\end{eqnarray}

\subsection{The different eras}
\label{sec_erapos}

In the fast oscillation regime, a SF with a repulsive self-interaction
($a_s\ge 0$)
undergoes two distinct eras. For $a\rightarrow 0$, the EOS
(\ref{posba7}) reduces to Eq. (\ref{posea5}) so the SF behaves as radiation. 
The scale factor increases like $a\propto t^{1/2}$. For $a\rightarrow +\infty$,
the EOS
(\ref{posba7}) reduces to Eq. (\ref{posea6}) so the SF
behaves essentially as pressureless matter (dust) like in the EdS
model.\footnote{The pressure of the SF is nonzero but since $P\propto
\epsilon^2\ll
\epsilon$ for $\epsilon\rightarrow 0$, everything happens
in the cosmological Friedmann equations (\ref{h4}) and (\ref{h5}) describing the
large scales as if the Universe were pressureless.
In particular, Eq. (\ref{h4}) implies $\epsilon\propto a^{-3}$ for $a\rightarrow
+\infty$ as when $P=0$. However, the 
nonzero pressure of the SF is important at small scales, i.e. at the scale of
dark matter
halos, because it can prevent singularities and
avoid the cusp problem and the
missing satellite problem as discussed in the Introduction (see also Appendix
\ref{sec_mas}).} The scale factor
increases like $a\propto t^{2/3}$.
 Therefore, the SF undergoes a radiationlike era
($w=1/3$)  followed by a matterlike era ($w=0$). Since $w>-1/3$, the Universe
is always decelerating.
As emphasized by Li {\it et al.} \cite{b36}, the radiationlike era is due
to the self-interaction of the SF ($a_s\neq 0$). There is no such phase for a
noninteracting SF ($a_s=0$). This remark will be made more precise in Sec.
\ref{sec_valp}. On the other hand, if we identify the SF as the source of DM, it
is possible to determine its charge $Q$ by considering its asymptotic behavior
in the matterlike era. It is given by Eq. (\ref{mrf2}) of Appendix
\ref{sec_mr}.

In conclusion, a SF with a repulsive self-interaction
behaves at early times as radiation and at late times as dust. We can estimate
the
transition between the radiationlike era and the
matterlike era of the SF as follows. First of all, using Eqs.
(\ref{posba1}) and (\ref{posba4b}), we find that the scale factor
corresponding to a value $w$ of the EOS parameter is
\begin{eqnarray}
a=\left (\frac{2\pi a_s \hbar^2 Q}{m^2c^2}\right
)^{1/3}\frac{(1-3w)^{1/2}}{w^{1/3}(1+w)^{1/6}}.
\label{differa1}
\end{eqnarray}
Interestingly, this equation provides an analytical expression of the function
$a(w)$, the inverse of the function $w(a)$ plotted in Fig. \ref{awpos}. If
we consider that the transition between the radiationlike era and the matterlike
era of the SF corresponds to $w_t=1/6$,\footnote{This value is obtained by
analogy with the standard  model (see Appendix \ref{sec_mr}). Since $P_{\rm
m}=0$ and $P_{\rm r}=\epsilon_{\rm r}/3$, the EOS
parameter of the standard model (neglecting here dark energy) is $w=P_{\rm
r}/(\epsilon_{\rm r}+\epsilon_{\rm m})=\epsilon_{\rm
r}/[3(\epsilon_{\rm
r}+\epsilon_{\rm m})]$. At the radiation-matter equality ($\epsilon_{\rm
r}=\epsilon_{\rm m}$), we get $w=1/6$. This transition value is also
the arithmetic mean of $w=1/3$ (radiation) and $w=0$ (dust).} we obtain
\begin{eqnarray}
a_t=\frac{\sqrt{3}}{7^{1/6}}\left (\frac{2\pi a_s \hbar^2 Q}{m^2c^2}\right
)^{1/3}.
\label{differa2}
\end{eqnarray}
This corresponds to $\epsilon_t=2\rho_t c^2=m^3c^4/3\pi a_s\hbar^2$. In order to
make numerical applications here and in the following sections, it
is convenient to introduce the reference scale factor $a_*$ defined in
Appendix \ref{sec_dv}. Using the
expression (\ref{mrf2}) of the charge of the SF, we get
\begin{eqnarray}
a_*\equiv \left (\frac{2\pi |a_s|\hbar^2 Q}{m^2c^2}\right
)^{1/3}=\left (\frac{2\pi |a_s|\hbar^2 \Omega_{\rm
dm,0}\epsilon_0}{m^3c^4}\right
)^{1/3}\nonumber\\
=6.76\times 10^{-7}\, \left(\frac{|a_s|}{\rm fm}
\right)^{ 1/3 } \frac{{\rm eV}/c^2}{m}.\qquad
\label{astar}
\end{eqnarray}
Therefore, $a_t=(\sqrt{3}/7^{1/6})a_*$. According to Eq. (\ref{astar}),
we note that $a_t$ depends only on
the ratio
$a_s/m^3$ (see Sec. \ref{sec_rasm}). For a SF with a ratio $a_s/m^3$
given by Eq. (\ref{mas5}), we get $a_t=1.26\times 10^{-5}$. For a SF with a
ratio $a_s/m^3$ given by Eq. (\ref{mas8}), we get $a_t=1.35\times
10^{-5}$. This analytical result is in good agreement with the
numerical result of Li {\it et al.} \cite{b36} (see their Fig. 1).

\subsection{Validity of the fast oscillation regime}
\label{sec_valp}

The previous results are valid in the fast oscillation regime $\omega\gg H$. In
this section, we determine the domain of validity of this regime.

\begin{figure}[h]
\scalebox{0.33}{\includegraphics{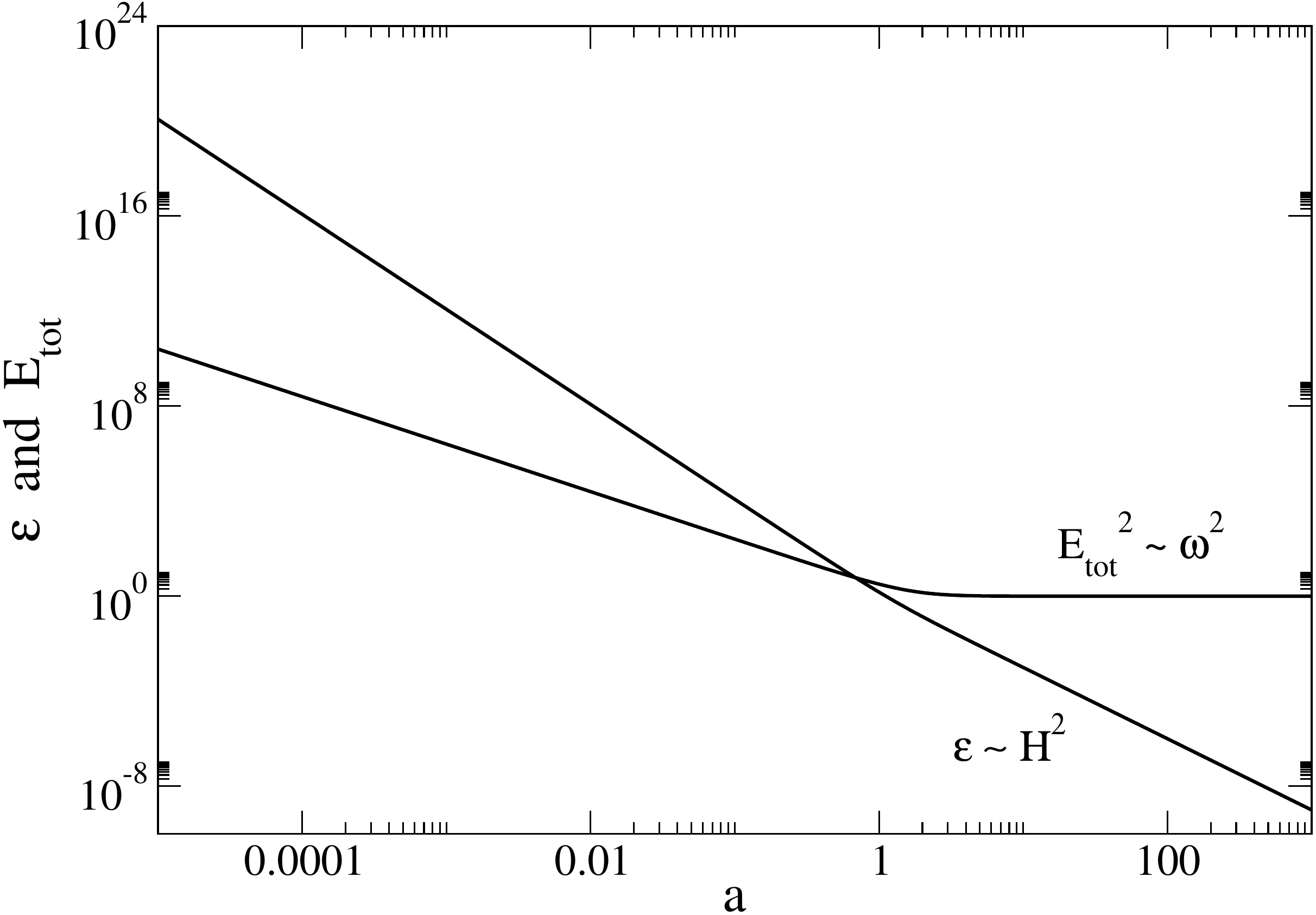}} 
\caption{Graphical construction determining the validity of the fast
oscillation regime. The transition scale $a_v$ corresponds to the intersection
of the curves $\sigma {\tilde E}_{\rm
tot}^2$ and   $\tilde\epsilon$.}
\label{validitypos}
\end{figure}

\begin{figure}[h]
\scalebox{0.33}{\includegraphics{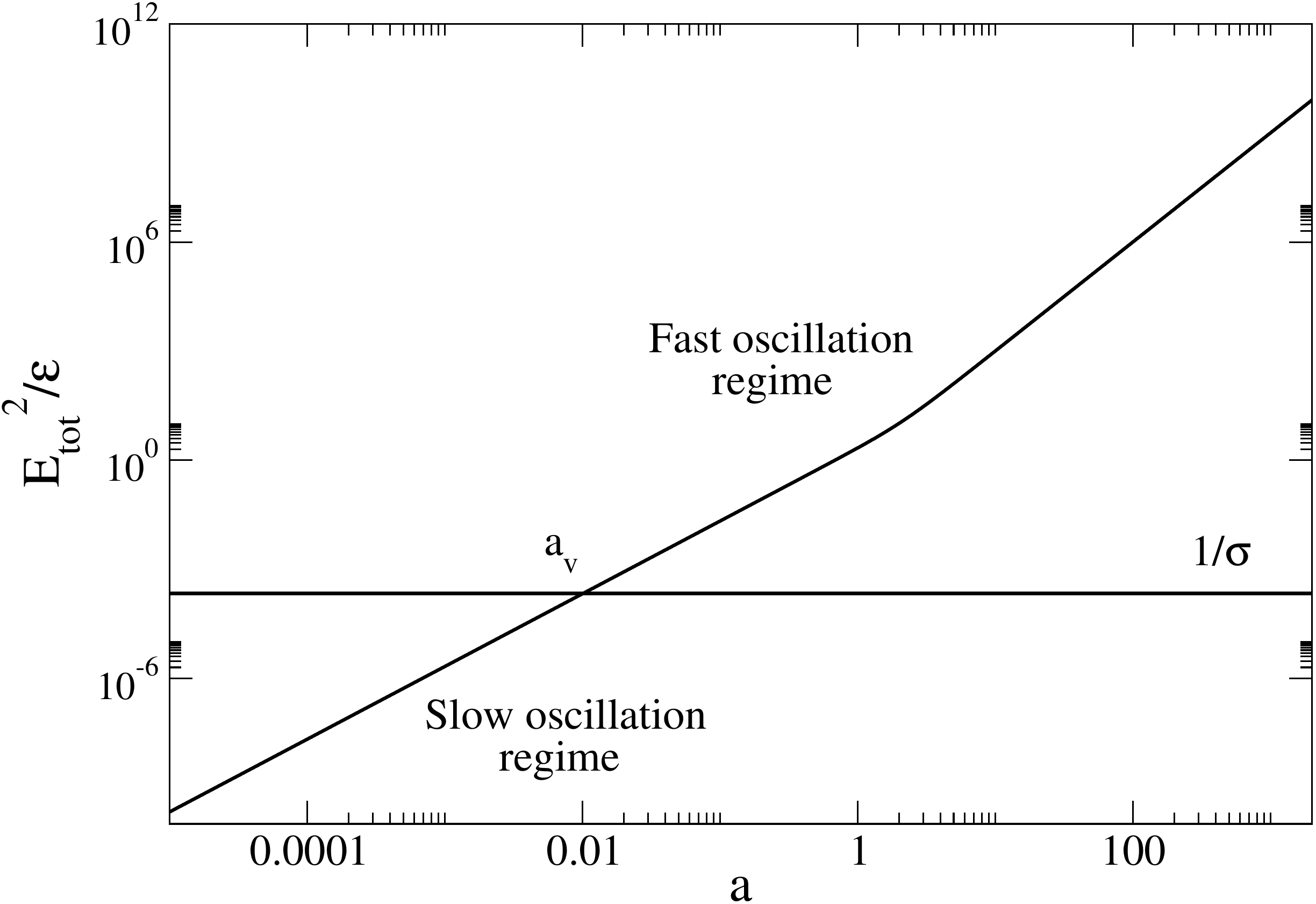}} 
\caption{Ratio $(\omega/H)^2$ as a function of the scale
factor $a$.}
\label{aEtot2sepspos}
\end{figure}

For a spatially homogeneous SF, the pulsation is given by
$\omega=d\theta/dt=(1/\hbar)dS_{\rm tot}/dt=-E_{\rm tot}/\hbar$ (see Appendix
\ref{sec_ext}) and the Hubble parameter is given by
$H^2=8\pi G\epsilon/3c^2$ (see Sec. \ref{sec_frid}). Therefore, the fast
oscillation regime
corresponds to
\begin{equation}
\frac{E_{\rm tot}^2}{\hbar^2}\gg \frac{8\pi G}{3c^2}\epsilon.
\label{valp1}
\end{equation}
Introducing the dimensionless variables of Appendix \ref{sec_dv}, this
condition can be
rewritten as
\begin{equation}
{\tilde E}_{\rm tot}^2\gg {\tilde\epsilon}/\sigma,
\label{valp2}
\end{equation}
where
\begin{equation}
\sigma=\frac{3 a_s c^2}{4Gm}
\label{valp3}
\end{equation}
is a new dimensionless parameter that can be interpreted as the ratio
$\sigma=3a_s/2r_S$ between the effective Schwarzschild radius $r_S=2Gm/c^2$ 
of the bosons (see Sec. \ref{sec_sr})
and their scattering length $a_s$. Introducing proper normalizations, we get
\begin{equation}
\sigma=5.67\times 10^{47}\, \frac{a_s}{\rm fm}\frac{{\rm
eV}/c^2}{m}. 
\label{valp4}
\end{equation}
The dimensionless variables ${\tilde E}_{\rm
tot}^2$ and $\tilde\epsilon$ are plotted as a function of $\tilde a$ in Fig.
\ref{validitypos}. Their ratio ${\tilde E}_{\rm tot}^2/\tilde\epsilon$ is
plotted  as a function of $\tilde a$ in Fig. \ref{aEtot2sepspos}. The
intersection of this curve with the
line ${\tilde E}_{\rm
tot}^2/\tilde\epsilon=1/\sigma$ determines the domain of validity of the  fast
oscillation regime.

Combining Eqs. (\ref{posba1}), (\ref{posba3}) and
(\ref{posba5}), we find that the fast oscillation regime is valid for $a\gg a_v$
with
\begin{eqnarray}
a_v= \left (\frac{2\pi a_s\hbar^2 Q}{m^2c^2}\right
)^{1/3}\, f\left (\frac{3 a_s c^2}{4 Gm}\right ),
\label{valp5}
\end{eqnarray}
where the function $f(\sigma)$ is defined by
\begin{eqnarray}
f(\sigma)=\frac{1}{r^{1/3}(1+4r)^{1/6}}
\label{valp6}
\end{eqnarray}
with
\begin{eqnarray}
r=\frac{4\sigma-1+\sqrt{(4\sigma-1)^2+12\sigma}}{6}.
\label{valp7}
\end{eqnarray}
For $\sigma\rightarrow 0$:
\begin{eqnarray}
f(\sigma)\sim \frac{1}{\sigma^{1/3}}.
\label{valp9}
\end{eqnarray}
For $\sigma\rightarrow +\infty$:
\begin{eqnarray}
f(\sigma)\sim \frac{\sqrt{3}}{2^{4/3}}\frac{1}{\sigma^{1/2}}.
\label{valp8}
\end{eqnarray}
These asymptotic results can be written more explicitly by restoring the
original variables. When $a_s=0$:
\begin{eqnarray}
a_v(0)=\left (\frac{8\pi G Q\hbar^2}{3mc^4}\right
)^{1/3}.
\label{valp11}
\end{eqnarray}
This correponds to $\epsilon_v(0)=\rho_v(0)c^2=3m^2c^6/8\pi G\hbar^2$.
Using the expression of the
charge given by Eq. (\ref{mrf2}), and introducing proper normalizations, we
obtain 
\begin{eqnarray}
a_v(0)\simeq 8.17\times 10^{-23} \left
(\frac{{\rm
eV}/c^2}{m}\right )^{2/3}.
\label{valp15}
\end{eqnarray}
This value corresponds to the begining of the fast
oscillation regime  in the
noninteracting case ($\omega=mc^2/\hbar\gg H$). When $a_s\gg r_S$:
\begin{eqnarray}
a_v\sim \left (\frac{\pi^2G^3
\hbar^4 Q^2}{a_s m c^{10}}\right
)^{1/6}.
\label{valp10}
\end{eqnarray}
Using the expression of the
charge given by Eq. (\ref{mrf2}), and introducing proper normalizations, we
obtain 
\begin{eqnarray}
a_v\simeq 6.17\times 10^{-31} \left (\frac{\rm fm}{a_s}\right )^{1/6}\left
(\frac{{\rm
eV}/c^2}{m}\right )^{1/2}.
\label{valp14}
\end{eqnarray}
This value corresponds to the begining of the fast oscillation regime in the
strongly self-interacting case.

For $a\gg a_v$, we are in the fast oscillation regime in which the SF behaves
successively as radiation and matter. For $a\ll a_v$, we are in the slow
oscillation regime in which the SF behaves as stiff matter. Therefore, $a_v$
marks the end of the stiff matter era (see Appendix \ref{sec_stiff}). A complex
SF generically
undergoes three successive eras: a stiff matter era for $a<a_v$, a
radiationlike era for $a_v<a<a_t$, and a matterlike era for $a>a_t$. The
transition scales $a_v$ and $a_t$ are given analytically by Eqs. (\ref{valp5})
and  (\ref{differa2}) respectively. Actually, the radiationlike era only exists
if $a_t>a_v$. This corresponds to
$f(\sigma)<\sqrt{3}/7^{1/6}$ leading to the condition
\begin{equation}
\sigma=\frac{3 a_s c^2}{4Gm}>\frac{2}{7},\quad  {\rm i.e.},\quad 
a_s>\frac{8}{21}\frac{Gm}{c^2}=\frac{4}{21}r_S.
\end{equation}
When $a_s<(4/21)r_S$, the SF undergoes only two successive eras:
a stiff matter era for $a<a_v$ and a
matterlike era for $a>a_v$. There is no radiationlike era even though the SF
is self-interacting. This generalizes the result of Li {\it et al.} \cite{b36} 
according to which a noninteracting SF ($a_s=0$) does not
present a radiationlike era. This result remains true as long as
$a_s<(4/21)r_S$. In this regime, the transition scale
$a_v$ depends very weakly on the scattering length $a_s$ of the bosons  (see
below). In
the
noninteracting case ($a_s=0$), $a_v$ is given by Eq. (\ref{valp11}). We note
that the transition between the stiff matter era and the matterlike era happens
later with decreasing mass. This is in agreement with the observation of Li
{\it et al.} \cite{b36} but Eq. (\ref{valp11}) provides an explicit analytical
formula refining this statement. This formula, together with Eq. 
(\ref{mrf2}), displays a $m^{-2/3}$ scaling for $a_v(0)$.

\subsection{Phase diagram}
\label{sec_pdp}

\begin{figure}[h]
\scalebox{0.33}{\includegraphics{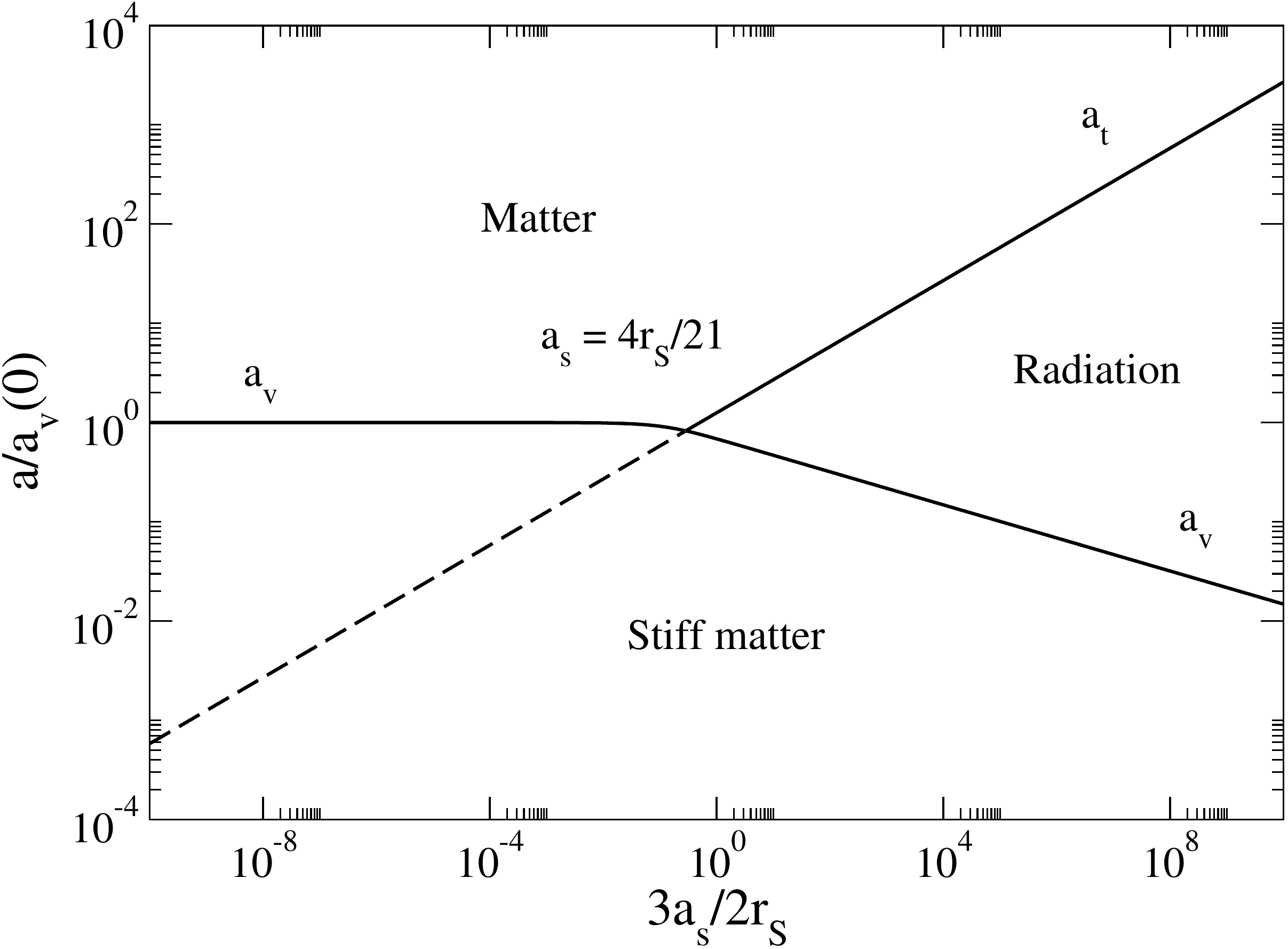}} 
\caption{Phase diagram showing the different eras of the SF during the
evolution of the Universe as a function of the scattering length of the bosons
in the case of a repulsive self-interaction.}
\label{fpd}
\end{figure}

We can represent the previous results on a phase diagram (see Fig.
\ref{fpd}) where we plot the transition scales $a_v$ and $a_t$ as a function of
the
scattering length $a_s$. To that purpose, it is convenient to normalize the
scale factor $a$ by the reference value $a_v(0)$ given by Eq. (\ref{valp11})
that is independent on $a_s$. The scattering length $a_s$ can be normalized by
the effective Schwarzschild radius $r_S$ using the parameter
$\sigma=3a_s/2r_S$
defined by
Eq. (\ref{valp3}). With
these normalizations, the transition scale $a_v$ between the slow and fast
oscillation regimes is given by 
\begin{eqnarray}
\frac{a_v}{a_{v}(0)}=f(\sigma)\sigma^{1/3}.
\label{pdp1}
\end{eqnarray}
For $\sigma=0$:
\begin{eqnarray}
\frac{a_v}{a_v(0)}=1.
\label{pdp2}
\end{eqnarray}
For $\sigma=2/7$:
\begin{eqnarray}
\frac{a_v}{a_v(0)}=\frac{\sqrt{3}}{7^{1/6}}\left (\frac{2}{7}\right
)^{1/3}\simeq 0.825.
\label{pdp2b}
\end{eqnarray}
For $\sigma\rightarrow +\infty$:
\begin{eqnarray}
\frac{a_v}{a_{v}(0)}\sim \frac{1}{4^{1/6}}\left (\frac{3}{4}\right
)^{1/2}\frac{1}{\sigma^{1/6}}.
\label{pdp3}
\end{eqnarray}
The transition scale $a_v$ starts from the value $a_v(0)$ given by Eq.
(\ref{valp11}) for $a_s=0$, decreases slowly up to
$a_v=(\sqrt{3}/7^{1/6})(2/7)^{1/3} a_v(0)$ when $a_s=(4/21)r_S$, and decreases
like $a_s^{-1/6}$ according to Eq.
(\ref{valp10}) for $a_s\gg r_S$.
Therefore, the domain of validity
of the fast oscillation regime is larger when the self-interaction is
stronger (the stiff matter era ends earlier). On the other hand, the transition
scale $a_t$ between the
radiationlike era and the matterlike era is given by
\begin{eqnarray}
\frac{a_t}{a_v(0)}= \frac{\sqrt{3}}{7^{1/6}}\sigma^{1/3}.
\label{valp12}
\end{eqnarray}
It starts from $0$ at $a_s=0$ and increases like $a_s^{1/3}$ according to Eq.
(\ref{differa2}). The transition scales $a_v$ and $a_t$
cross each other at
$a_s=(4/21)r_S$. At that point $a_v=a_t=(\sqrt{3}/7^{1/6})({2}/{7})^{1/3}
a_v(0)$.

We can now describe the phase diagram (see Fig.
\ref{fpd}). When $a_s\le (4/21)r_S$, the SF is in
the
stiff matter era for $0\le a\le a_v$ and in the
matterlike era for $a\ge a_v$.\footnote{For $a_s=0$, the stiff matter era may
be connected to the matterlike era by a short period of inflation with a
constant energy density (plateau) as argued in \cite{arbeycosmo}.} When $a_s\ge
(4/21)r_S$,
the SF is in
the
stiff matter era for $0\le a\le a_v$, in the radiationlike era for $a_v\le a\le
a_t$, and in the matterlike era for $a\ge a_t$. Therefore, when $a_s\le
(4/21)r_S$,
$a_v$ determines the transition scale between the stiff matter
era and the matterlike era. When $a_s \ge (4/21)r_S$, $a_v$ determines the
transition scale between the stiff matter era and the
radiationlike era.  We note that the radiationlike era
starts earlier and lasts longer as the self-interaction strength $a_s$
increases (the stiff matter era ends earlier and the matterlike era starts
later). This is in agreement with Fig. 1 of Li {\it et al.}
\cite{b36}.

Let us make a numerical application. We
first consider a
noninteracting SF. Using the value of $m$ given by Eq. (\ref{mas2}), we
obtain $a_v(0)=1.86\times 10^{-8}$. This is the transition scale between the
stiff matter era and the matterlike era. For a self-interacting SF, using the
values of ($m,a_s$) given by Eq. (\ref{mas6}), we obtain 
$\sigma=2.27\times 10^{45}$ and  $a_v=1.45\times 10^{-28}$. In that case, the
stiff matter era (if it really physically exists) ends very early. On the other
hand,
using the values
of ($m,a_s$) given by Eq. (\ref{mas7}), we obtain $\sigma=2.10\times 10^{10}$
and $a_v=5.14\times 10^{-11}$. In the two cases $\sigma\gg 2/7$
so the SF is deep in the strongly self-interacting regime and
there is a
radiationlike era. Therefore, $a_v$ determines the transition between
the stiff matter era and the radiationlike era (see Sec. \ref{sec_erapos} for
the determination of the transition scale between the radiationlike era and the
matterlike era). Our analytical result
$a_v=5.14\times 10^{-11}$ is in good agreement with the numerical result
obtained by Li
{\it et al.} \cite{b36} (see their Fig. 1).

{\it Particular cases:} When $m=a_s=0$, Eqs. (\ref{h2}) and (\ref{h3}) imply
$P=\epsilon$, so there is only a stiff matter era. When $m\neq 0$ and
$a_s=0$ (noninteracting case), there is only a stiff matter era and 
a pressureless matterlike era. The transition takes place at $a_v(0)$ given by
Eq.
(\ref{valp11}). It scales as $m^{-2/3}$ and tends to $+\infty$ when
$m\rightarrow 0$. When $m=0$ and
$a_s>0$ (massless case), there is only a stiff matter era and 
a radiationlike era. The transition takes place at
\begin{eqnarray}
a_v=\left (\frac{8\pi^3G^3\hbar^3Q^2}{\lambda c^9}\right )^{1/6}
\end{eqnarray}
obtained from Eq. (\ref{valp10}) by replacing $a_s$ by $\lambda$, using Eq.
(\ref{sic1}). It scales as $\lambda^{-1/6}$ (assuming $Q$ independent on
$\lambda$) and tends to $+\infty$ when $\lambda\rightarrow 0$. 

\subsection{Inflation era?}
\label{sec_ie}

It is well-known that a massive real SF with a quartic potential (or a
sufficiently flat
potential) undergoes a stiff matter era followed by an inflation era which is an
attractor of the KGE equations \cite{belinsky,piran}. Finally, it oscillates and
behaves in average as
radiation and pressureless matter. We may wonder
whether
a complex SF also experiences an inflation era. This would be the case if
$S_{\rm tot}\simeq 0$ in the early Universe because, in that case, it
would behave as a real SF. However,
because of the charge conservation constraint (\ref{b6}), the condition $S_{\rm
tot}\simeq 0$ implies $E_{\rm tot}\simeq Q\simeq 0$. Therefore, the
charge of the SF should be extremely small \cite{sj,js} which may be considered
artificial.

\subsection{Weakly and strongly self-interacting regimes}
\label{sec_sr}

We note that the phase diagram of Fig. \ref{fpd} depends on a dimensionless
control parameter $\sigma$ which is, up to a factor $3/2$, the ratio between the
scattering length $a_s$ of the bosons and their effective Schwarzschild
radius
\begin{eqnarray}
r_S=\frac{2Gm}{c^2}=2.65\times 10^{-48}\, \frac{m}{{\rm
eV}/c^2}\, {\rm fm}. 
\label{sr1}
\end{eqnarray}
The strongly self-interacting regime corresponds to $a_s\gg r_S$ and the weakly
self-interacting regime corresponds to $a_s\ll r_S$. In general $r_S$ is
very small. For example, for bosons with $m=2.92\times 10^{-22}\, {\rm eV}/c^2$
(see Appendix \ref{sec_mas}),
we have $r_S=7.74\times 10^{-70}\, {\rm fm}$. Therefore, even when $a_s\sim
10^{-68}\, {\rm fm}$ we are in the strongly self-interacting regime, not in the
weakly self-interacting regime, although this value of $a_s$ may seem very
``small'' at
first sight. 

We note that the effective
Schwarzschild radius of the bosons is much smaller than their
Compton
wavelength 
\begin{eqnarray}
\lambda_C=\frac{\hbar}{mc}=0.197\, \frac{{\rm
GeV}/c^2}{m}\, {\rm fm}
\label{sr2}
\end{eqnarray}
because their mass $m$ is much smaller than the Planck mass $M_P=(\hbar
c/G)^{1/2}=1.22\times 10^{19}\, {\rm GeV}/c^2$. For
example, for bosons with $m=2.92\times 10^{-22}\, {\rm eV}/c^2$,
we have $\lambda_C=6.75\times 10^{29}\, {\rm fm}$.

The condition of validity of the  strongly self-interacting
regime can also be expressed in terms of the variables introduced in Appendix
\ref{sec_sic}. Using the dimensionless self-interaction constant (\ref{sic1}),
we find
that the strongly self-interacting regime $a_s\gg r_S$ corresponds to
\begin{equation}
\frac{\lambda}{8\pi}\gg \frac{r_S}{\lambda_C}=2\left (\frac{m}{M_P}\right )^2.
\label{sr3}
\end{equation}
For bosons with $m=2.92\times 10^{-22}\, {\rm eV}/c^2$, we get
${\lambda}/{8\pi}\gg 1.15\times 10^{-99}$. Therefore, even when
${\lambda}/{8\pi}\sim 10^{-98}$ we are in the strongly self-interacting regime,
not in
the noninteracting regime $\lambda=0$ (!). A similar remark was made in Appendix
A.3 of \cite{cd} using different arguments. Therefore, it is
important to take the self-interaction of the bosons into account even if the
self-interaction constant seems to be very small. Many works
(see, e.g., \cite{marshrev,nature}) neglect the self-interaction of the bosons.
Their results may  substantially change if it is taken into
account.

Finally,  using the dimensional self-interaction constant
(\ref{sic3}), we
find that the strongly self-interacting regime $a_s\gg r_S$ corresponds
to
\begin{equation}
\lambda_s\gg \frac{8\pi G\hbar^2}{c^2}=1.295\times 10^{-69}\, {\rm
eV}\, {\rm
cm}^3.
\label{sr3b}
\end{equation}
We note that this bound is independent of the mass of the bosons.

According to the previous results, the dimensionless parameter $\sigma$ that
measures the strength of the
self-interaction for our problem can be written as
\begin{equation}
\sigma=\frac{3a_s}{2r_S}=\frac{3}{4}\frac{\lambda}{8\pi}\left
(\frac{M_P}{m}\right )^2=\frac{3\lambda_s c^2}{16\pi G\hbar^2}.
\label{sr4}
\end{equation}
The weakly self-interaction regime corresponds to $\sigma\ll 1$ and the
strongly self-interaction regime corresponds to $\sigma\gg 1$. The
dimensionless self-interaction constant $\lambda$ has a different meaning. We
can be in the strongly self-interaction regime $\sigma\gg 1$ for our problem
even when $\lambda\ll 1$ (weak
self-interaction in quantum field theory) due to
the large factor $(M_P/m)^2$ when $m\ll M_P$.

\subsection{The ratio $a_s/m^3$}
\label{sec_rasm}

Using the expression (\ref{mrf2}) of the charge $Q$ of the SF, we see that the
equations of the problem (\ref{posba1})-(\ref{posba5})
depend on the mass $m$ of the SF and on its scattering length $a_s$ only through
the ratio $a_s/m^3$.\footnote{This is because, as noted in
Sec. \ref{sec_rm}, the simplified equations (\ref{posba1})-(\ref{posba5}) are
obtained
in a
TF, or semiclassical, approximation where the quantum potential is
neglected ($\hbar=0$). By contrast, the scale $a_v$
marking the transition between the slow and fast oscillation regimes
is due to quantum mechanics ($\hbar\neq
0$) so it depends on the two individual parameters $a_s$ and $m$, or
equivalently $a_s/m^3$ and $a_s/m$ as is apparent on Eq. (\ref{valp5}) with Eq.
(\ref{mrf2}). To
see the effect of $\hbar$ in the equations, it is better to use the
parameter $\lambda_s/(mc^2)^2$ instead of $4\pi a_s\hbar^2/m^3c^4$ [see Eq.
(\ref{sic4})] because the appearance of $\hbar$ is the latter does not
correspond to the quantum potential and can be absorbed in the self-interaction
constant. Equations (\ref{posba1})-(\ref{posba5}) can then be written in terms
of $\lambda_s/(mc^2)^2$ only, in which $\hbar$ does not  appear (TF
approximation). By
contrast, $a_v$ given by Eq. (\ref{pdp1}),
depends on $m/\hbar$ [through Eqs. (\ref{valp11}) and (\ref{mrf2})] and on
$\lambda_s/\hbar^2$
[through Eq. (\ref{sr4})], in which $\hbar$ appears explicitly.} In
this section, we show how cosmological (large scale) observations  can constrain
the ratio $a_s/m^3$. We then compare these constraints with the value of
$a_s/m^3$ obtained from astrophysical (small scale) observations (see Appendix
\ref{sec_mas}).

Following Li {\it et al.} \cite{b36}, we impose that, at the epoch of
matter-radiation
equality, corresponding to the scale factor $a_{\rm eq}=2.95\times 10^{-4}$
(see Appendix \ref{sec_mr}), the
SF should be nonrelativistic, i.e., it should behave as pressureless
matter (CDM-like phase).
This is a constraint imposed by CMB.
This condition can be expressed by the inequality 
\begin{eqnarray}
w(a_{\rm eq})\le \chi,
\label{con1}
\end{eqnarray}
where $\chi$ is a small constant that Li {\it et al.} \cite{b36}  take
equal (somehow arbitrarily) to $\chi=10^{-3}$. In the matterlike era where $w\ll
1$, the function
$w(a)$ can be approximated by Eq. (\ref{posea9}) so that
\begin{eqnarray}
w(a_{\rm eq})\simeq \frac{2\pi a_s\hbar^2 Q}{m^2c^2a_{\rm eq}^3}.
\label{con2}
\end{eqnarray}
Using the expression (\ref{mrf2}) of the charge $Q$ of the SF and the
expression (\ref{mrs6}) of
the scale factor at the epoch of matter-radiation
equality, we obtain
\begin{eqnarray}
w(a_{\rm eq})\simeq \frac{2\pi a_s\hbar^2\epsilon_0  \Omega_{\rm
dm,0}\Omega_{\rm m,0}^3}{m^3c^4\Omega_{\rm r,0}^3}.
\label{con3}
\end{eqnarray}
Introducing proper normalizations, we get
\begin{eqnarray}
w(a_{\rm eq})=1.20\times 10^{-8}\, \frac{a_s}{\rm fm}\left (\frac{{\rm
eV}/c^2}{m}\right )^3.
\label{con7a}
\end{eqnarray}
The condition of Eq. (\ref{con1}) implies
\begin{eqnarray}
\frac{a_s}{m^3}\le  \frac{\chi c^4\Omega_{\rm r,0}^3}{2\pi
\hbar^2 \Omega_{\rm dm,0}\Omega_{\rm m,0}^3\epsilon_0},
\label{con4q}
\end{eqnarray}
i.e.,
\begin{eqnarray}
\frac{a_s}{\rm fm}\left (\frac{{\rm eV}/c^2}{m}\right )^3\le 8.31\times
10^4.
\label{con7}
\end{eqnarray}
Using the results of Appendix \ref{sec_sic}, we
 analytically recover the result
${\lambda_s}/{(mc^2)^2}\le 4.07\times 10^{-17}\, {\rm cm}^3/{\rm eV}$
of
Li {\it et al.} \cite{b36} [see their Eq. (38)].

Using astrophysical considerations related to the minimum
size of DM halos (Fornax) observed in the Universe  and interpreted as the
ground
state of a self-gravitating BEC (see Appendix \ref{sec_mas}), we find
that the ratio $a_s/m^3$ has the value $({a_s}/{\rm fm})(({{\rm
eV}/c^2})/{m})^3=3.28\times 10^3$ leading to $w(a_{\rm
eq})=3.94\times
10^{-5}$. These values are much smaller than the bounds implied by Eqs.
(\ref{con1}) and (\ref{con7}) for $\chi=10^{-3}$. These inequalities are
fulfilled by two orders of magnitude. The same remark applies to the values 
$({a_s}/{\rm fm})(({{\rm
eV}/c^2})/{m})^3=4.10\times 10^3$ and $w(a_{\rm
eq})=4.92\times 10^{-5}$ corresponding to the fiducial model of Li
{\it et al.} \cite{b36}.

Actually, we can relate the EOS parameter $w(a_{\rm eq})$ at the epoch of
matter-radiation
equality (cosmology/large scales) to the minimum size $R$ of the DM halos
observed in the Universe (astrophysics/small scales). Indeed, combining Eqs.
(\ref{con3}) and (\ref{mas3}), we get
\begin{eqnarray}
w(a_{\rm eq})=\frac{2\epsilon_0 GR^2\Omega_{\rm
dm,0}\Omega_{\rm m,0}^3}{\pi c^4\Omega_{r,0}^3}.
\label{con10}
\end{eqnarray}
Introducing
proper normalizations, Eq. (\ref{con10}) can be rewritten as
\begin{eqnarray}
w(a_{\rm eq})=3.94\times 10^{-5}\, \left (\frac{R}{\rm kpc}\right )^2.
\label{con11}
\end{eqnarray}
The condition $w(a_{\rm eq})\le 10^{-3}$ corresponds to a minimum halo size less
than $R=5.04\, {\rm kpc}$, a condition which is observationaly realized.
Inversely, taking $R=1\, {\rm kpc}$
(Fornax), we get $w(a_{\rm eq})=3.94\times 10^{-5}$. Taking
$R=1.12\, {\rm kpc}$, corresponding to the fiducial model of Li {\it et
al.} \cite{b36}, we
get $w(a_{\rm
eq})=4.92\times 10^{-5}$.

We can also compare the scale $a_t$ corresponding to the transition between the
radiationlike era and the matterlike era of the SF (see Sec. \ref{sec_valp})
with the scale $a_{\rm eq}$
corresponding to the matter-radiation equality. Combining Eqs.
(\ref{differa2}) and (\ref{con2}), we obtain 
\begin{eqnarray}
\frac{a_t}{a_{\rm eq}}=\frac{\sqrt{3}}{7^{1/6}}w(a_{\rm eq})^{1/3}.
\label{con12}
\end{eqnarray}
We first note that the condition $w(a_{\rm eq})\ll 1$ is equivalent to $a_t\ll
a_{\rm eq}$, i.e., the transition between the radiationlike era and the
matterlike era of the SF must take place long before the standard
radiation-matter
equality. Taking $w(a_{\rm eq})\le 10^{-3}$, we obtain the constraint
${a_t}/{a_{\rm
eq}}\le 0.125$. Taking $w(a_{\rm eq})=3.94\times 10^{-5}$, corresponding to the
model of Appendix \ref{sec_mas} (Fornax),  we obtain ${a_t}/{a_{\rm
eq}}=4.26\times 10^{-2}$. Taking $w(a_{\rm eq})=4.92\times
10^{-5}$, corresponding to the
fiducial model of Li {\it et al.} \cite{b36},
we obtain ${a_t}/{a_{\rm
eq}}=4.59\times 10^{-2}$. Since $a_t$ is much below $a_{\rm eq}$, we confirm
that,
at the radiation-matter equality epoch, the SF behaves as pressureless
matter, i.e., it is nonrelativistic.

Finally, we can obtain the value of the ratio $\mu=\epsilon_{\rm
SF}/\epsilon_{\rm r}$  between the radiation of the SF
and the standard radiation in the radiationlike era of the SF (see Appendix
\ref{sec_mr}).
Combining Eqs. (\ref{mrf5})
and (\ref{con3}), we obtain 
\begin{eqnarray}
\mu=\left (\frac{27}{16}\right )^{1/3}\frac{\Omega_{\rm
dm,0}}{\Omega_{\rm
m,0}}w(a_{\rm eq})^{1/3}.
\label{conn13}
\end{eqnarray}
We first note that the condition $w(a_{\rm eq})\ll 1$ is equivalent to
$\mu\ll 1$ i.e. the radiation of the SF must be much smaller than the standard
radiation. Taking  $w(a_{\rm eq})\le 10^{-3}$, we obtain the
constraint $\mu\le 0.100$. Therefore, the energy of
SF radiation must be about one order of magnitude smaller than the energy of
standard
radiation.  Taking $w(a_{\rm eq})=3.94\times 10^{-5}$, corresponding to the
model of Appendix \ref{sec_mas} (Fornax), we obtain $\mu=3.42\times 10^{-2}$.
Taking $w(a_{\rm
eq})=4.92\times
10^{-5}$, corresponding to the
fiducial model of Li {\it et al.} \cite{b36},
we obtain $\mu=3.68\times 10^{-2}$. This is in good agreement with the value
infered from their Fig. 3. We also find that the effective temperature of the
SF [see Eq. (\ref{mrf11})] is $T_{\rm eff}\simeq 0.43\, T$.

\begin{figure}[h]
\scalebox{0.33}{\includegraphics{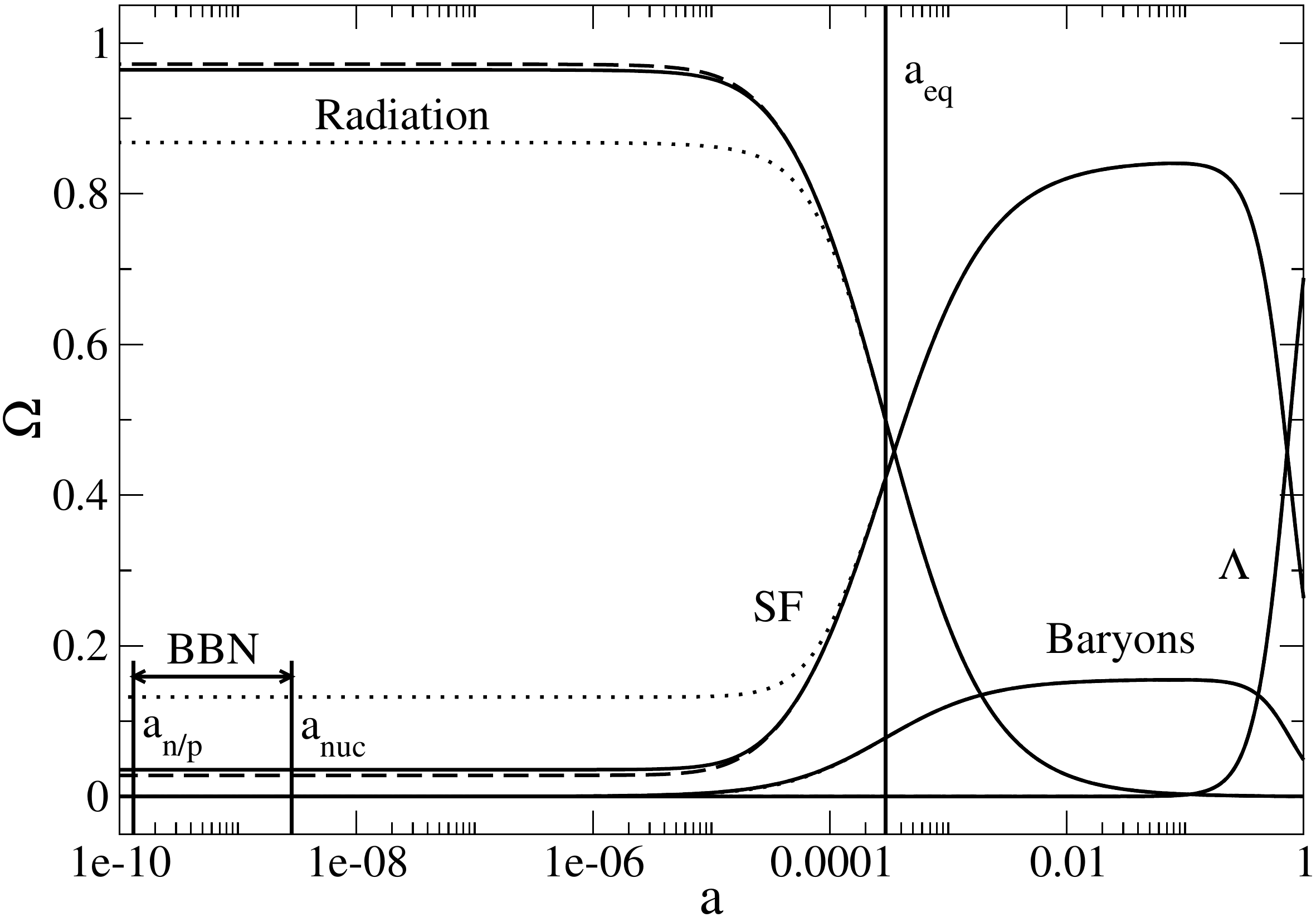}} 
\caption{Evolution of the fraction of the energy density of each component
(standard radiation, SF, baryons, DE) during the fast oscillation regime of the
SF (see Appendix \ref{sec_compl}). We have taken the values of $(m,a_s)$
corresponding to the fiducial model of Li {\it et al.} \cite{b36} (full lines).
The dashed and dotted lines correspond to the models leading to the bounds of
Eq. (\ref{conn15}). In this figure, $a$ is the true scale factor  (not $\tilde
a$). Figure 3 of Li {\it et al.} \cite{b36} is more general since
it takes into account the stiff matter era that prevails for
$a\lesssim 10^{-10}$.}
\label{universe}
\end{figure}

We can be more precise by introducing the
fraction of standard radiation $\Omega_{\rm r}=\epsilon_{\rm r}/\epsilon$ and
the fraction
of SF $\Omega_{\rm SF}=\epsilon_{\rm SF}/\epsilon$. During the radiationlike era
of the SF, i.e. for $a_v\le a\le a_t$, since $\epsilon_{\rm r}$ and
$\epsilon_{\rm
SF}$ both decay as $a^{-4}$, the fraction of SF has a  constant value
\begin{eqnarray}
\Omega_{\rm SF}({\rm plateau})=\frac{\epsilon_{\rm SF}}{\epsilon_{\rm
SF}+\epsilon_{\rm r}}=\frac{\mu}{\mu+1}.
\label{conn14}
\end{eqnarray}
For the fiducial model of Li {\it et al.} \cite{b36},
we obtain $\Omega_{\rm SF}({\rm plateau})=3.55\times 10^{-2}$ in good agreement
with the value infered from their Fig. 3. More generally, using their
constraint coming from BBN \cite{b36}:
\begin{eqnarray}
0.028\le \Omega_{\rm SF}({\rm plateau})\le 0.132,
\label{conn15}
\end{eqnarray}
we obtain $2.88\times 10^{-2}\le \mu\le 0.152$, giving [see Eq. (\ref{mrf6b})]:
\begin{eqnarray}
1.95\times 10^3\le \frac{a_s}{\rm fm}\left (\frac{{\rm eV}/c^2}{m}\right )^3\le
2.87\times 10^5.
\label{conn16}
\end{eqnarray}
Using the results of Appendix \ref{sec_sic}, we
 analytically recover the result
$9.54\times 10^{-19}{\rm
eV}^{-1}{\rm
cm}^3\le {\lambda_s}/{(mc^2)^2}\le 1.40 \times 10^{-16}{\rm
eV}^{-1}{\rm
cm}^3$ of
Li {\it et al.} \cite{b36} [see their Eq. (43)]. The cosmological constraints
corresponding to the bounds of Eq. (\ref{conn15}) are illustrated in Fig.
\ref{universe}. 

The approach of 
Li {\it et al.} \cite{b36} is  more general than ours because
they make precisely the matching between the slow oscillation regime and
the fast oscillation regime. This allows them to obtain precise bounds on 
$m$ and $a_s$ from BBN. We can, however, obtain a bound on $m$ by the following
(rough) argument. We require that the stiff matter era is over at
the
beginning of the neutron-proton ratio freeze-out $a_{\rm n/p}\sim
10^{-10}$, i.e., when BBN begins.
This leads to the constraint $a_v\le  a_{\rm n/p}$. Using Eq. (\ref{valp14}), we
obtain the condition
\begin{eqnarray}
\frac{m}{{\rm eV}/c^2}\ge \frac{6.17\times 10^{-31}}{a_{\rm n/p}}\left
(\frac{\rm fm}{a_s}\right )^{1/6}\left (\frac{m}{{\rm eV}/c^2}\right )^{1/2}.
\label{conn17}
\end{eqnarray}
Combining this equation with Eq. (\ref{conn16}), we obtain the constraint
\begin{eqnarray}
m\ge 1.75\times 10^{-21} \, {\rm eV}/{c^2},\quad ({\rm approx.})
\label{conn18}
\end{eqnarray}
which is very close to the exact constraint  $m\ge 2.4\times 10^{-21} \,
{\rm eV}/{c^2}$ obtained by Li {\it et al.} \cite{b36}.

In conclusion, we confirm the important results of Li {\it et
al.} \cite{b36}. An interest of our approach is that we obtain all
the relevant quantities analytically, so we can understand better 
where they come from. This also allows us to play more easily with the
parameters. The case of a SF at nonzero temperature ($T_{\rm SF}\neq 0$) will be
considered in a future work \cite{victor}.

{\it Remark:} As shown by Li {\it et al.} \cite{b36},
cosmological constraints from CMB and BBN exclude the possibility that the
bosons are noninteracting. Indeed, according to the inequality of Eq.
(\ref{conn16}) resulting from the constraint (\ref{conn15}) coming from BBN,
the SF must be self-interacting. If we ignore the constraint (\ref{conn15}),
take $a_s=0$, and impose the constraint $a_v(0)\le
a_{\rm n/p}$, we find from Eq. (\ref{valp15}) that $m\ge
7.38\times 10^{-19}\, {\rm eV/c^2}$. This
cosmological constraint is in contradiction with the astrophysical constraint
of Eq. (\ref{mas2}). This confirms that the SF must be self-interacting.

\section{The case of a quartic potential with a negative scattering
length}
\label{sec_neg}

We now consider the case of a SF with a negative scattering length $a_s<0$
corresponding to an attractive self-interaction. This is the case, for example, 
of the axion field that has been proposed as a dark matter candidate.

\subsection{The basic equations}
\label{sec_negba}

The equations of the problem are
\begin{eqnarray}
\rho
\sqrt{1-\frac{8\pi |a_s|\hbar^2}{m^3c^2}\rho}=\frac{Qm}{a^3},
\label{negba1}
\end{eqnarray}
\begin{eqnarray}
\frac{3H^2}{8\pi
G}=\rho\left (1-\frac{6\pi
|a_s|\hbar^2}{m^3c^2}\rho\right ),
\label{negba2}
\end{eqnarray}
\begin{eqnarray}
\epsilon=\rho c^2\left (1-\frac{6\pi
|a_s|\hbar^2}{m^3c^2}\rho\right ),
\label{negba3}
\end{eqnarray}
\begin{eqnarray}
P=-\frac{2\pi |a_s|\hbar^2}{m^3}\rho^2,
\label{negba4}
\end{eqnarray}
\begin{eqnarray}
w=\frac{-\frac{2\pi |a_s|\hbar^2}{m^3c^2}\rho}{1-\frac{6\pi
|a_s|\hbar^2}{m^3c^2}\rho},
\label{negba4b}
\end{eqnarray}
\begin{eqnarray}
E_{\rm tot}=mc^2\sqrt{1-\frac{8\pi
|a_s|\hbar^2}{m^3c^2}\rho}.
\label{negba5}
\end{eqnarray}
Solving Eq. (\ref{negba3}) for $\rho$, we find two acceptable solutions 
\begin{eqnarray}
\rho=\frac{m^3c^2}{12\pi |a_s|\hbar^2}\left (1\pm \sqrt{1-\frac{24\pi
|a_s|\hbar^2}{m^3c^4}\epsilon}\right ).
\label{negba6}
\end{eqnarray}
Substituting Eq. (\ref{negba6}) into Eq. (\ref{negba4}), we obtain  the EOS
\begin{eqnarray}
P=-\frac{m^3c^4}{72\pi |a_s|\hbar^2}\left (1\pm\sqrt{1-\frac{24\pi
|a_s|\hbar^2}{m^3c^4}\epsilon}\right )^2.
\label{negba7}
\end{eqnarray}

\subsection{The evolution of the parameters with the scale factor $a$}
\label{sec_negea}

\begin{figure}[h]
\scalebox{0.33}{\includegraphics{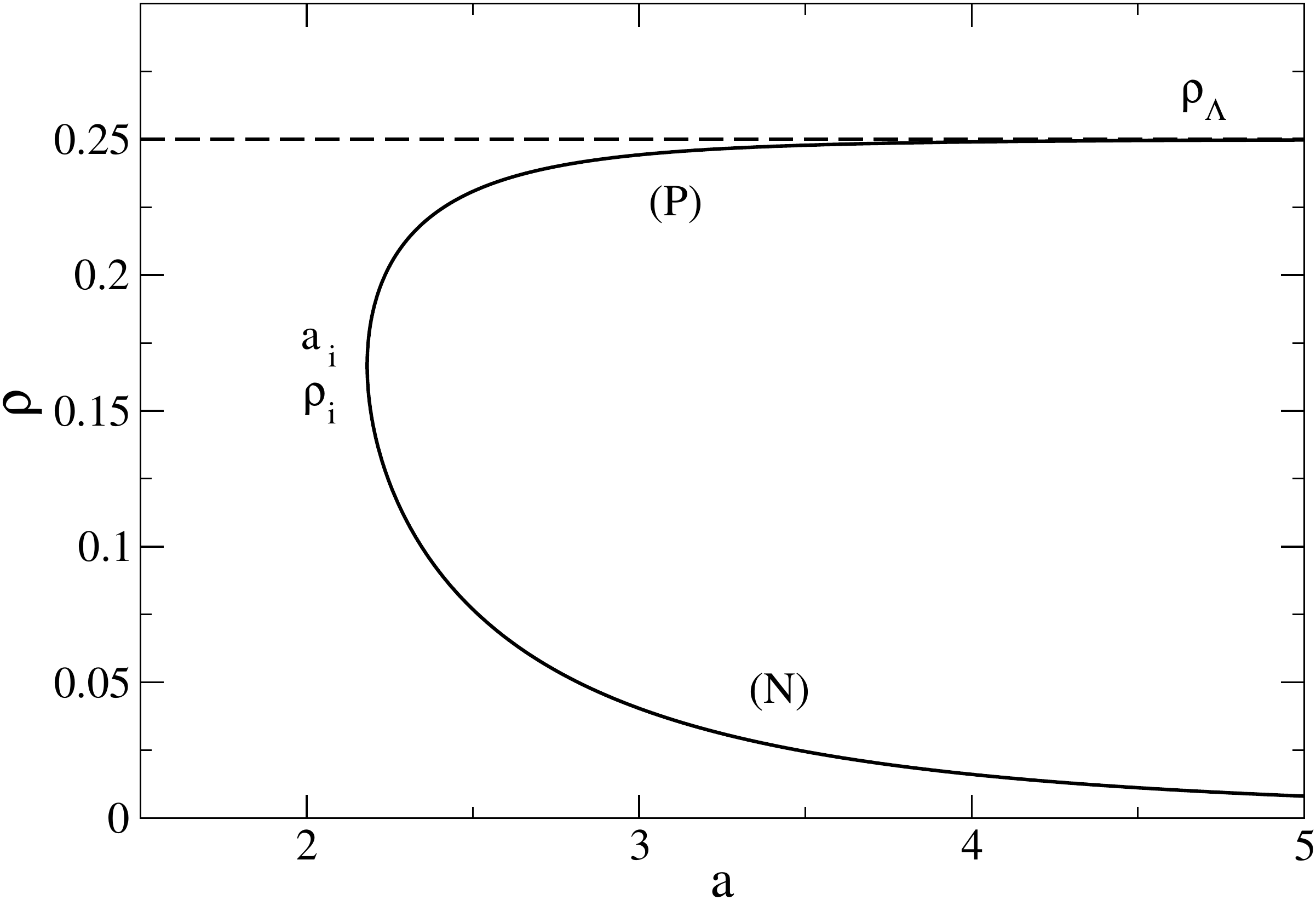}} 
\caption{Pseudo rest-mass density $\rho$ as a function of the scale factor
$a$.}
\label{arhoneg}
\end{figure}

The evolution of the pseudo rest-mass density $\rho$ with the scale
factor $a$ is plotted in Fig. \ref{arhoneg}. The curve $\rho(a)$ has two
branches. These two branches start from the same point corresponding to the
minimum scale factor 
\begin{equation}
a_{i}=\left(\frac{12\sqrt{3}\pi|a_s|\hbar^2Q}{m^2c^2}\right)^{1/3}
\label{negea1}
\end{equation}
and to the density
\begin{equation}
\rho_{i}=\frac{m^3c^2}{12\pi|a_s|\hbar^2}.
\label{negea2}
\end{equation}
For $a\rightarrow a_i$:
\begin{eqnarray}
\rho\simeq \rho_i\left\lbrack 1\pm\sqrt{2\left (\frac{a}{a_{i}}-1\right
)}\right\rbrack.
\label{negea3}
\end{eqnarray}
On the ``normal'' branch (sign $-$), the pseudo rest-mass density decreases as
the scale factor increases and asymptotically tends to $0$. 
For $a\rightarrow +\infty$:
\begin{eqnarray}
\rho\sim \frac{Qm}{a^3}.
\label{negea4}
\end{eqnarray}
On the ``peculiar'' branch (sign $+$), the rest-mass density increases as the
scale
factor increases\footnote{Although the pseudo
rest-mass density increases with the scale factor, the Universe is not phantom,
contrary to our claim in Ref. \cite{b40}. Indeed, as shown below, the energy
density $\epsilon$ always decreases with the scale factor $a$, even on the
peculiar branch.}
 and asymptotically tends to a maximum density
\begin{equation}
\rho_{\Lambda}=\frac{m^3c^2}{8\pi|a_s|\hbar^2}.
\label{negea5}
\end{equation}
For $a\rightarrow +\infty$:
\begin{eqnarray}
\rho\simeq \rho_{\Lambda} \left\lbrack 1-\left (\frac{8\pi Q
|a_s|\hbar^2}{m^2 c^2 a^3}\right )^2\right\rbrack.
\label{negea6}
\end{eqnarray}

\begin{figure}
\scalebox{0.33}{\includegraphics{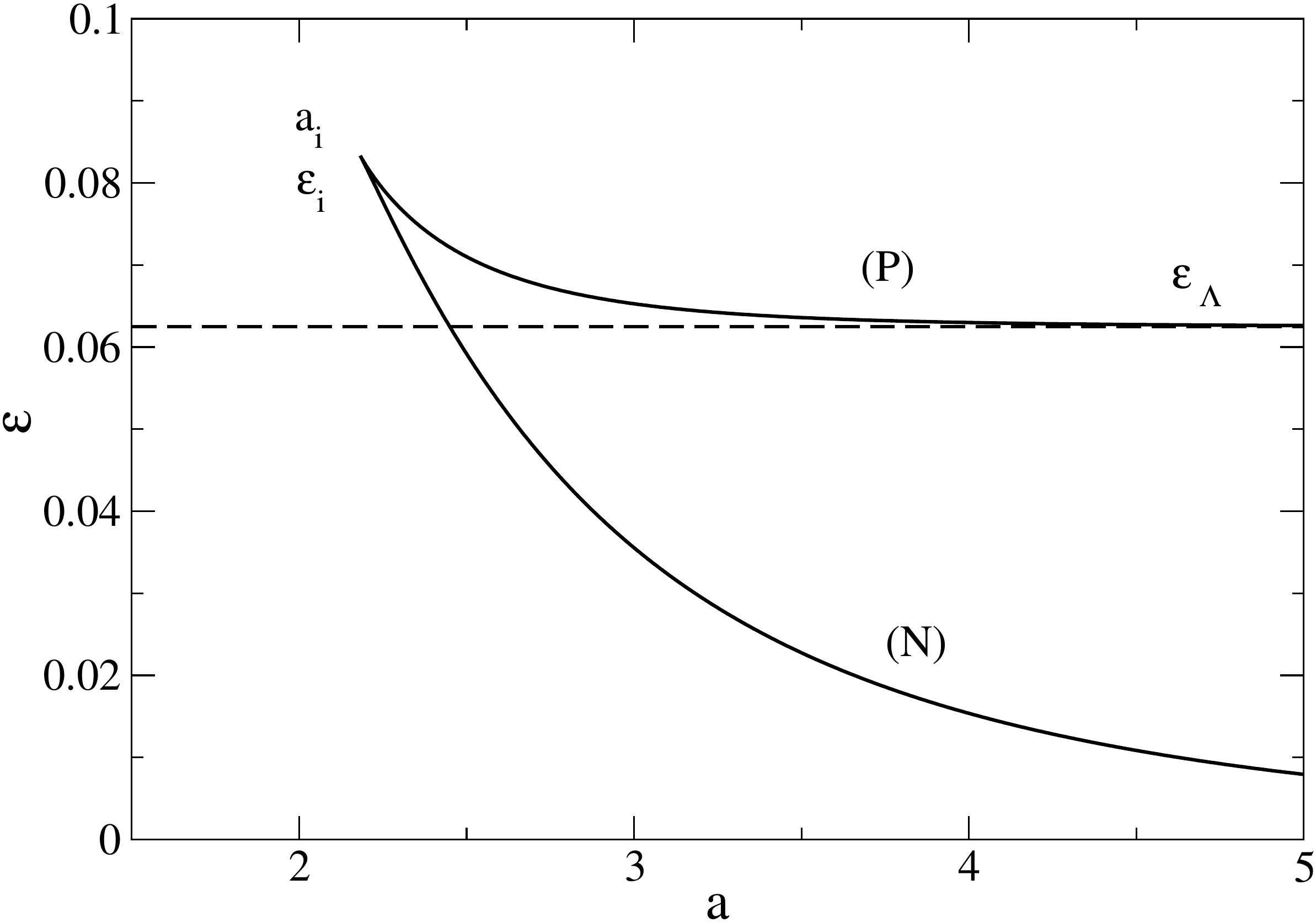}} 
\caption{Energy density $\epsilon$ as a function of the scale factor
$a$.}
\label{aepsneg}
\end{figure}

The evolution of the energy density $\epsilon$ with the scale factor $a$ is
plotted in Fig. \ref{aepsneg}. It starts at  $a=a_{i}$ from 
\begin{equation}
\epsilon_{i}=\frac{1}{2}\rho_i
c^2=\frac{m^3c^4}{24\pi |a_s|\hbar^2}.
\label{negea7}
\end{equation}
For $a\rightarrow a_i$:
\begin{eqnarray}
\epsilon\simeq  \epsilon_i \left\lbrack 1-\left
(\frac{\rho}{\rho_i}-1\right )^2\right\rbrack\simeq\nonumber\\
\epsilon_i\biggl\lbrack 1-2\left
(\frac{a}{a_{i}}-1\right )
\pm \frac{4\sqrt{2}}{3}\left (\frac{a}{a_{i}}-1\right )^{3/2} \biggr\rbrack.
\label{negea8}
\end{eqnarray}
On the
normal branch, the energy density decreases as the scale
factor increases and asymptotically tends to $0$. For  $a\rightarrow +\infty$:
\begin{eqnarray}
\epsilon\sim \rho c^2\sim  \frac{Qmc^2}{a^3}.
\label{negea9}
\end{eqnarray}
On the peculiar branch, the energy density decreases as the scale factor
increases and  asymptotically tends to a minimum density given by
\begin{equation}
\epsilon_{\Lambda}=\frac{1}{4}\rho_{\Lambda}
c^2=\frac{m^3c^4}{32\pi|a_s|\hbar^2}.
\label{negea10}
\end{equation}
For $a\rightarrow +\infty$:
\begin{eqnarray}
\epsilon\simeq \frac{3}{4} \rho_{\Lambda} c^2 -\frac{1}{2}\rho c^2
\simeq \epsilon_{\Lambda} \left\lbrack 1+2\left
(\frac{8\pi Q
|a_s|\hbar^2}{m^2 c^2 a^3}\right )^2\right\rbrack .
\label{negea11}
\end{eqnarray}

\begin{figure}[h]
\scalebox{0.33}{\includegraphics{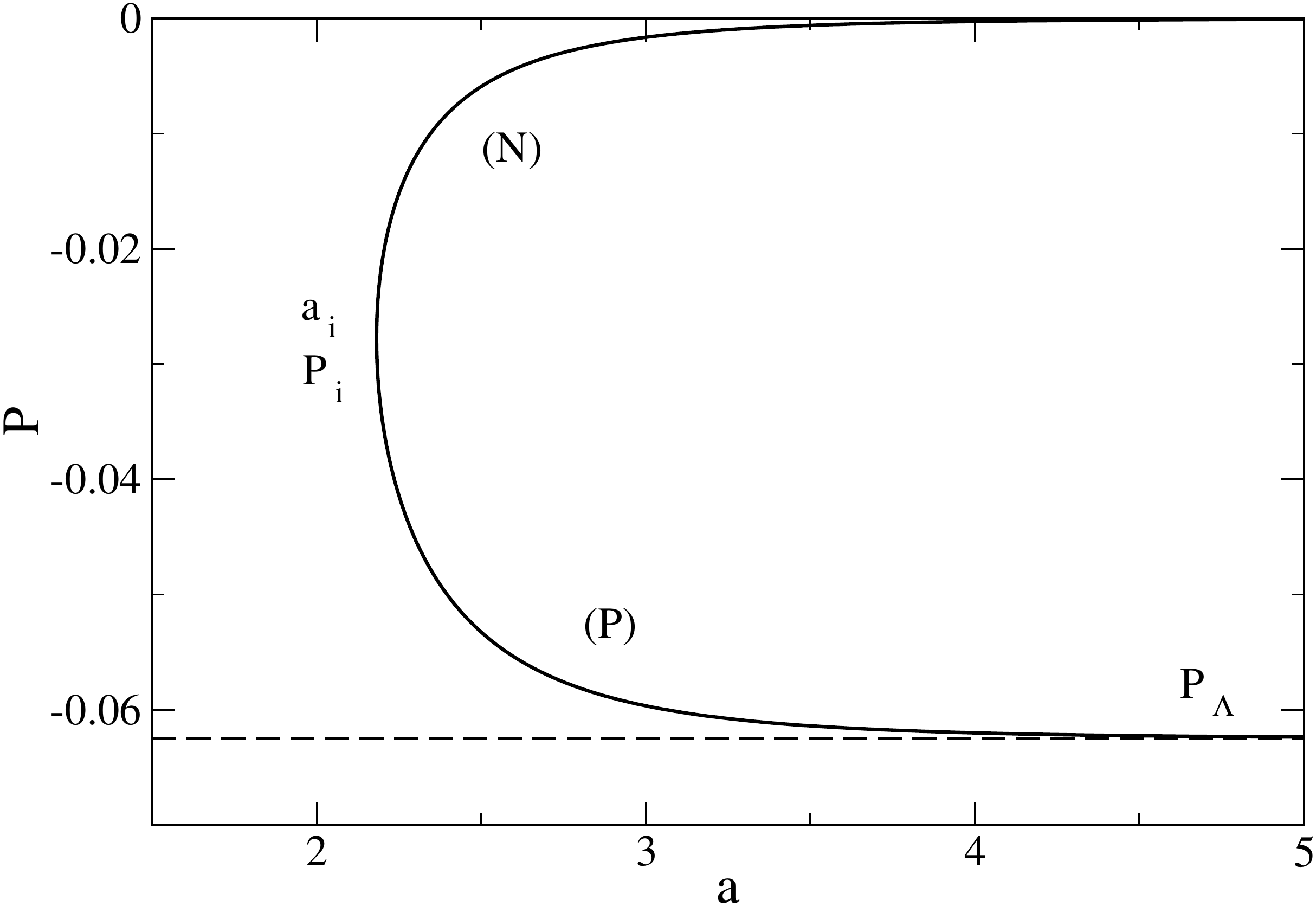}} 
\caption{Pressure $P$ as a function of the scale factor
$a$.}
\label{aPneg}
\end{figure}

\begin{figure}[h]
\scalebox{0.33}{\includegraphics{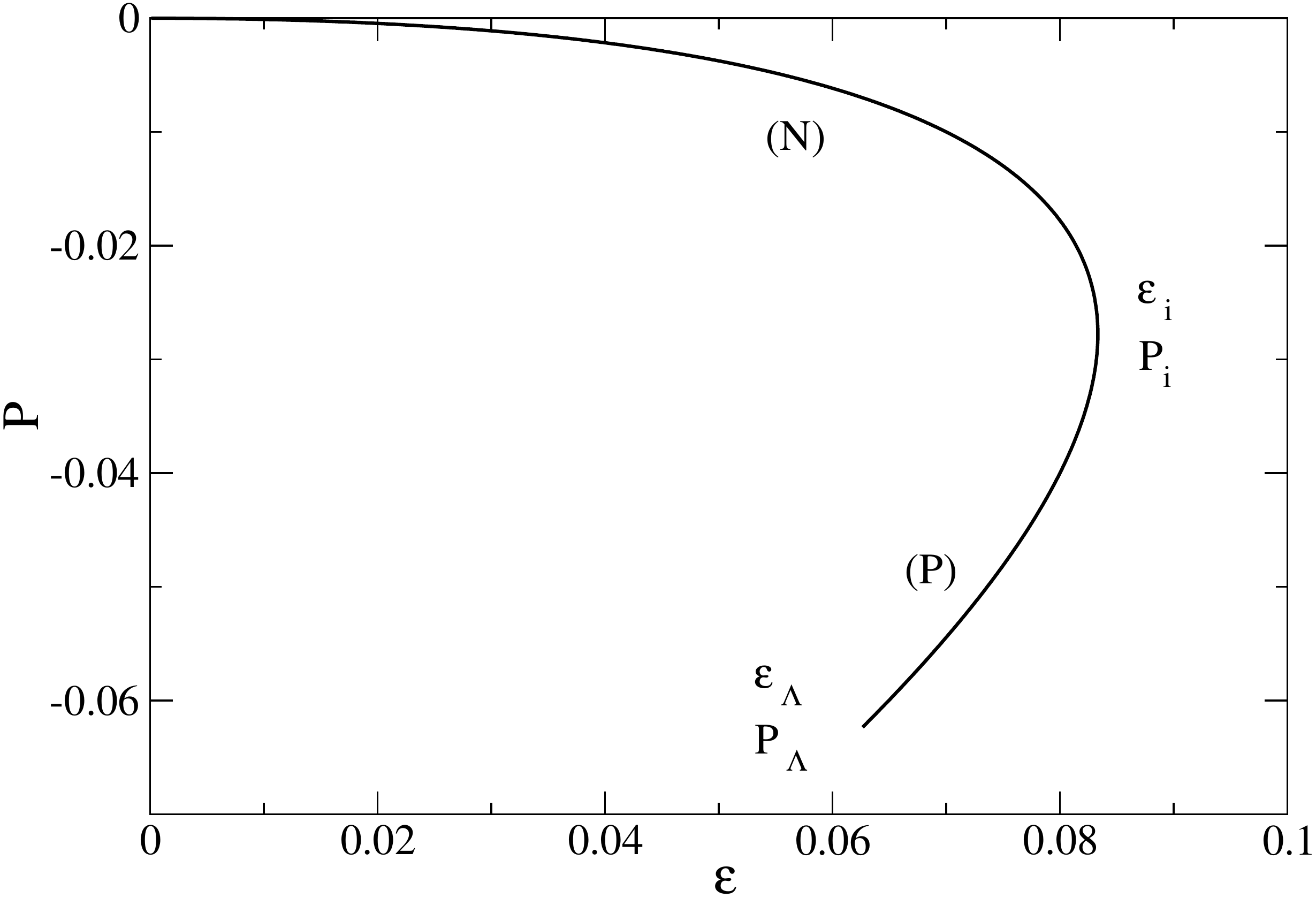}} 
\caption{Pressure $P$ as a function of the energy density
$\epsilon$.}
\label{epsPneg}
\end{figure}

The evolution of the pressure $P$ with the scale factor $a$ is
plotted in Fig. \ref{aPneg}. The pressure is always negative. It starts at
$a=a_{i}$  from 
\begin{equation}
P_{i}=-\frac{1}{6}\rho_i c^2=-\frac{m^3c^4}{72\pi |a_s|\hbar^2}.
\label{negea12}
\end{equation}
For  $a\rightarrow a_i$:
\begin{eqnarray}
 P\simeq  P_i \left (1\pm
2\sqrt{1-\frac{\epsilon}{\epsilon_i}}\right )
\simeq
P_i\left\lbrack 1\pm 2\sqrt{2\left
(\frac{a}{a_{i}}-1\right
)}\right\rbrack.\nonumber\\
\label{negea13}
\end{eqnarray}
On the normal branch, the pressure increases as the
scale factor increases and asymptotically tends to $0$. For $a\rightarrow
+\infty$:
\begin{eqnarray}
P\sim -\frac{2\pi |a_s|\hbar^2}{m^3 c^4}\epsilon^2\sim -\frac{2\pi
|a_s|\hbar^2 Q^2}{m a^6}\simeq 0.
\label{negea14}
\end{eqnarray}
On the
peculiar branch, the pressure decreases as the scale factor
increases and asymptotically  tends to the minimum value 
\begin{equation}
P_{\Lambda}=-\epsilon_{\Lambda}.
\label{negea15}
\end{equation}
For $a\rightarrow +\infty$:
\begin{eqnarray}
 P\simeq \epsilon -2\epsilon_{\Lambda} \simeq
-\epsilon_{\Lambda} \left\lbrack 1-2\left
(\frac{8\pi Q
|a_s|\hbar^2}{m^2 c^2 a^3}\right )^2\right\rbrack.
\label{negea16}
\end{eqnarray}
The relationship between the
pressure and the energy density is
plotted in Fig. \ref{epsPneg}. The normal branch corresponds to $0\le \rho\le
\rho_{i}$, $0\le
\epsilon\le
\epsilon_{i}$, and $P_i\le P\le 0$. The peculiar branch corresponds
to $\rho_{i}\le \rho\le
\rho_{\Lambda}$,
$\epsilon_{\Lambda}\le \epsilon\le
\epsilon_{i}$, and $P_{\Lambda}\le P\le P_i$.

\begin{figure}[h]
\scalebox{0.33}{\includegraphics{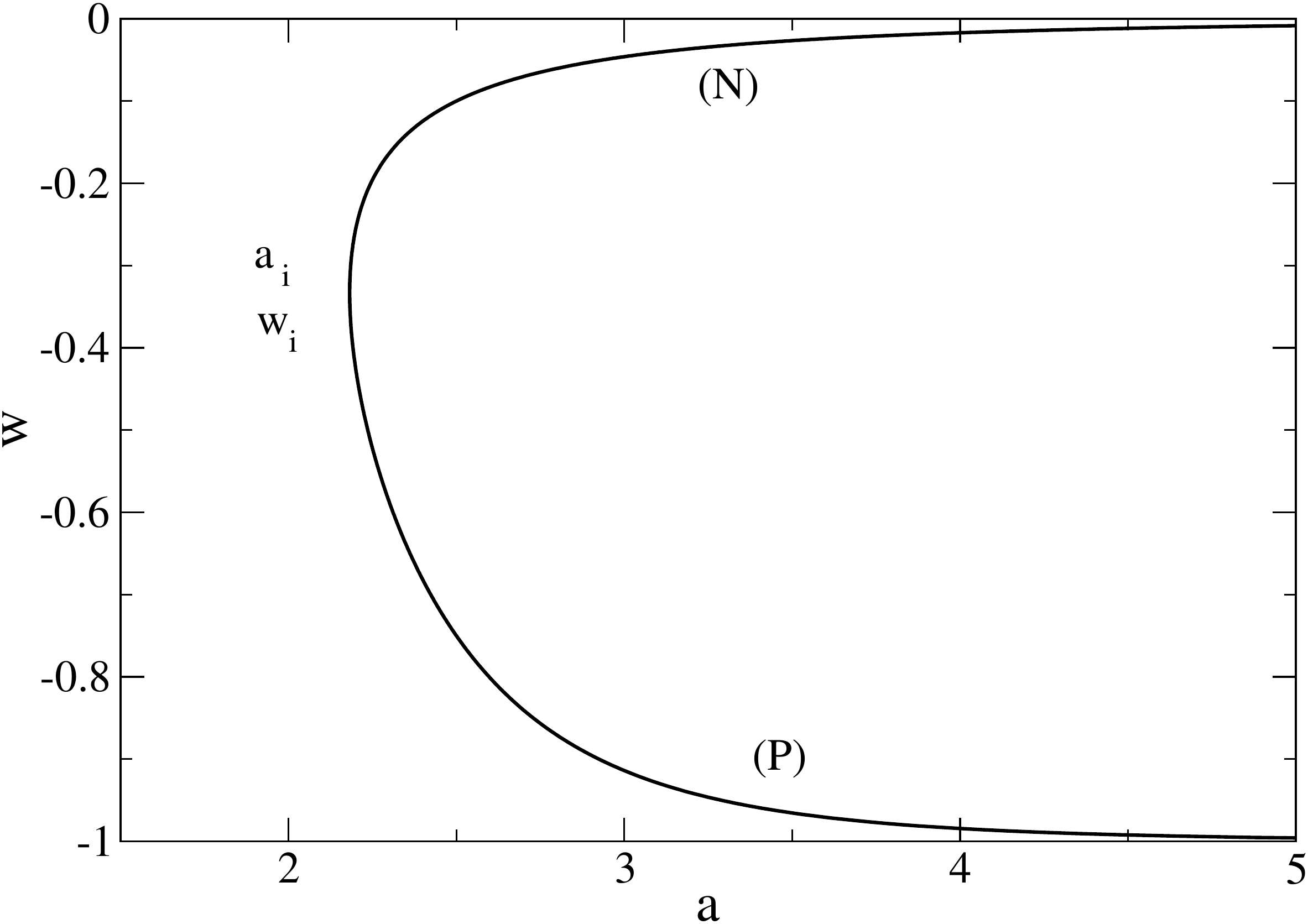}} 
\caption{EOS parameter $w$ as a function of the scale factor
$a$.}
\label{awneg}
\end{figure}

The evolution of the EOS parameter $w=P/\epsilon$ with the scale
factor $a$ is plotted in Fig. \ref{awneg}. The EOS parameter is
always negative. It starts at $a=a_{i}$ from
\begin{equation}
w_{i}=-\frac{1}{3}.
\label{negea17}
\end{equation}
For $a\rightarrow a_i$:
\begin{equation}
w\simeq -\frac{1}{3} \left\lbrack 1\pm 2\sqrt{2\left (\frac{a}{a_i}-1\right
)}\right\rbrack.
\label{negea18}
\end{equation}
On the normal branch, $w$ increases as the
scale factor increases and asymptotically tends to $0$. For  $a\rightarrow
+\infty$:
\begin{equation}
w\sim -\frac{2\pi
|a_s|\hbar^2Q}{m^2c^2a^3}.
\label{negea19}
\end{equation}
On the peculiar branch, $w$ decreases as the scale factor
increases and asymptotically tends to
\begin{equation}
w_{\Lambda}=-1.
\label{negea20}
\end{equation}
For $a\rightarrow +\infty$:
\begin{equation}
w\simeq -1+4 \left (\frac{8\pi Q|a_s|\hbar^2}{m^2c^2a^3}\right )^2.
\label{negea21}
\end{equation}

The total energy $E_{\rm tot}/mc^2$ starts at $a=a_{i}$ from
\begin{equation}
\frac{(E_{\rm tot})_i}{mc^2}=\frac{1}{\sqrt{3}}.
\label{negea22}
\end{equation}
For $a\rightarrow a_i$:
\begin{eqnarray}
\frac{E_{\rm tot}}{mc^2}\simeq \frac{1}{\sqrt{3}}\mp\sqrt{\frac{2}{3}\left
(\frac{a}{a_{i}}-1\right )}.
\label{negea23}
\end{eqnarray}
On the normal branch,  $E_{\rm tot}/mc^2$ increases as the
scale factor increases and asymptotically tends to $1$. For $a\rightarrow
+\infty$:
\begin{eqnarray}
\frac{E_{\rm tot}}{mc^2}\simeq 1-\frac{4\pi
|a_s|\hbar^2 Q}{m^2c^2a^3}.
\label{negea24}
\end{eqnarray}
On the peculiar branch, $E_{\rm tot}$ decreases as the
scale factor increases and asymptotically  tends to $0$. For  $a\rightarrow
+\infty$:
\begin{eqnarray}
 \frac{E_{\rm tot}}{mc^2}\simeq \frac{8\pi Q
|a_s|\hbar^2}{m^2 c^2 a^3}.
\label{negea25}
\end{eqnarray}

\subsection{The temporal evolution of the parameters}
\label{sec_negt}

In this section, we determine the temporal evolution of the parameters
assuming that the Universe contains only the SF. For a quartic potential with
$a_s<0$, the differential equation (\ref{b16})
becomes
\begin{eqnarray}
\left (\frac{d\rho}{dt}\right )^2=24\pi G\rho^3\frac{\left (1-\frac{6\pi
|a_s|\hbar^2}{m^3c^2}\rho\right )\left (1-\frac{8\pi
|a_s|\hbar^2}{m^3c^2}\rho\right )^2}{\left (1-\frac{12\pi
|a_s|\hbar^2}{m^3c^2}\rho\right )^2}.\nonumber\\
\label{negt1}
\end{eqnarray}
The solution of this differential equation which takes the value $\rho_i$ at
$t=0$ is
\begin{eqnarray}
\int_{\frac{2\pi
|a_s|\hbar^2}{m^3c^2}\rho}^{1/6}\frac{(1-6x)\,
dx}{x^{3/2}(1-3x)^{1/2}(1-4x)}=\sqrt{\frac{12 G
m^3c^2}{|a_s|\hbar^2}}t.\nonumber\\
\label{negt2}
\end{eqnarray}
The integral can be computed analytically: 
\begin{eqnarray}
\int\frac{(1-6x)\,
dx}{x^{3/2}(1-3x)^{1/2}(1-4x)}=\nonumber\\
-2\sqrt{\frac{1-3x}{x}}+2\ln\left
(\frac{2+\sqrt{1-3x}+3\sqrt{x}}{2+\sqrt{1-3x}-3\sqrt{x}}\right
)\nonumber\\
+2\ln\left (\frac{1-2\sqrt{x}}{1+2\sqrt{x}}\right ).
\label{negt3}
\end{eqnarray}

\begin{figure}[h]
\scalebox{0.35}{\includegraphics{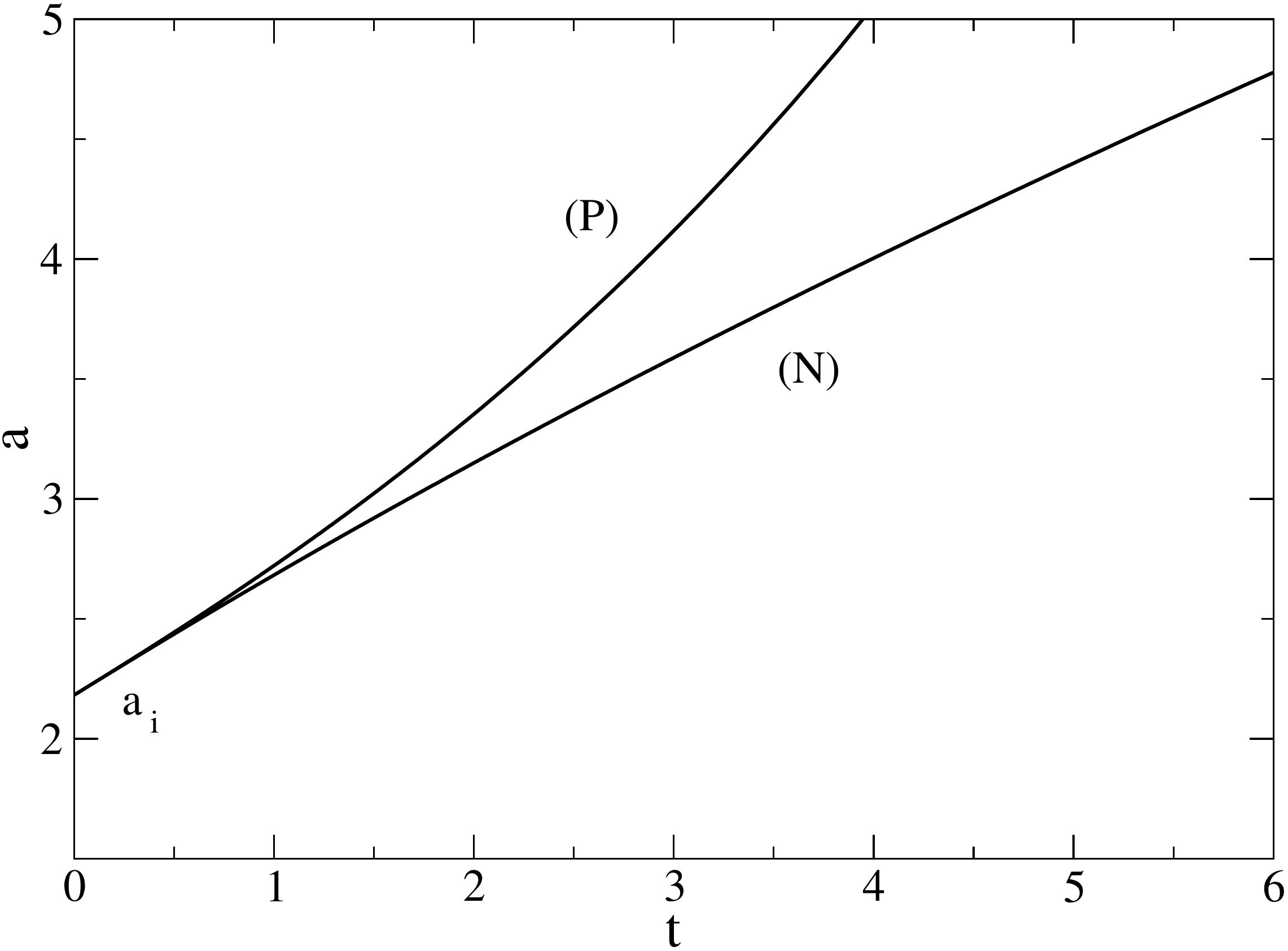}} 
\caption{Temporal evolution of the scale factor $a$.}
\label{taneg}
\end{figure}

From these equations, we can obtain the temporal evolution of the
pseudo rest-mass density $\rho(t)$. Then, using Eqs.
(\ref{negba1})-(\ref{negba5}), we can obtain the temporal evolution of the
all the parameters. The temporal evolution of the scale factor $a$ is plotted in
Fig. \ref{taneg}. It starts from $a=a_i$ at $t=0$ and increases to $+\infty$ as
$t\rightarrow +\infty$ on the two branches. We do
not show the other curves because they can be easily deduced from Figs.
\ref{arhoneg}, \ref{aepsneg}, \ref{aPneg} and \ref{awneg}
since $a$ is a monotonic function of time on the two branches. However, we
provide below the
asymptotic behaviors of all the parameters.

For $t\rightarrow 0$: 
\begin{eqnarray}
a\simeq a_{i}\left \lbrack 1+\left (\frac{4\pi G\rho_i}{3}\right
)^{1/2}t\pm \frac{4\sqrt{2}}{15}\left (\frac{4\pi G\rho_i}{3}\right
)^{5/4}t^{5/2} \right\rbrack,\nonumber\\
\label{negt4}
\end{eqnarray}
\begin{eqnarray}
\rho=\rho_i\left\lbrack 1\pm \left (\frac{16\pi G\rho_i}{3}\right
)^{1/4}\sqrt{t}\right\rbrack,
\label{negt7}
\end{eqnarray}
\begin{eqnarray}
\epsilon\simeq \epsilon_i \left\lbrack 1-2\left (\frac{4\pi
G\rho_i}{3}\right
)^{1/2} t\right\rbrack,
\label{negt9}
\end{eqnarray}
\begin{eqnarray}
P\simeq  P_i \left\lbrack 1\pm 2\left (\frac{16\pi
G\rho_i}{3}\right
)^{1/4}\sqrt{t}\right\rbrack,
\label{negt11}
\end{eqnarray}
\begin{eqnarray}
w\simeq -\frac{1}{3}\left \lbrack 1\pm 2\left (\frac{16\pi G\rho_i}{3}\right
)^{1/4}\sqrt{t} \right\rbrack,
\label{negt13}
\end{eqnarray}
\begin{eqnarray}
\frac{E_{\rm tot}}{mc^2}\simeq \frac{1}{\sqrt{3}}\mp\sqrt{\frac{2}{3}\left
(\frac{a}{a_{i}}-1\right )}.
\label{negt15}
\end{eqnarray}
On the normal branch, for $t\rightarrow +\infty$: 
\begin{eqnarray}
 a\sim  (6\pi
GQmt^2)^{1/3},
\label{negt5}
\end{eqnarray}
\begin{eqnarray}
\rho\sim  \frac{1}{6\pi Gt^2},
\label{negt8}
\end{eqnarray}
\begin{eqnarray}
\epsilon\sim \frac{c^2}{6\pi G t^2},
\label{negt10}
\end{eqnarray}
\begin{eqnarray}
P \sim -\frac{|a_s|\hbar^2}{18\pi m^3 G^2 t^4},
\label{negt12}
\end{eqnarray}
\begin{eqnarray}
w\sim -\frac{|a_s|\hbar^2}{3 m^3c^2  Gt^2},
\label{negt14}
\end{eqnarray}
\begin{eqnarray}
 \frac{E_{\rm tot}}{mc^2}\sim 1-\frac{2|a_s|\hbar^2}{3 m^3c^2  Gt^2}.
\label{negt16}
\end{eqnarray}
On the peculiar branch, for $t\rightarrow +\infty$:  
\begin{eqnarray}
a\sim  \left (\frac{8\pi Q|a_s|\hbar^2}{m^2c^2}\right )^{1/3}e^{-A/12}e^{-1/6}
e^{\left (\frac{2}{3}\pi G\rho_{\Lambda}\right )^{1/2}t},
\label{negt6}
\end{eqnarray}
where
\begin{equation}
A=-2\sqrt{3}+2\ln\left
\lbrack\frac{(2\sqrt{6}+\sqrt{3}+3)(\sqrt{6}-2)}{(2\sqrt{6}+\sqrt{3}-3)(\sqrt{6}
+2) } \right \rbrack.
\label{negt6v}
\end{equation} 
Numerically, $A=-6.09802...$. The other parameters converge
towards their asymptotic
values determined in Sec. \ref{sec_negea} exponentially rapidly.

{\it Remark:} if we assume that the Universe contains only a SF with an
attractive self-interaction, and if we assume that the fast oscillation regime
is always valid (see, however, Sec. \ref{sec_valn}), we find that the Universe
emerges from an initial state in which the scale factor is nonzero and the
energy
density is finite. This initial state is nonsingular for what
concerns the values of  $\rho_i$, $\epsilon_i$ and $a_i$. However, it is 
singular because the time derivative  of the pseudo rest mass
density $\rho$ is infinite: $\dot\rho_i=\infty$. This is different from the Big
Bang
singularity in
which the scale
factor vanishes and the energy density is infinite. There are important claims
that support the idea of a nonsingular Universe, one example being the case of
bouncing Universes where the big bang is taken as the
beginning
of a period of expansion that followed a period of contraction. These kinds of
behaviors are often referred to as a (nonsingular) big crunch followed by
a (nonsingular) big
bang, or more simply, 
a  big bounce \cite{ashtekar,poplawski,cai} (see also \cite{stiff} in a
different context). Our solution valid for
$t\ge 0$ can be extended to $t\le 0$ by symmetry leading to a big bounce.
However, this assumes that the Universe contains only the SF although this is
not the case in reality.

\subsection{The different eras}

In the fast oscillation regime, a SF with an attractive self-interaction
($a_s<0$) undergoes two distinct eras.  It emerges from an
initial state where the scale factor $a_i$ is nonzero and the energy
density $\epsilon_i$ is finite. The SF does not exist before
$a_i$ (see, however, the limitations of our approximations in Sec.
\ref{sec_valn}). For $a\rightarrow a_i$, the EOS parameter tends to $w_i=-1/3$.
This value
is the same as for a
gas of cosmic strings\footnote{Cosmic strings are a type of topological defects
which may have formed during a symmetry
breaking phase transition in the early Universe.  The phase
transitions leading
to the production of cosmic strings are likely to have occurred during the
earliest moments of the Universe's evolution, just after cosmological inflation,
and are a fairly generic prediction in both quantum field theory and string
theory models of the early Universe
\cite{schild,fraisse,planckstring}.} described by
the EOS $P=-\epsilon/3$.
The EOS parameter $w=-1/3$ marks the transition between accelerating and
decelerating Universes. Indeed, for the EOS $P=-\epsilon/3$,
using the Friedmann equations (\ref{h4}) and (\ref{h5}), we find
that $\epsilon=\epsilon_{\rm s,0}/a^2$ so that the scale factor increases
linearly with time as $a=(8\pi G\epsilon_{s,0}/3c^2)^{1/2} t$.
In our model, the evolution of the scale factor is different but we note that
the leading term in Eq. (\ref{negt4}) valid for short times also scales linearly
with $t$.
Therefore, it is possible that the early evolution of our model shares some
analogies with a gas of cosmic strings.  At later times, two evolutions
are possible. The normal branch is
asymptotically similar to the evolution of a pressureless Universe like in the
EdS model. Indeed, for $a\rightarrow +\infty$, the EOS (\ref{negba7})
reduces to Eq. (\ref{negea14}) and the EOS
parameter tends to $0$ [see Eq. (\ref{negea19})]  so the SF behaves essentially
as pressureless DM (dust) with an EOS
$P=0$.\footnote{As explained in footnote 5, the pressure of
the SF is nonzero but since $P\propto -\epsilon^2\ll \epsilon$ for
$\epsilon\rightarrow 0$, everything happens at large scales as if the
Universe were pressureless. However, the nonzero pressure of the SF is important
at the scale of DM halos. Contrary to a repulsive self-interaction that
stabilizes the halos, an attractive self-interaction has the tendency to
destabilize the halos. Stable halos with $a_s<0$ can exist only below a maximum
mass \cite{b34,cd,bectcoll}. For QCD axions, this mass is too small to
account for the mass of DM halos. It becomes of the order of DM halos in the
case of ultralight axions with a very small self-interaction (see
\cite{b34,cd,bectcoll} and Appendix
\ref{sec_mas}).} In that case, the scale
factor increases algebraically with time like $a\propto t^{2/3}$. On the other
hand, the peculiar branch is asymptotically similar to the evolution of a de
Sitter Universe. Indeed, the energy density tends to a constant given by Eq.
(\ref{negea10})
and the EOS parameter tends to $w_{\Lambda}=-1$ [see Eq.
(\ref{negea20})] similar to the EOS parameter of DE
with an EOS $P=-\epsilon$ [see Eq. (\ref{negea15})]. In that case, the
scale
factor increases exponentially rapidly with time like $a\propto
e^{(2\pi G\rho_{\Lambda}/3)^{1/2}t}$. Therefore, the SF undergoes a cosmic
stringlike era ($w=-1/3$) followed by a matterlike era ($w=0$) on the
normal branch or a de Sitterlike era ($w=-1$) on the peculiar branch. On the
normal branch, since $w\ge -1/3$,
the Universe is always decelerating. On the peculiar branch, since $w\le -1/3$, 
the Universe is always accelerating. In conclusion, a SF with an attractive 
self-interaction behaves at early times as a gas of cosmic strings and at late
times as DM (normal branch) or as DE (peculiar branch). We
note that the cosmic stringlike era and the peculiar branch are due to the
attractive self-interaction of the SF.

In the fast oscillation regime, the SF exists only for $a>a_i$ where
$a_i$ is given by Eq. (\ref{negea1}). Using Eq.
(\ref{astar}) relying on the expression (\ref{mrf2})  of the charge of the
SF,\footnote{In principle, this expression is only valid for the SF on the
normal branch which asymptotically behaves as DM. The SF on the peculiar
branch which behaves as DE is treated in Sec. \ref{sec_effl}.} we can obtain
$a_i=(6\sqrt{3})^{1/3}a_*$ as a function of the ratio $a_s/m^3$. In order to be
consistent with the
observations, we must require that
$a_i\ll a_{\rm eq}=2.95\times 10^{-4}$. Using Eq. (\ref{mrs6}), we obtain
the
constraint
\begin{equation}
\frac{|a_{s}|}{m^3}\ll \frac{c^4\Omega_{\rm
r,0}^3}{12\sqrt{3}\pi\hbar^2\epsilon_0\Omega_{\rm dm,0}\Omega_{\rm m,0}^3}.
\label{nb3}
\end{equation}
Introducing proper normalization, we get
\begin{equation}
\frac{|a_s|}{\rm fm} \left (\frac{{\rm eV}/c^2}{m}\right )^3\ll 8.00\times 10^6.
\label{nb4}
\end{equation}
For a QCD axion field [see Eq. (\ref{mas12})], we
obtain $({|a_s|}/{\rm fm})(({{\rm
eV}/c^2})/{m})^3=5.8\times 10^{-26}$ and $a_i=5.69\times 10^{-15}$. For an
ultralight axion [see Eq. (\ref{mas13})], we
obtain $({|a_s|}/{\rm fm})(({{\rm
eV}/c^2})/{m})^3=1.06\times 10^3$ and $a_i=1.50\times 10^{-5}$. Therefore, the
constraint
$a_i\ll
a_{\rm eq}$  is verified.

We can estimate the transition between the cosmic stringlike era and the
matterlike era (normal branch) or the de Sitterlike era (peculiar branch) of
the SF as follows. First of all, using Eqs.
(\ref{negba1}) and (\ref{negba4b}), we find that the scale factor
corresponding to a value $w$ of the EOS parameter is
\begin{eqnarray}
a=\left (\frac{2\pi |a_s| \hbar^2 Q}{m^2c^2}\right
)^{1/3}\frac{(1-3w)^{1/2}}{|w|^{1/3}(1+w)^{1/6}}.
\label{differa1b}
\end{eqnarray}
Interestingly, this equation provides an analytical expression of the function
$a(w)$, the inverse of the function $w(a)$ plotted in Fig. \ref{awneg}. If
we consider that the transition between the cosmic stringlike era and the
matterlike
era of the SF corresponds to $w_t^{({\rm N})}=-1/6$ (the arithmetic mean of
$w=-1/3$ and $w=0$), we obtain
\begin{eqnarray}
\frac{a_t^{(\rm N)}}{a_i}=\frac{\sqrt{3}}{2^{1/3}5^{1/6}}=1.05129...
\end{eqnarray}
Similarly, if we consider that the transition between the cosmic stringlike era
and the de Sitterlike era of the SF corresponds to $w_t^{(\rm P)}=-2/3$ (the
arithmetic mean of
$w=-1/3$ and $w=-1$), we
obtain
\begin{eqnarray}
\frac{a_t^{(\rm P)}}{a_i}=\frac{\sqrt{3}}{2^{2/3}}=1.09112...
\end{eqnarray}
The transition scales $a_t^{(\rm N)}$ and $a_t^{(\rm P)}$ are very close to
$a_i$ so that the duration of the cosmic stringlike era is extremely short. On
the
other hand, for QCD and ultralight axions, the transition scales $a_t^{(\rm
N)}$ and $a_t^{(\rm P)}$ are below $a_{\rm eq}=2.95\times 10^{-4}$ by
several orders of magnitude so that, at the equality epoch, the axionic SF
already behaves as DM (normal branch) or DE (peculiar branch).

\subsection{Validity of the fast oscillation regime}
\label{sec_valn}

The previous results are valid in the fast oscillation regime $\omega\gg H$.
In this section, we determine the domain of validity of this regime.

\begin{figure}
\scalebox{0.33}{\includegraphics{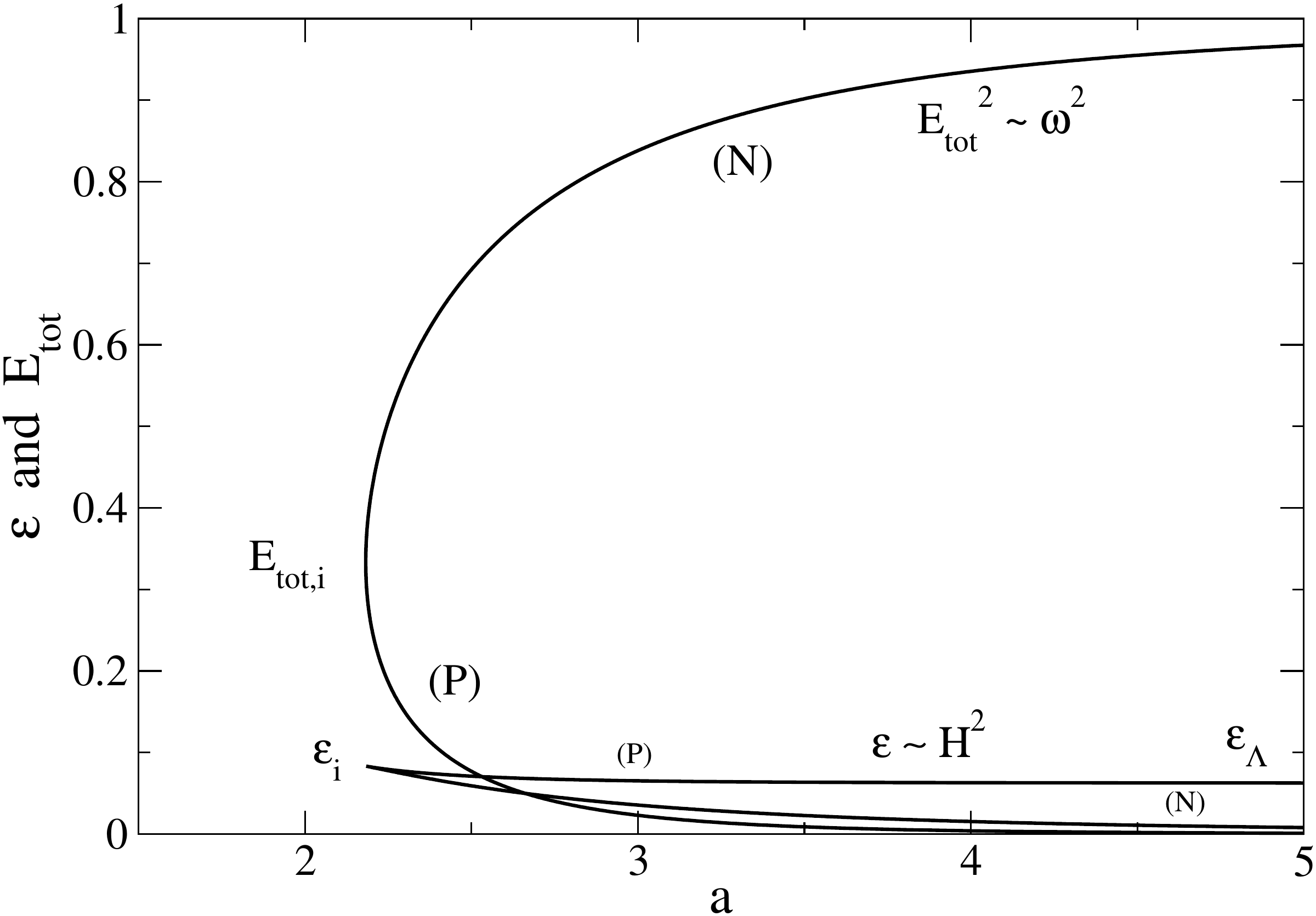}} 
\caption{Graphical construction
determining the validity of the fast oscillation regime.  The transition scale
$a'_v$ corresponds to the intersection
of the curves $\sigma {\tilde E}_{\rm
tot}^2$ and   $\tilde\epsilon$.}
\label{validityneg}
\end{figure}

\begin{figure}
\scalebox{0.33}{\includegraphics{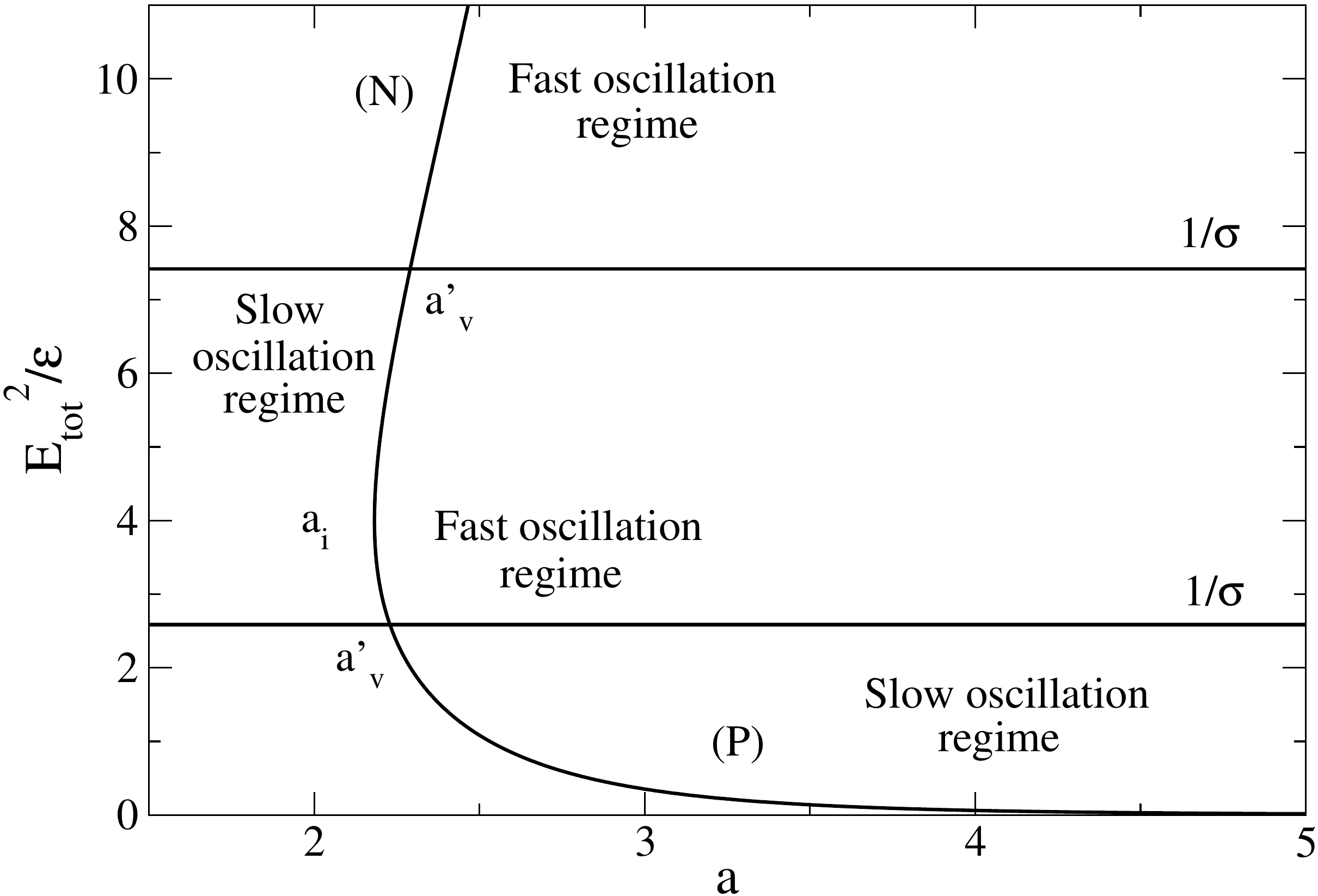}} 
\caption{Ratio $\omega/H$ as a function of the scale
factor $a$. }
\label{aEtot2sepsneg}
\end{figure}

In terms of dimensionless variables, the condition $\omega\gg H$ can be
expressed by Eq. (\ref{valp2}) where $\sigma$ is given by Eq. (\ref{valp3}) in
which $a_s$ is replaced by $|a_s|$. The
dimensionless variables ${\tilde E}_{\rm
tot}^2$ and $\tilde\epsilon$ are plotted as a function of $\tilde a$ in Fig.
\ref{validityneg}. Their ratio ${\tilde E}_{\rm tot}^2/\tilde\epsilon$ is
plotted  as a function of $\tilde a$ in Fig. \ref{aEtot2sepsneg}. The
intersection of this curve with the
line ${\tilde E}_{\rm
tot}^2/\tilde\epsilon=1/\sigma$ determines the domain of validity of the  fast
oscillation regime.

We first consider the normal branch. Since
${\tilde E}_{\rm tot}^2$ (corresponding to $\omega^2)$ increases with the
scale factor $a$ up to $1$ while $\tilde\epsilon$ (corresponding to $H^2$)
decreases to $0$, the fast oscillation regime ${\tilde E}_{\rm
tot}^2\gg \tilde\epsilon/\sigma$ (corresponding to  $\omega\gg
H$) will be valid for any $a\ge a_i$ if it is valid at $a=a_i$. Since 
$({\tilde E}_{\rm tot}^2)_i=1/3$ and ${\tilde \epsilon}_i=1/12$, we find that
the fast oscillation regime is valid for any $a\ge a_i$  when
\begin{eqnarray}
\sigma> \frac{1}{4},\quad{\rm i.e.}\quad  |a_s|>
\frac{Gm}{3c^2}=\frac{1}{6}r_S.
\label{vip1}
\end{eqnarray}
When $\sigma< 1/4$, the fast oscillation regime starts to be valid for $a\gg
a'_v$
with
\begin{eqnarray}
a'_v=\left (\frac{2\pi |a_s|\hbar^2 Q}{m^2c^2}\right
)^{1/3} g\left (\frac{3|a_s|c^2}{4Gm}\right ),
\label{vip2}
\end{eqnarray}
where the function $g(\sigma)$ is defined by
\begin{eqnarray}
g(\sigma)=\frac{1}{r^{1/3}(1-4r)^{1/6}}
\label{vip3}
\end{eqnarray}
with
\begin{eqnarray}
r=\frac{4\sigma+1-\sqrt{(4\sigma+1)^2-12\sigma}}{6}.
\label{vip4}
\end{eqnarray}
For $\sigma\rightarrow 0$:
\begin{eqnarray}
g(\sigma)\sim \frac{1}{\sigma^{1/3}}.
\label{vip6}
\end{eqnarray}
For $\sigma\rightarrow +\infty$:
\begin{eqnarray}
g(\sigma)\sim 2^{4/3}\sigma^{1/6}.
\label{vip5}
\end{eqnarray}
The function $g(\sigma)$ takes its 
minimum value $g_{\rm min}=2^{1/3}\sqrt{3}\simeq
2.18225...$ at $\sigma=1/4$. These asymptotic results can be written more
explicitly by restoring the original variables. When $a_s=0$, we find that
$a'_v$ is given by Eq. (\ref{valp11}) corresponding
to the begining of the fast oscillation regime in the
noninteracting case. When $|a_s|=r_S/6$, we get $a'_v=a_i$. When
$|a_s|\ge r_S/6$, the fast oscillation regime is valid for any
$a\ge a_i$.\footnote{Actually, the fast oscillation condition $\omega\gg H$ is a
necessary but not a
sufficient condition for the validity of Eqs. (\ref{negba1})-(\ref{negba5}). We
must also require that $\omega\gg \dot\rho/\rho$ (see Appendix
\ref{sec_ext}). This condition is never realized by the solution of
Eqs. (\ref{negba1})-(\ref{negba5}) close to $a_i$ since $\dot\rho_i$ is
infinite. This means that Eqs. (\ref{negba1})-(\ref{negba5}) are never
valid close to $a_i$, even when $\omega\gg H$.
If we remember that Eqs. (\ref{negba1})-(\ref{negba5}) are obtained from the
exact equations (\ref{b4})-(\ref{b3}) by
neglecting the terms involving $\hbar$ (see Sec. \ref{sec_rm}),
i.e. by neglecting the quantum potential, we come to the conclusion that quantum
mechanics is important at early times and must be taken into account. In
particular, it will prevent the divergence of $\dot\rho$ at $a_i$ and regularize
the
evolution of the SF in the early Universe.} We may wonder what happens to the SF
in the early Universe, before the fast
oscillation regime. Is there a
stiff matter era? Does the stiff matter era exist for
any value of $a_s<0$? A stiff matter era must exist  in the early
Universe if $|a_s|$ is sufficiently small since it exists during
a finite period of time $0\le a\le a_v(0)$ when $|a_s|=0$ and it cannot
disappear suddenly when $a_s<0$. In Sec. \ref{sec_tp} we
argue that the stiff
matter era exists for $a_i \le a\le a'_v$ when $|a_s|\le r_S/6$ and does not
exist anymore when $|a_s|\ge r_S/6$. We also argue that, in the very early
Universe, for $0\le a\le a_i$, the SF undergoes an inflation era. However, in
order to investigate these regimes, we need to solve the exact equations
(\ref{b4})-(\ref{b3}), taking quantum terms into account. This study would be
particularly important 
to determine the duration of the inflation era (if there is really one), and its
connection to the stiff
matter era when $|a_s|\le r_S/6$ or its connection to the matter era when
$|a_s|\ge r_S/6$ (in particular, the inflation era is expected to stop long
before $a_i$). However, this study is
beyond the scope of this paper.

We now consider the peculiar branch. Since ${\tilde E}_{\rm
tot}^2$ (corresponding to $\omega^2$) decreases to $0$ with the scale
factor $a$ while $\tilde\epsilon$ (corresponding to $H^2$) decreases to
$\tilde{\epsilon}_{\Lambda}=1/16$, the fast oscillation regime $\omega\gg H$
is not valid for large $a$. If $\sigma< 1/4$, the fast oscillation regime
is
never valid. If $\sigma> 1/4$, the fast oscillation regime is only valid for
$a\ll a'_v$ where $a'_v$ is given by Eq. (\ref{vip2}). The asymptotic
results can be written more explicitly by restoring the original variables. When
$|a_s|\le r_S/6$, the fast oscillation regime is never valid. When
$|a_s|=r_S/6$, we
get $a'_v=a_i$. When $|a_s|\gg r_S$:
\begin{eqnarray}
a'_v\simeq \left (\frac{768\pi^2|a_s|^3\hbar^4 Q^2}{Gm^5c^{2}}\right
)^{1/6}.
\label{vip6b}
\end{eqnarray}
Using the expression of the
charge given by Eq. (\ref{mrf2}) (see, however, footnote 12), 
and introducing proper normalizations, we
obtain 
\begin{eqnarray}
a'_v\simeq 1.55\times 10^{2} \left (\frac{a_s}{\rm fm}\right )^{1/2}\left
(\frac{{\rm
eV}/c^2}{m}\right )^{7/6}.
\label{valp14qw}
\end{eqnarray}
This value corresponds to the end of the fast oscillation regime in the
strongly self-interacting regime. We may wonder what happens to the SF in the
early ($a<a_i$) and late ($a>a'_v$) Universe when the fast oscillation regime is
not valid. In that case, we have to take quantum mechanics into account and
solve the exact equations (\ref{b4})-(\ref{b3}). This is beyond the scope
of the present paper but we can make the following remarks:

(i) It is shown in Sec. \ref{sec_tpn} that the peculiar branch requires very
particular initial conditions, so it is not clear how it
can be connected to another, more primordial, era before $a_i$. Probably, the
SF emerges suddenly at a nonzero scale factor $a_i$ whose exact value may be
affected by quantum mechanics as discussed in footnote 13. 

(ii) It is argued in Sec. \ref{sec_tpn} that the de Sitter
regime
stops after $a'_v$ and that the SF eventually enters in a
matterlike era (so that the Universe passes from acceleration to deceleration).
In between, quantum
mechanics must be taken into account. It is
curious, but not excluded, that quantum mechanics (i.e., the quantum
potential) becomes important at
late times, after $a'_v$.

\subsection{Phase diagrams}
\label{sec_pdn}

\begin{figure}[h]
\scalebox{0.33}{\includegraphics{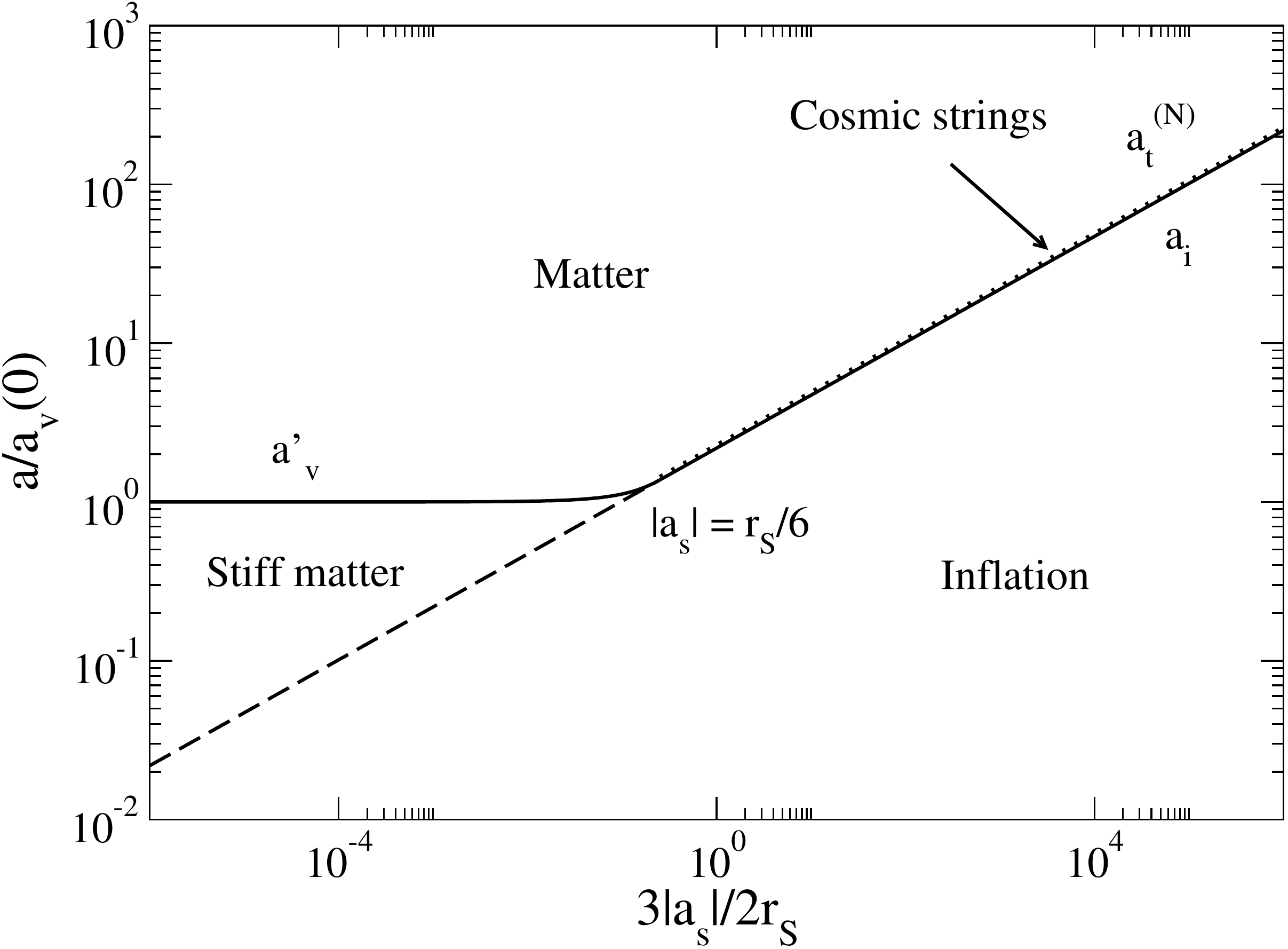}} 
\caption{Phase diagram (mainly hypothetical) showing the different eras of the
SF during the
evolution of the Universe as a function of the scattering length of the
bosons in the case of an attractive self-interaction (normal branch).}
\label{gpd}
\end{figure}

\begin{figure}[h]
\scalebox{0.33}{\includegraphics{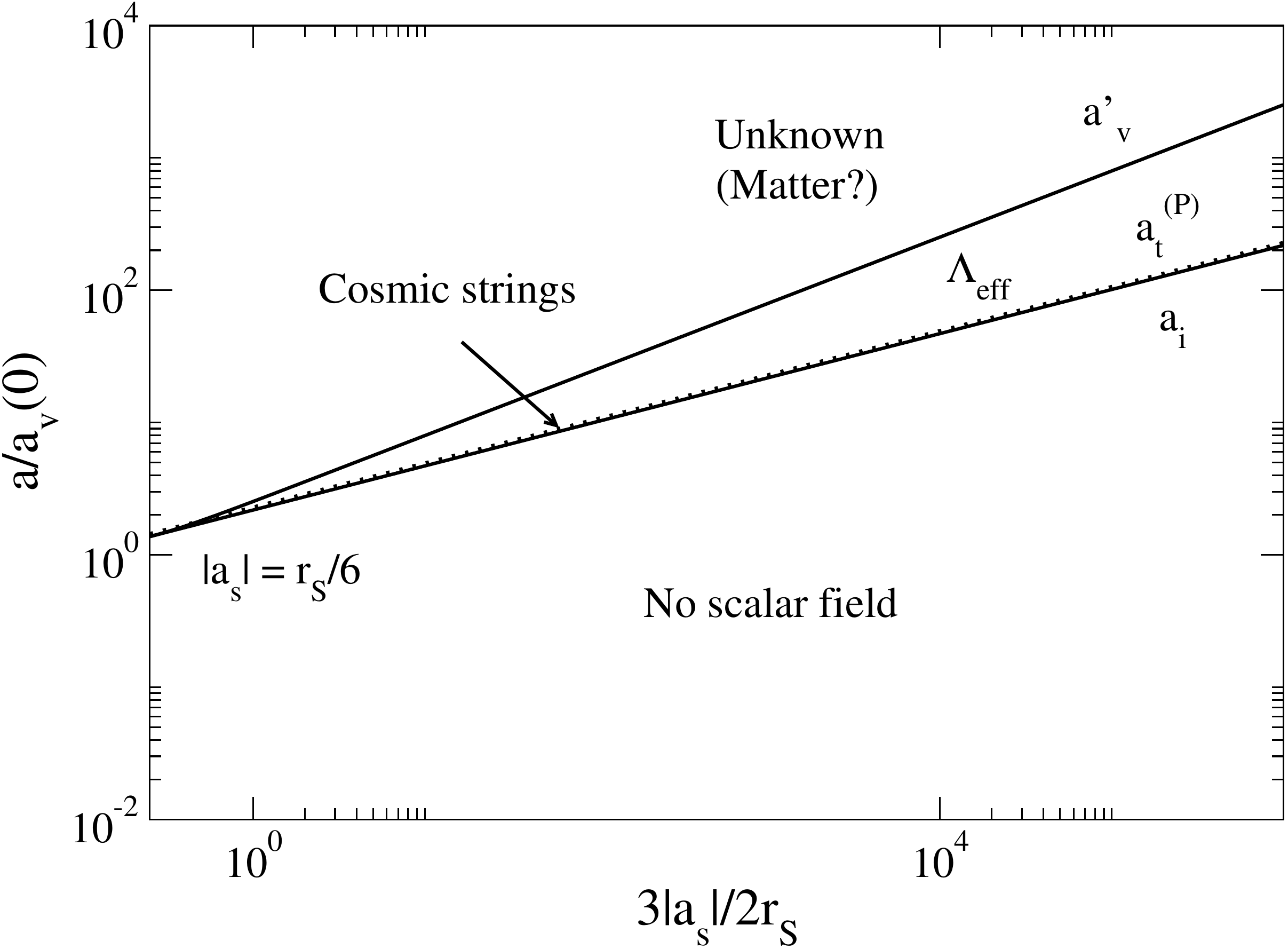}} 
\caption{Same as Fig. \ref{gpd} for the peculiar branch.}
\label{gpdpeculiar}
\end{figure}

We can represent the previous results on a phase diagram (see Figs.
\ref{gpd} and \ref{gpdpeculiar}) where we plot the transition scales 
$a'_v$, $a_i$, $a^{\rm (N)}_t$ and $a^{\rm (P)}_t$ as a function of the
scattering length $a_s$. To that purpose, it is convenient to normalize the
scale factor $a$ by the reference value $a_v(0)$ given by Eq. (\ref{valp11})
that is independent on $a_s$. The scattering length $|a_s|$ can be normalized by
the effective Schwarzschild radius $r_S$ using the parameter
$\sigma=3|a_s|/2r_S$ defined
by Eq. (\ref{valp3}). With these normalizations, the scale $a_i$ marking the
emergence of the SF is given by
\begin{eqnarray}
\frac{a_i}{a_v(0)}=(6\sqrt{3}\sigma)^{1/3}.
\label{vip6ca}
\end{eqnarray}
It increases as
$|a_s|^{1/3}$ according to Eq. (\ref{negea1}). It starts from $0$ at $a_s=0$ and
takes the value $a_i=(\sqrt{3}/2^{1/3}) a_v(0)$ at $|a_s|=r_S/6$.  Therefore,
the SF appears later when $|a_s|$ is larger. On the other hand, the transition
scale $a_v'$ determining the begining (normal branch) or the end (peculiar
branch) of the fast oscillation regime is given by
\begin{eqnarray}
\frac{a'_v}{a_v(0)}= g(\sigma)\sigma^{1/3}.
\label{vip6cb}
\end{eqnarray}
For $\sigma=0$:
\begin{eqnarray}
\frac{a'_v}{a_v(0)}=1.
\label{vip6cc}
\end{eqnarray}
For $\sigma=1/4$:
\begin{eqnarray}
\frac{a'_v}{a_v(0)}=\frac{\sqrt{3}}{2^{1/3}}\simeq 1.37.
\label{vip6cd}
\end{eqnarray}
For $\sigma\rightarrow +\infty$:
\begin{eqnarray}
\frac{a'_v}{a_v(0)}\sim 2^{4/3}\sigma^{1/2}.
\label{vip6ce}
\end{eqnarray}
The transition scale $a_v'$ starts from the value $a_v(0)$ given by Eq.
(\ref{valp11}) for $a_s=0$, increases slowly up to $a_i=(\sqrt{3}/2^{1/3})
a_v(0)$ when $|a_s|=r_S/6$, and increases like $|a_s|^{1/2}$ according to Eq.
(\ref{vip6b}) for $|a_s|\gg r_S$.

We can now describe the phase diagrams. We first consider the normal branch
(see Fig. \ref{gpd}). When $|a_s|=0$, the SF experiences a
stiff matter era for $0\le a\le a_v(0)$ (slow oscillation regime) and a
matterlike era for $a\ge a_v(0)$ (fast oscillation regime). When $0<|a_s|<
r_S/6$, the SF experiences an inflation era for $0\le a\le a_i$, a stiff
matter era for $a_i\le a\le a'_v$ (slow oscillation regime) and a matterlike
era for $a\ge a'_v$ (fast oscillation regime). When $|a_s|\gg r_S/6$, the SF
experiences an inflation era for $0\le a\le a_i$, a cosmic stringlike era
for $a_i\le a\le a_t^{\rm (N)}$ and a matterlike era for $a\ge a_t^{\rm (N)}$.
We
now consider the peculiar branch (see Fig. \ref{gpdpeculiar}). When
$|a_s|< r_S/6$, the fast oscillation regime is never valid. When $|a_s|>
r_S/6$, the SF appears suddenly at $a_i$ (presumably). It experiences a cosmic
stringlike era for $a_i\le a\le a_t^{\rm (P)}$ and a de Sitter-like era,
equivalent
to
an effective cosmological constant $\Lambda_{\rm eff}$ (see Sec.
\ref{sec_effl}), for $a_t^{\rm (P)}\le
a\le a'_v$. For $a>a'_v$, the fast oscillation regime is not valid and the
behavior of the SF is unknown. It may finally enter in
a matterlike era (see Sec. \ref{sec_tpn}).

Let us make a numerical application. For a QCD axion field [see Eq.
(\ref{mas12})], we obtain $\sigma=3.29\times
10^{14}$ and $a_v'=1.73\times 10^{-12}=304\, a_i$.
For an ultralight
axion [see Eq. (\ref{mas13})], we obtain $\sigma=2.87\times 10^{7}$
and $a_v'=3.03\times 10^{-4}=20.2\, a_i$. Since $\sigma\gg 1/4$,
the SF is strongly self-interacting and
the fast
oscillation regime is always valid on
the normal branch (see, however, footnote
13). By contrast, the fast oscillation regime is valid on the
peculiar branch only in a very small range of scale factors. Remember,
however, that the expression of the charge (\ref{mrf2}) used in the
calculations is valid only for the normal branch (see footnote
12) so the numerical application may not be relevant for the peculiar branch
(the peculiar branch is treated in the next section).

\subsection{Effective cosmological constant}
\label{sec_effl}

A striking, and relatively mysterious, result of our study is the discovery
that, under certain conditions, a complex SF with an attractive self-interaction
may behave as DE. Indeed, on the peculiar branch of Fig. \ref{aepsneg}, the
energy density
asymptotically tends to a constant $\epsilon_{\Lambda}$. Furthermore, the final
value of the energy density is not very different from its initial value
$\epsilon_i$. According to Eqs. (\ref{negea7}) and (\ref{negea10}), we have
$\epsilon_{\Lambda}=(3/4)\epsilon_i$. Therefore, a SF with a negative scattering
length naturally generates a cosmological model with an approximately constant
energy density $\sim\epsilon_{\Lambda}$. This may be a physical mechanism to
produce a cosmological constant leading to a de Sitter evolution in which  the
scale factor increases exponentially rapidly with
time.\footnote{Usually, one accounts for DE (or for
a cosmological
constant) by adding a constant term $V_0=\epsilon_\Lambda$, called the vacuum
energy, in the SF potential $V(|\varphi|^2)$. This introduces a constant term
$+\epsilon_\Lambda$ in the energy density $\epsilon$ and a constant term
$-\epsilon_\Lambda$ in the pressure $P$. However, particle physics predicts that
the vacuum energy is of the order of the Planck energy that differs from $123$
orders of magnitude from the cosmological energy. This is the cosmological
constant problem \cite{b3,b1h,martin}. Our model is very different in this
respect
since $V_0$ is equal to zero.
Our
effective cosmological constant comes from the properties of a complex SF
with an attractive self-interaction that can maintain an almost constant energy
density because of the centrifugal force resulting from its fast rotation (see
Sec. \ref{sec_tp} and
Appendix \ref{sec_ext}). This
solution corresponds to a
particular case of spintessence \cite{spintessence} . There is no such solution
for a real SF, nor for a
repulsive self-interaction.} This exponential
growth
of the scale factor may account for the early inflation or for the late
acceleration of the Universe. Furthermore, the attractive self-interaction of
the bosons ($a_s<0$) could justify that the pressure is negative during these
periods and that $w\simeq -1$. In this section, we try to constrain the
parameters of the SF in order to make the value of the effective cosmological
constant
consistent with observations. This section is highly speculative so that
only orders of magnitude will be considered.

The asymptotic value of the energy density of a SF with
an attractive
self-interaction ($a_s<0$) on the peculiar branch is 
\begin{eqnarray}
\epsilon_{\Lambda}=\frac{m^3c^4}{32\pi |a_s|\hbar^2}.
\label{pb1}
\end{eqnarray}
On the other hand, the energy density produced by a
cosmological constant $\Lambda$ is
\begin{eqnarray}
\epsilon_{\Lambda}=\frac{\Lambda c^2}{8\pi G}.
\label{pb2}
\end{eqnarray}
Comparing Eqs. (\ref{pb1}) and  (\ref{pb2}), we find that a SF with
an attractive self-interaction is equivalent
to an effective cosmological constant 
\begin{eqnarray}
\Lambda=\frac{Gm^3c^2}{4|a_s|\hbar^2}.
\label{pb3}
\end{eqnarray}

In the very early Universe, the cosmological constant
may account for the phase of inflation. In that case, the energy density is of
the order of the Planck energy density $\epsilon_P=\rho_P c^2$ where
$\rho_P=c^5/\hbar G^2=5.16\times 10^{99}\, {\rm g}\, {\rm m}^{-3}$.
Substituting this value into Eq.
(\ref{pb1}), we obtain
\begin{eqnarray}
\frac{|a_s|}{m^3}=\frac{c^2}{32\pi\hbar^2\rho_P}=\frac{G^2}{32\pi\hbar c^3}.
\label{pb6}
\end{eqnarray}
Introducing proper normalizations, we get
\begin{eqnarray}
\frac{|a_s|}{\rm fm}\left (\frac{{\rm eV}/c^2}{m}\right )^3= 8.83\times
10^{-107}.
\label{pb7}
\end{eqnarray}

In the late Universe, the cosmological constant may account for the phase of
acceleration ($\Lambda$CDM model).  In that
case, the 
energy density is equal to the cosmological
density $\epsilon_{\Lambda}=\Omega_{\Lambda,0}\epsilon_0=5.25\times
10^{-7}{\rm g}\, {\rm m}^{-1}\, {\rm s}^{-2}$
where $\Omega_{\Lambda,0}=0.687$ is the present fraction of DE and
$\epsilon_0=3c^2H_0^2/8\pi G=7.64\times
10^{-7}{\rm g}\, {\rm m}^{-1}\, {\rm s}^{-2}$ is the present energy density of
the Universe. We introduce
$\rho_{\Lambda}=\epsilon_{\Lambda}/c^2=5.84\times
10^{-24}\, {\rm g}\, {\rm m}^{-3}$.
Substituting this value into Eq. (\ref{pb1}), we obtain
\begin{eqnarray}
\frac{|a_s|}{m^3}=\frac{c^2}{32\pi\hbar^2\rho_\Lambda}=\frac{
Gc^2}{12\hbar^2\Omega_{\Lambda,0}H_0^2}.
\label{pb4}
\end{eqnarray}
Introducing proper normalizations, we get
\begin{eqnarray}
\frac{|a_s|}{\rm fm}\left (\frac{{\rm eV}/c^2}{m}\right )^3= 7.80\times
10^{16}.
\label{pb5}
\end{eqnarray}

Let us make a numerical application. For a QCD axion field
[see Eq. (\ref{mas12})], we obtain
$|a_s|/m^3=5.8\times 10^{-26}\, {\rm fm}/({\rm eV}/c^2)^3$ and 
$\rho_{\Lambda}=7.85\times 10^{18}\, {\rm g}\,  {\rm m}^{-3}$.
The value of $|a_s|/m^3$ is very
different from the one given by Eqs. (\ref{pb7})
and (\ref{pb5}). We conclude that QCD axions cannot
account for the value of the cosmological constant
during the early inflation or the late acceleration of the Universe. 
The early inflation and the late acceleration of the Universe could be produced
by another self-attractive complex SF with a ratio $|a_s|/m^3$ given by Eqs.
(\ref{pb7})
and (\ref{pb5}).

Let us try to determine the parameters of this hypothetical SF. The value of the
cosmological constant $\Lambda$ associated with the energy density
$\epsilon_\Lambda=\rho_\Lambda c^2$ determines the ratio $|a_s|/m^3$ according
to Eq. (\ref{pb1}), i.e.,
\begin{eqnarray}
\frac{|a_s|}{m^3}=\frac{c^2}{32\pi \hbar^2\rho_{\Lambda}}.
\label{spec1}
\end{eqnarray}
The begining of the inflation era identified with $a_i$ determines the charge of
the SF according to Eq. (\ref{negea1}). Using Eq. (\ref{spec1}), we get
\begin{eqnarray}
Qm=\frac{8}{3\sqrt{3}}\rho_\Lambda a_i^3.
\label{spec2}
\end{eqnarray}
Finally, the end of the inflation era identified (somehow arbitrarily) with
$a'_v$
determines the ratio $|a_s|/m$ from the relation
\begin{eqnarray}
\frac{a'_v}{a_i}=\frac{1}{(6\sqrt{3})^{1/3}}g\left (\frac{3|a_s|c^2}{4Gm}\right
)
\label{spec3}
\end{eqnarray}
obtained by combining Eqs. (\ref{negea1}) and (\ref{vip2}). In order to have
sufficient inflation, we need $a'_v\gg a_i$. This requires $\sigma\gg 1/4$
allowing
us to use the approximate expression (\ref{vip5}) of $g(\sigma)$. In that case,
Eq.
(\ref{spec3}) gives
\begin{eqnarray}
\frac{|a_s|c^2}{Gm}= \frac{9}{16}\left (\frac{a'_v}{a_i}\right )^6.
\label{spec4}
\end{eqnarray}
From Eqs. (\ref{spec1}) and (\ref{spec4}), we obtain
\begin{eqnarray}
m= \frac{3}{4}\left (\frac{a'_v}{a_i}\right )^3
m_\Lambda,\quad |a_s|=\frac{27}{2048\pi}\left
(\frac{a'_v}{a_i}\right )^9 r_\Lambda, 
\label{spec5}
\end{eqnarray}
where the mass and length scales $m_\Lambda$ and $r_\Lambda$ are defined in
Appendix \ref{sec_c}. Equation (\ref{spec5}) determines the mass $m$ and the
scattering length $a_s$ of the hypothetical SF producing the early inflation or
the late acceleration of the Universe in our model. Their precise values depend
on the ratio $a'_v/a_i$.

\section{The total potential}
\label{sec_tp}

\subsection{Spintessence}
\label{sec_spin}

The total potential of the SF including the rest-mass term and the
self-interaction
term is given by 
\begin{equation}
V_{\rm tot}(|\varphi|^2)=\frac{m^2c^2}{2\hbar^2}|\varphi|^2+\frac{2\pi
a_sm}{\hbar^2}|\varphi|^4.
\label{tp1}
\end{equation}
Using the relation from Eq. (\ref{kge7}) between the modulus of the SF and the
pseudo rest-mass density, we can rewrite it as
\begin{equation}
V_{\rm tot}=\frac{1}{2}\rho c^2+\frac{2\pi
a_s\hbar^2}{m^3}\rho^2=\frac{1}{2}\rho c^2 \left (1+\frac{4\pi
a_s\hbar^2}{m^3c^2}\rho\right ).
\label{tp3}
\end{equation}

We can study the evolution of the SF in the total potential $V_{\rm
tot}(|\varphi|^2)$ by using a mechanical analogy. To that
purpose, we write $\varphi=R e^{i\theta}$ where $R=|\varphi|$ (see Appendix
\ref{sec_ext}). The KG equation (\ref{suna1}) takes the form
\begin{equation}
\frac{1}{c^2}\frac{d^2R}{dt^2}+\frac{3H}{c^2}\frac{dR}{dt}=-\frac{dV_{\rm
tot}}{dR}+R\omega^2
\label{spin1}
\end{equation}
with
\begin{equation}
\omega^2=\frac{Q^2\hbar^2c^4}{R^4a^6}.
\label{spin2}
\end{equation}
This is similar to the equation describing the axisymmetric motion of a damped
particle in polar coordinates, where $R$ plays the role of the radial
distance, $\theta$ the angle, and $\omega=\dot\theta$ the angular velocity. The
fictive particle is submitted to a friction force $-(3H/c^2)\dot R$ (Hubble
drag) that tends to slow it down, a radial force $-dV_{\rm
tot}/dR$ that  tends to decrease $R$  and a
centrifugal force $R\omega^2$ that tends to increase $R$. This
centrifugal force is a specificity of a complex SF called spintessence
\cite{spintessence}. For a real SF, there is just the radial force so the SF
descends the potential towards $R=0$ and displays damped oscillations about it.
Because of the
presence of the centrifugal force, the evolution of a complex SF is richer.
The fast oscillation regime that we have considered corresponds to a quasistatic
equilibrium between the radial force and the centrifugal force:
\begin{equation}
\frac{dV_{\rm
tot}}{dR}=R\omega^2.
\label{spin3}
\end{equation} 
A complex SF has the tendency to spin with angular velocity $\omega$ at a
fixed radial
distance $R$. However, according to Eq. (\ref{spin2}) the angular velocity
decreases as the scale factor increases. Therefore, the centrifugal force
becomes less and less effective at time goes on. We can nevertheless maintain a
quasistatic equilibrium at any time if $dV_{\rm
tot}/dR$ decreases as the scale factor increases. As a result, the SF moves
towards an extremum of ${V_{\rm
tot}}$, either the minimum (when $a_s\ge 0$ or $a_s<0$)
or the maximum (when $a_s<0$). We now describe more specifically the evolution
of the SF in the total potential $V_{\rm
tot}(|\varphi|^2)$ for the  solutions obtained in Secs. \ref{sec_pos} and
 \ref{sec_neg}.

\subsection{The case $a_s\ge 0$}

\begin{figure}
\scalebox{0.33}{\includegraphics{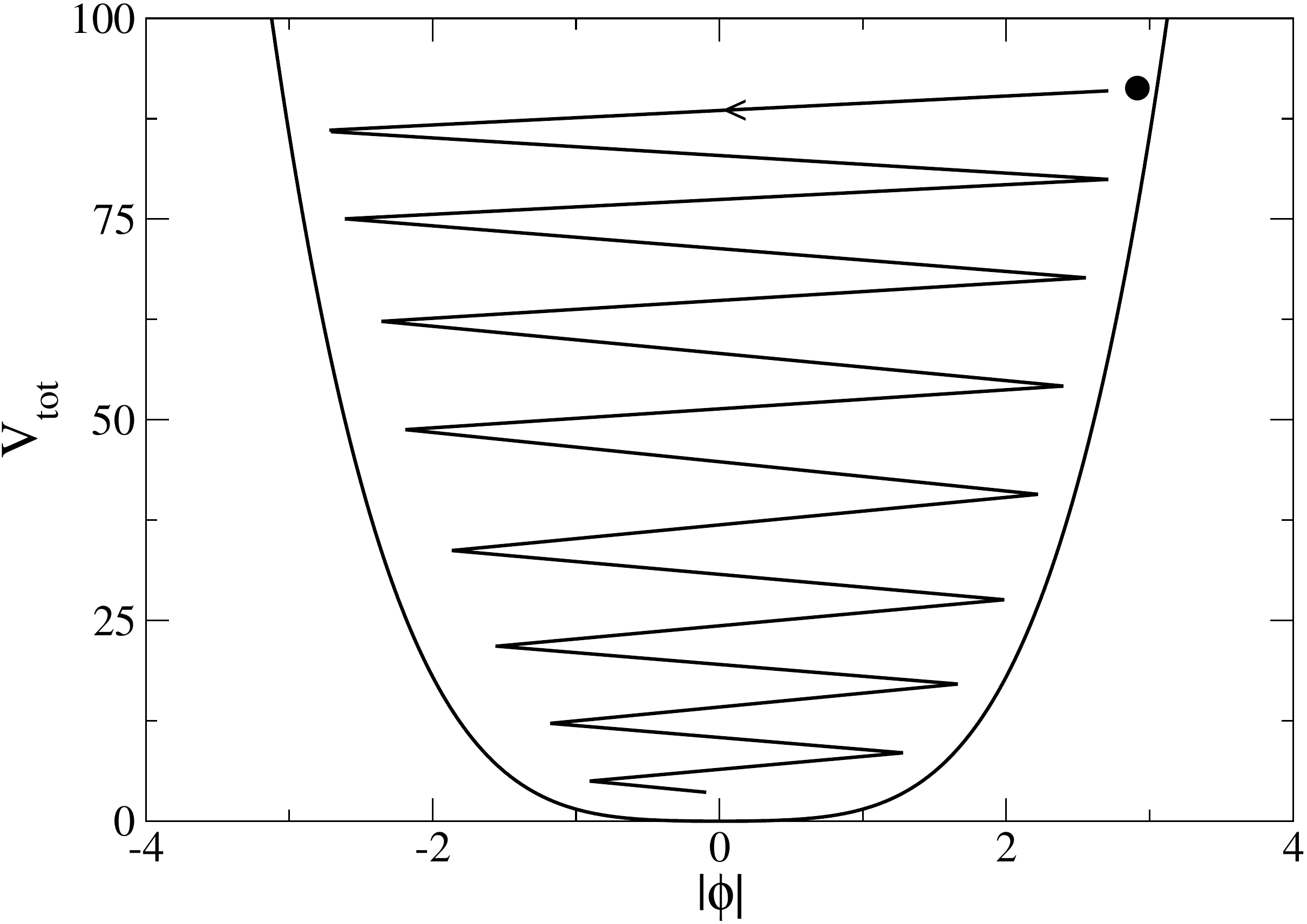}} 
\caption{Motion of the SF in the total potential $V_{\rm
tot}$ when $a_s\ge 0$. The zig-zag is a rough representation of the
spiralling motion of the SF in the (3D) potential.}
\label{rvtot}
\end{figure}

When $a_s\ge 0$, the total potential (see Fig. \ref{rvtot}) has a single
minimum $V_{\rm tot}=0$ at $|\varphi|=\rho=0$. In the
fast oscillation regime studied in  Sec. \ref{sec_pos}, the SF descends the
potential from $+\infty$ to $0$. The initial value of
$\dot{|\varphi|}_i$ (``velocity'') is $-\infty$ (see Appendix \ref{sec_ic}).

The total potential $V_{\rm tot}$ starts  at
$a=0$ from $+\infty$ and decreases to
$0$ as $a\rightarrow +\infty$. For $a\rightarrow 0$:
\begin{equation}
V_{\rm tot}\sim\frac{1}{2}(Q^4m\pi a_s\hbar^2c^4)^{1/3}\frac{1}{a^4}.
\end{equation}
For $a\rightarrow +\infty$:
\begin{equation}
V_{\rm tot}\sim\frac{Qmc^2}{2a^3}.
\end{equation}
Assuming that the Universe contains only the SF and using the results of Sec.
\ref{sec_post}, we can obtain the temporal evolution of $V_{\rm tot}$. For
$t\rightarrow 0$:
\begin{equation}
V_{\rm tot}\sim\frac{c^2}{32\pi Gt^2}.
\end{equation}
For $t\rightarrow +\infty$:
\begin{equation}
V_{\rm tot}\sim\frac{c^2}{12\pi G t^2}.
\end{equation}

In the stiff matter era, corresponding to the slow oscillation regime, using Eq.
(87) of \cite{b40}, we find that the SF behaves as
\begin{equation}
|\varphi|\sim \left (\frac{3c^4}{4\pi G}\right )^{1/2}(-\ln a).
\label{sti}
\end{equation}
The stiff matter era precedes the radiation and matter eras.  The SF $|\varphi|$
starts from $+\infty$ and decreases with time. The SF descends the
potential. As shown in Appendix \ref{sec_stiff}, the stiff
matter era (slow oscillation regime) connects smoothly the radiation and
matter eras (fast oscillation regime) at $a\sim a_v$ where $a_v$ is given by Eq.
(\ref{valp5}).

\subsection{The case $a_s<0$}
\label{sec_tpn}

\begin{figure}
\scalebox{0.33}{\includegraphics{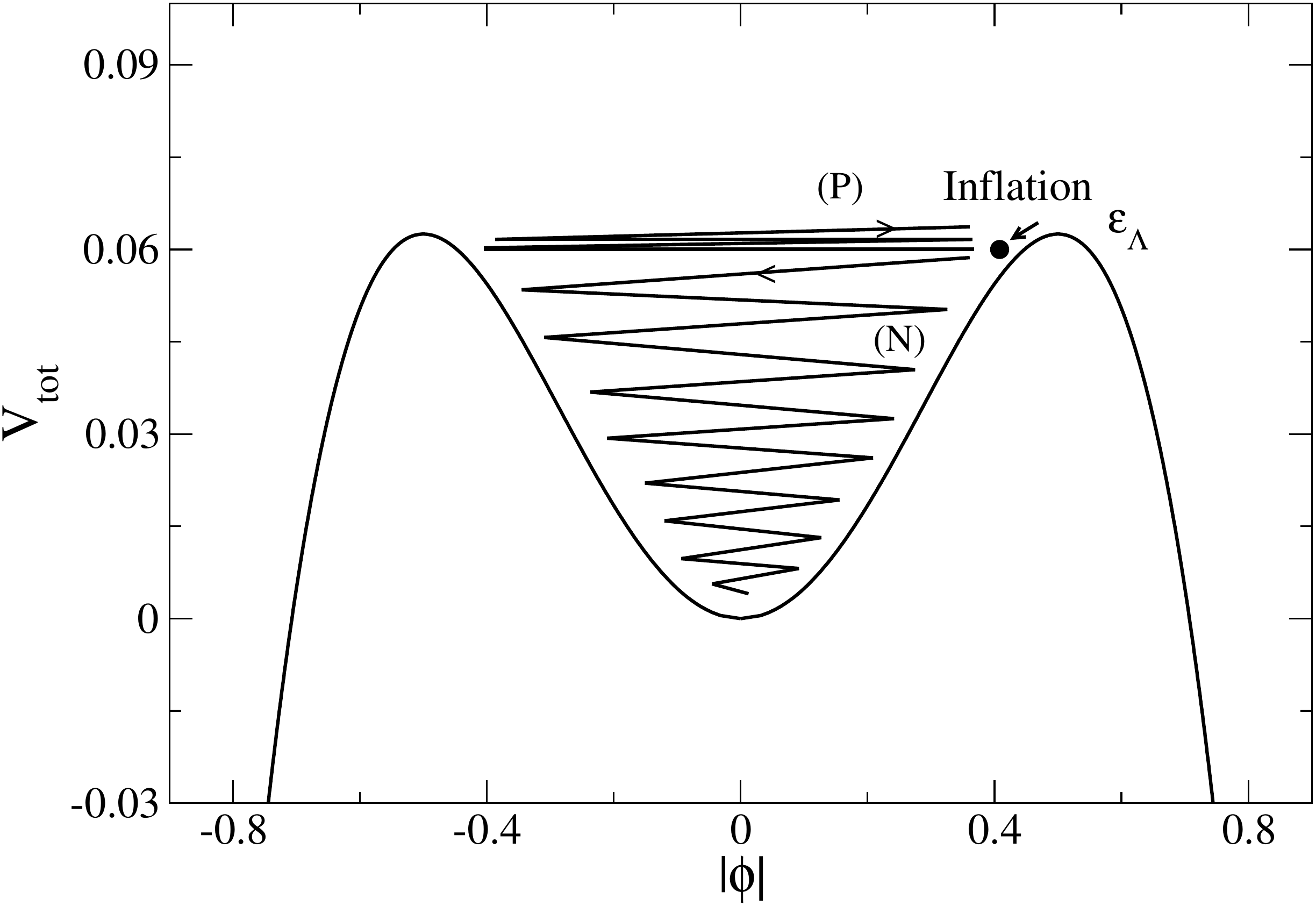}} 
\caption{Motion of the SF in the total potential $V_{\rm
tot}$ when $a_s<0$.}
\label{rvtotneg}
\end{figure}

When $a_s<0$, the total potential (see Fig. \ref{rvtotneg}) has a local
minimum $V=0$ at
$|\varphi|=\rho=0$ and a maximum  at 
\begin{equation}
|\varphi_{\Lambda}|=\left (\frac{m c^2}{8\pi
|a_s|}\right )^{1/2},\qquad \rho_{\Lambda}=\frac{m^3c^2}{8\pi |a_s|\hbar^2},
\label{tp4}
\end{equation}
whose value is
\begin{equation}
(V_{\rm tot})_{\Lambda}=\frac{1}{4}\rho_{\Lambda}
c^2=\epsilon_{\Lambda}=\frac{m^3c^4}{32\pi|a_s|\hbar^2}.
\label{tp4b}
\end{equation}
In the fast oscillation regime studied in Sec.
\ref{sec_neg}, the SF starts from
\begin{equation}
|\varphi_{i}|=\left (\frac{m c^2}{12\pi
|a_s|}\right )^{1/2},\qquad \rho_{i}=\frac{m^3c^2}{12\pi |a_s|\hbar^2},
\label{tp5}
\end{equation}
corresponding to a total potential
\begin{equation}
(V_{\rm tot})_i=\frac{1}{3}\rho_ic^2=\frac{m^3c^4}{36\pi|a_s|\hbar^2}.
\end{equation}
We note that $|\varphi_{i}|$ differs from the inflexion point  ($d^2V_{\rm
tot}/d|\varphi|^2=0$) of the potential given by
\begin{equation}
|\varphi_{\rm I}|=\left (\frac{m c^2}{24\pi
|a_s|}\right )^{1/2},\qquad \rho_{\rm I}=\frac{m^3c^2}{24\pi
|a_s|\hbar^2},
\label{tp6}
\end{equation}
corresponding to a total potential
\begin{equation}
(V_{\rm tot})_I=\frac{5}{12}\rho_I c^2=\frac{5m^3c^4}{288\pi|a_s|\hbar^2}.
\end{equation}
We find $\rho_i=2\rho_{\rm I}$, $|\varphi_i|=\sqrt{2}|\varphi_{\rm I}|$, and
$(V_{\rm tot})_i=(8/5) (V_{\rm tot})_I$. On the normal branch, the SF descends
the potential from $|\varphi_{i}|$ to
$0$. The initial value of $\dot{|\varphi|_i}$ (``velocity'') is $-\infty$ (see
Appendix \ref{sec_ic}). On the peculiar branch, the SF ascends the potential
from $|\varphi_{i}|$
to $|\varphi_{\Lambda}|$.  The initial value of $\dot{|\varphi|_i}$
(``velocity'') is $+\infty$ (see
Appendix \ref{sec_ic}).  Since $|\varphi_{\Lambda}|$ corresponds to the maximum
$(V_{\rm tot})_\Lambda=\epsilon_{\Lambda}$ of the
potential, we understand why the SF reaches a de
Sitter regime $\epsilon\simeq\epsilon_\Lambda$ at late
times.

\begin{figure}
\scalebox{0.33}{\includegraphics{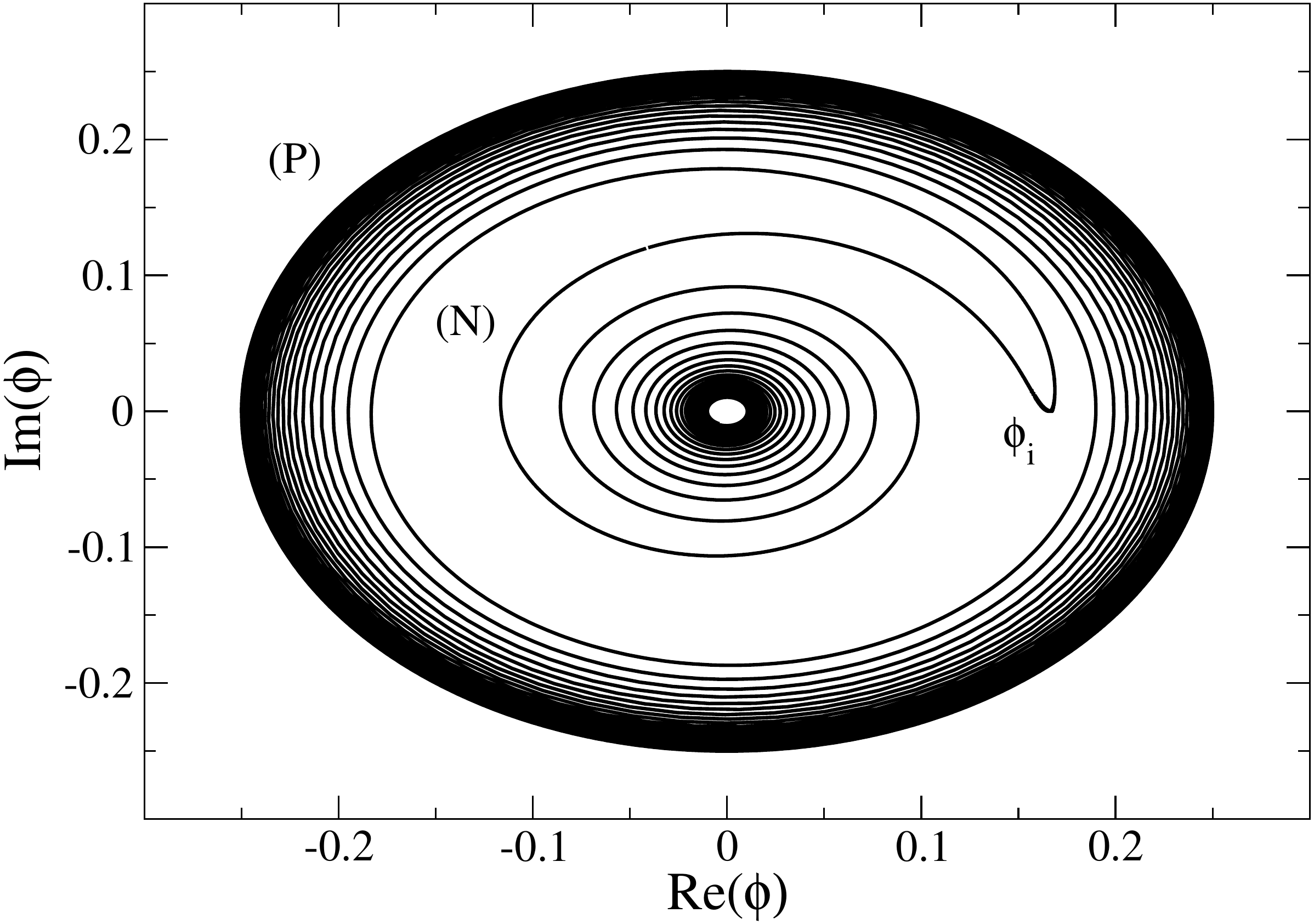}} 
\caption{Temporal evolution of the complex SF (spintessence). On the normal
branch, it spirals towards the center giving rise to a matterlike era. On
the peculiar branch, it reaches a limit cycle giving rise to a de Sitter era.}
\label{spintessence}
\end{figure}

It
is unusual, but not impossible, that the SF ascends the potential.
In the present case, the SF ascends the potential, and maintains an almost
constant value of $|\varphi|$, because of the centrifugal force that is
specific to a complex SF. Indeed, the SF
is in a quasistatic equilibrium between the ``attractive'' radial force and the
``repulsive'' centrifugal force [see Eq. (\ref{spin3})]. As the scale factor $a$
increases, the SF slowly moves towards the
maximum of
$V_{\rm tot}(|\varphi|^2)$ so as to
decrease $dV_{\rm tot}/d|\varphi|$  (see Sec. \ref{sec_spin}).
Therefore, the almost constant
value of $|\varphi|$ giving rise
to a de Sitter era and to an effective cosmological constant is a manifestation
of
spintessence for a complex SF with an attractive self-interaction. We now
understand the origin of the two branches (N) and (P) corresponding to DM and
DE. A SF with an attractive self-interaction can have two possible evolutions
because the total SF potential has two extrema: a minimum at $|\varphi|=0$ and a
maximum at $|\varphi|=|\varphi_{\Lambda}|$. According to the discussion of Sec.
\ref{sec_spin}, a SF in quasistatic equilibrium moves towards an extremum of
$V_{\rm tot}$ (see Fig. \ref{spintessence}). If the SF starts from $|\varphi_i|$
with a velocity $\dot{|\varphi|_i}=+\infty$ (or from $|\varphi_0|>|\varphi_i|$
with a finite positive velocity), it will ascend the potential towards the
maximum in order to decrease $dV_{\rm tot}/d|\varphi|$. In that case, it will
behave as DE. If the SF starts from $|\varphi_i|$ with a velocity 
$\dot{|\varphi|_i}=-\infty$ (or from
$|\varphi_0|<|\varphi_i|$ with a finite negative velocity), it will descend the
potential towards the minimum in order to decrease $dV_{\rm tot}/d|\varphi|$. In
that case, it will
behave as DM.
Therefore, depending on the initial condition, the SF may behave either as DM
or as DE. Note that a SF with a repulsive self-interaction can only have one
possible
evolution because the total SF potential has only one extremum: a minimum at
$|\varphi|=0$. It can only descend the potential towards the
minimum in order to decrease $dV_{\rm tot}/d|\varphi|$.

\begin{figure}
\scalebox{0.33}{\includegraphics{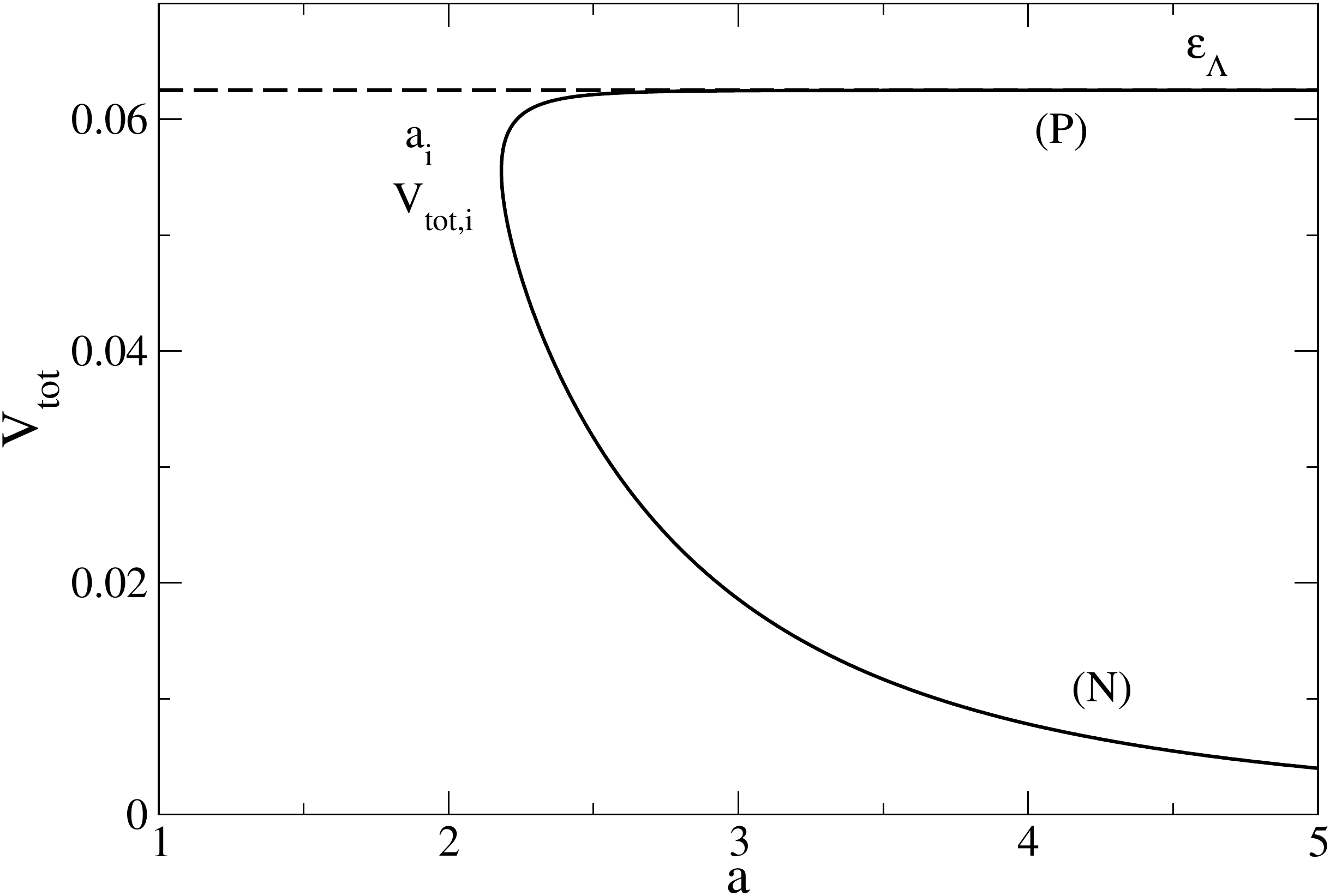}} 
\caption{Total SF potential $V_{\rm tot}$ as a function of the scale factor
$a$ when $a_s<0$.}
\label{avtotneg}
\end{figure}

The  evolution of the total potential $V_{\rm tot}$ with the
scale factor $a$ is
plotted in Fig. \ref{avtotneg}. The potential starts at $a=a_i$ from $(V_{\rm
tot})_i$. For
$a\rightarrow a_i$:
\begin{equation}
V_{\rm
tot}\simeq\frac{\rho_ic^2}{3}\left\lbrack
1\pm\sqrt{\frac{1}{2}\left(\frac{a}{a_i }
-1\right)}\right\rbrack.
\end{equation}
On the normal branch, the potential decreases as the scale factor increases and
asymptotically tends to $0$. For $a\rightarrow +\infty$:
\begin{equation}
V_{\rm tot}\sim\frac{Qmc^2}{2a^3}.
\end{equation}
On the peculiar branch, the potential increases as the scale factor increases
and asymptotically tends to its maximum value
$(V_{\rm tot})_\Lambda=\epsilon_{\Lambda}$. For $a\rightarrow +\infty$:
\begin{equation}
V_{\rm tot}\simeq\epsilon_\Lambda\left[1-\left(\frac{8\pi
Q|a_s|\hbar^2}{m^2c^2a^3}\right)^4\right].
\end{equation}
Assuming that the Universe contains only the SF and
using
the results of Sec. \ref{sec_negt}, we can obtain the temporal evolution of
$V_{\rm tot}$.
For $t\rightarrow
0$:
\begin{equation}
V_{\rm
tot}\simeq\frac{\rho_ic^2}{3}\left\lbrack 1\pm\frac{1}{\sqrt2}\left(\frac{4\pi
G\rho_i}{3}\right)^{1/4}t^{1/2}\right\rbrack.
\end{equation}
On the normal branch, for $t\rightarrow +\infty$:
\begin{equation}
V_{\rm tot}\sim\frac{c^2}{12\pi G t^2}.
\end{equation}
On the peculiar branch, for $t\rightarrow +\infty$, $V_{\rm tot}$ converges to
its asymptotic value $\epsilon_{\Lambda}$ exponentially rapidly.

We now comment on the early evolution of the SF, before the fast oscillation
regime. We first consider the normal branch. In the stiff matter era,
corresponding to the slow oscillation regime, the evolution of the SF
$|\varphi|$ is given by Eq.
(\ref{sti}). It starts from $+\infty$ and decreases with time. When
$a_s<0$, this solution implies that the SF should climb the outer branch of
the
potential (see Fig. \ref{rvtotneg}). This is very unlikely, if not impossible
(when $a_s=0$, there is no
problem because this branch is rejected at infinity). This suggests that
the stiff matter era is not valid at very early times when $a_s<0$.  On the
other hand, it is shown in Appendix \ref{sec_stiff} that the stiff matter era
can be connected smoothly to the matterlike era (at $a_v$) only when
$|a_s|\lesssim
r_S$ (briefly, this is because $|\tilde\varphi|_{\rm stiff}\sim \sigma$
according to Eq. (\ref{stiff1b}) so we need $\sigma\lesssim 1$ to have
$|\tilde\varphi|_{\rm stiff}\lesssim
|\tilde\varphi|_i=1/\sqrt{6}$). Therefore, the duration of the stiff matter era
should decrease as
$|a_s|$ increases up to $\sim r_S$. These results suggest that the stiff matter
era exists only for $a_i\le a\le a_v'$ when $|a_s|< r_S/6$ and does not exist
when 
$|a_s|> r_S/6$. We can now wonder what happens at very early times, before
$a_i$.
When $a_s<0$, the SF could start from the top of the potential and descend the
potential along the inner branch until it connects the solution described
previously at $a_i$. The initial motion $0\le a\le a_i$ is not in the fast
oscillation regime, nor in the slow oscillation regime. Therefore, quantum
mechanics must be taken into account, and we must solve the
exact equations (\ref{b4})-(\ref{b3}). Since the SF starts from
the maximum of the potential at $(V_{\rm tot})_\Lambda=\epsilon_{\Lambda}$, the
initial motion of the SF corresponds to a phase of inflation with a constant
energy density given by Eq. (\ref{pb1}). In our model, the inflation era is due
to the negative scattering length of the SF ($a_s<0$). This could give
a physical justification of why the pressure is negative during inflation.
 Expanding the total potential close to the maximum,
we get 
\begin{equation}
V_{\rm tot}\simeq
\epsilon_\Lambda-\frac{m^2c^2}{\hbar^2}(|\varphi|-|\varphi|_\Lambda)^2.
\end{equation}
This corresponds to an inverted $|\varphi|^2$ potential with an effective
mass $m_*=\sqrt{2}m$  interpreted as the inflaton mass (actually the mass is
imaginary). In our model, the same SF describes the inflation in the
early Universe (top of the potential) and the formation of DM halos in the
matterlike era (bottom of the potential). In order to account for the size of DM
halos, a SF with an attractive self-interaction must
have a mass given by Eq. (\ref{mas2}) and a very small scattering length $|a_s|$
satisfying the inequality (\ref{ggg}). If this same SF experiences
an inflation era with a constant energy density equal to the Planck density,
it must fulfill the constraint (\ref{pb7}). Combining Eqs. (\ref{mas2}) and
(\ref{pb7}), we obtain
\begin{equation}
m=2.92\times 10^{-22}\, {\rm eV}/c^2,\quad a_s=-2.20\times 10^{-171}\, {\rm fm},
\end{equation}
corresponding to $\lambda/8\pi=-3.26\times 10^{-201}$.

We now consider the peculiar branch. It is shown in Appendix
\ref{sec_stiff} that the stiff matter era can never be connected smoothly to
the peculiar branch. In addition, it is not clear what kind of phase could
appear before $a_i$ because, in the most likely scenario, the SF should first
descend the potential (for $a<a_i$) and suddenly reverse its motion and ascend
it (for $a>a_i$). We conclude that, if the peculiar branch ever makes sense, the
SF should emerge suddenly at $a_i$ from a very particular initial condition.
For what concerns its evolution after $a'_v$, when the fast
oscillation regime ceases to be valid, we may argue that the centrifugal force
becomes inefficient to maintain an almost constant energy density and that the
de Sitter regime comes to an end. The motion of the SF in the total potential is
reversed. The SF descends the potential, connects the normal branch for
$|\varphi|<|\varphi_i|$, and ultimately behaves as pressureless matter.
In that case, there will be a transition from acceleration to
deceleration in the late Universe. However,
the de Sitter regime can be sufficiently long  to be
physically relevant (see Sec. \ref{sec_effl}).

\section{Conclusion}
\label{sec_con}

In this paper, we have studied the cosmological evolution of a complex SF with 
repulsive or attractive self-interaction using a fully general relativistic
treatment. The SF may be interpreted as the wave function of a BEC. Although a
SF
is generally not a fluid, it can be studied through the hydrodynamic
representation of the KGE equations \cite{b38,b39,b40,b41,chavmatos}. For a 
$|\varphi|^4$ self-interaction, the parameters of the SF are the mass $m$ of the
bosons and their scattering length $a_s$. We have introduced a new length scale 
$r_S={2Gm}/{c^2}$, which can be interpreted as the effective Schwarzschild
radius of
the bosons. The evolution of the SF depends on how the scattering
length of
the bosons $a_s$ compares with their effective Schwarzschild radius $r_S$. Our
results can
be summarized as follows.

In the case of repulsive self-interaction ($a_s\ge 0$), we have confirmed and
complemented the results of Li {\it et al.} \cite{b36}. We have given many
analytical formulae that allow us to understand the results better and play more
easily with the parameters. When
$a_s<(4/21)r_S$, the SF undergoes a stiff matter era ($w=1$) followed by a
matter
era ($w=0$). There is no radiationlike era, even though $a_s>0$. For a
noninteracting SF with $m=2.92\times 10^{-22}\, {\rm eV}/c^2$,  the transition
takes place at $a_v=1.86\times 10^{-8}$. When
$a_s>(4/21)r_S$, the SF undergoes a stiff matter era ($w=1$) followed by a
radiationlike era ($w=1/3$), and finally a matterlike era ($w=0$). 
For a SF
with $m=3\times 10^{-21}\, {\rm
eV}/c^2$ and $a_s=1.11\times 10^{-58}\, {\rm fm}$, corresponding to the fiducial
model of Li {\it et al.} \cite{b36}, the transition between the stiff matter era
and the
radiationlike era takes place at $a_v=5.14\times 10^{-11}$ and the transition
between the radiationlike era and the matterlike era takes place at
$a_t=1.35\times 10^{-5}$. For a SF with $m=1.10\times 10^{-3}\, {\rm
eV}/c^2$ and $a_s=4.41\times 10^{-6}\, {\rm fm}$ (see Appendix \ref{sec_mas}),
we get $a_v=1.45\times 10^{-28}$ and $a_t=1.26\times
10^{-5}$. In both cases, the SF behaves at large scales as pressureless matter
(like the CDM model) at, and after, the epoch of radiation-matter equality
$a_{\rm
eq}=2.95\times 10^{-4}$. However, its intrinsic nonzero pressure (either due to
the scattering of the bosons or to the quantum potential taking into
account the Heisenberg uncertainty principle) manifests itself at small scales
and can balance the gravitational attraction. This leads to DM halos that
present a core (BEC/soliton) instead of a cusp. These cores are surrounded by a
halo with a NFW profile made of scalar radiation resulting from gravitational
cooling. Therefore, a SF with $a_s\ge 0$ has a lot of nice properties and is a
serious DM candidate that could solve the CDM small-scale
crisis.

The case of attractive self-interaction ($a_s<0$) has been studied in our paper
for the first time. We have found that the SF can
evolve along two different
branches, a normal branch where it behaves as DM and a peculiar
branch where it behaves as DE. We first consider the normal branch.
When $|a_s|=0$, the SF undergoes a stiff matter
era ($w=1$) followed by a matter era ($w=0$). When $0<|a_s|< r_S/6$, the SF
undergoes an inflation era, a stiff matter
era ($w=1$), and a matter era ($w=0$). 
The duration of the stiff
matter era decreases as the self-interaction $|a_s|$ increases.
When $|a_s|> r_S/6$, there is no stiff matter era anymore. The 
SF undergoes an inflation era, a very short cosmic stringlike
era ($w=-1/3$), and a matterlike era ($w=0$).  For QCD axions with 
$m=10^{-4}\, {\rm eV}/c^2$ and $a_s=-5.8\times 10^{-53}\, {\rm m}$, the
matterlike era starts at $a_i=5.69\times 10^{-15}$. For ultralight axions  with 
$m=2.19\times 10^{-22}\, {\rm eV}/c^2$ and $a_s=-1.11\times 10^{-62}\, {\rm
fm}$, we get $a_i=1.50\times 10^{-5}$. In each
case
$a_i\ll a_{\rm
eq}=2.95\times 10^{-4}$ so the axionic SF behaves at large scales as
pressureless matter (like the
CDM model) at, and after, the epoch of radiation-matter equality. However, its
intrinsic nonzero pressure  manifests itself at small scales. The 
quantum pressure arising from the Heisenberg uncertainty principle is always
repulsive but the negative pressure due to the self-interaction is attractive
and adds to the gravitational attraction. This can destabilize the halo. Stable
DM halos exist only below a maximum mass \cite{b34,cd,bectcoll}.
For QCD axions with $m=10^{-4}\, {\rm
eV}/c^2$ and $a_s=-5.8\times 10^{-53}\, {\rm m}$ this mass  $M_{\rm
max}=6.5\times 10^{-14}\, M_{\odot}$ is too small to account for the mass of DM
halos.
Therefore, QCD axions cannot form DM halos. They rather form mini axion
stars that could be the constituents of  DM halos in the form of
mini massive compact halo objects (mini-MACHOs) \cite{bectcoll}. However,
they would essentially behave as CDM and would not solve the CDM small-scale
crisis. The maximum
mass of self-gravitating axions
becomes of the order of the mass of DM halos $M \sim 10^8\, M_{\odot}$ in the
case of
ultralight axions with a mass $m=2.19\times 10^{-22}\, {\rm eV}/c^2$ and a very
weak self-interaction $a_s=-1.11\times
10^{-62}\, {\rm fm}$ \cite{bectcoll}. Such ultralight axions could  solve the
CDM small-scale
crisis. We now consider
the peculiar branch on
which the SF
behaves as DE  ($w=-1$) with an almost constant energy density. This peculiar
branch is valid only when $|a_s|> r_S/6$. It starts at $a_i$ and ceases to be
valid at $a_v'$. On
this branch, the SF is equivalent to an effective cosmological constant given
by 
$\Lambda_{\rm eff}={Gm^3c^2}/{4|a_s|\hbar^2}$. A complex SF with a negative
scattering
length could be a new mechanism to
produce a cosmological constant. Cosmic acceleration could arise from the
attractive self-interaction term present in the
SF potential. That could justify why the pressure of DE is
negative.

To our knowledge, the effective Schwarzschild radius of the bosons $r_S=2Gm/c^2$
has not been introduced before. It arises naturally in the
equations of the problem in order to separate the weakly self-interacting regime
$\sigma=3a_s/2r_S\ll 1$ from
the strongly self-interacting regime $\sigma\gg 1$.  Actually, for
ultralight bosons with a mass $m=2.92\times 10^{-22}\, {\rm eV}/c^2$, even for
a
very small value of $a_s\sim 10^{-68}\, {\rm fm}$ (or, equivalently, for a value
of the dimensionless
self-interaction constant $\lambda/8\pi$ as small as $10^{-98}$) we
are in
the strongly self-interacting regime, not in the weakly self-interacting regime
(see Sec.
\ref{sec_sr} and Appendix A.3 of \cite{cd}). This is because $\sigma\gg 1$
while $\lambda\ll 1$. Therefore, it is
important to
take into account the nonzero value of the self-interaction constant in the
problem even if it looks extremely small. This feature has been
overlooked in previous works that often consider a noninteracting SF (see,
e.g., \cite{marshrev,nature}). In this connection, we recall that the
cosmological bounds obtained
by Li {\it et al.} \cite{b36} exclude noninteracting SFs.

Although observations tend to favor the $\Lambda$CDM model, other cosmological
models cannot be rejected. The SF model is extremely rich and can have important
implications concerning the nature of DM and DE in the Universe. Therefore, SFs
should be considered as serious alternatives to find an answer to these
paradigms. We have obtained very intriguing results that deserve to be developed
in future works. For example, it is important to study what happens at very
early times, before the fast oscillation regime, when $a_s<0$. This requires to
go beyond the TF approximation and take into account quantum mechanics (i.e. the
quantum potential) by solving
the exact equations (\ref{b4})-(\ref{b3}). The analytical results
obtained in the present paper, valid in the fast oscillation regime, may be
useful to make the matching with this primordial era.  An important suggestion
of our work that needs to be confirmed is that a SF with an attractive
self-interaction ($a_s<0$) can produce a phase of early inflation followed by a
stiff matter era and/or a matter era. It is important to
determine whether this model can account for the observations because,
in that case, we could describe different phases of the
Universe with a single SF  (see the still
``hypothetical'' phase diagrams of Figs. \ref{gpd} and \ref{gpdpeculiar}). Many
other
developments are also possible. For example, we have assumed that the SF has a
$|\varphi|^4$ self-interaction potential and that this potential remains the
same during the whole history of the Universe. Of course, if the
SF has a different potential $V(|\varphi|^2)$, or if its potential changes
during the history of the Universe (for example in the very early Universe), our
conclusions must be revised. For simplicity, we have
focused on a $|\varphi|^4$ interaction but we could consider more general
potentials (see Appendix \ref{sec_plsf}). For example,
when
$|\varphi|$ is large (early Universe), the
$|\varphi|^4$ approximation may not be valid anymore and higher order terms in
the expansion of the potential should be considered. We have established
the
general equations to perform these studies. They will be considered in future
works. 

\acknowledgments{A. S. acknowledges CONACyT for the postdoctoral grant
received.}

\appendix

\section{The fast oscillation regime $\omega\gg H$ from the field theoretic
approach}
\label{sec_ext}

In this Appendix, we consider the fast oscillation regime $\omega\gg H$ from the
field theoretic approach based on the KG equation. We generalize the results
of Li {\it et al.} \cite{b36} to an arbitrary potential of interaction
$V(|\varphi|^2)$ and show the equivalence with the hydrodynamic
approach of Sec. \ref{sec_tb}.

\subsection{General equations}

The KG equation for a spatially homogeneous SF is given by Eq. (\ref{h1}).
Decomposing the complex SF as
\begin{eqnarray}
\varphi=|\varphi|e^{i\theta},
\label{ext2}
\end{eqnarray}
inserting this decomposition into the KG equation (\ref{h1}), and separating the
real and imaginary parts, we obtain
\begin{eqnarray}
\frac{1}{c^2}\left\lbrack \frac{d^2|\varphi|}{dt^2}-|\varphi| \left
(\frac{d\theta}{dt}\right
)^2\right\rbrack+\frac{3H}{c^2}\frac{d|\varphi|}{dt}\nonumber\\
+ \frac{m^2
c^2}{\hbar^2}|\varphi|+2\frac{dV}{d|\varphi|^2}|\varphi|=0,
\label{ext3}
\end{eqnarray}
\begin{eqnarray}
\frac{1}{c^2}\left (2\frac{d|\varphi|}{dt}\frac{d\theta}{dt}+|\varphi|
\frac{d^2\theta}{dt^2}\right
)+\frac{3H}{c^2}|\varphi|\frac{d\theta}{dt}=0. 
\label{ext4}
\end{eqnarray}
Equation (\ref{ext4}) can be exactly integrated once giving
\begin{eqnarray}
\frac{d}{dt}\left (a^3|\varphi|^2\frac{d\theta}{dt}\right )=0.
\label{ext6}
\end{eqnarray}
This can be rewritten as
\begin{eqnarray}
a^3|\varphi|^2\frac{d\theta}{dt}=-Q\hbar c^2,
\label{ext7}
\end{eqnarray}
where $Q$ is the charge of the SF \cite{arbeycosmo,gh,b36,b40}.

In the fast oscillation regime $H=\dot a/a\ll d\theta/dt$, introducing
the pulsation $\omega={d\theta}/{dt}$,  Eq. (\ref{ext3}) reduces to
\begin{eqnarray}
\omega^2=\frac{m^2c^4}{\hbar^2}+2c^2\frac{dV}{d|\varphi|^2}.
\label{ext5}
\end{eqnarray}
As
pointed out in \cite{b36}, this approximation also requires that
$|\varphi|^{-1}d|\varphi|/dt\ll d\theta/dt$, a condition that is not always
satisfied.
For a free field ($V=0$), the pulsation $\omega$ is proportional
to the mass of the SF ($|\omega|=mc^2/\hbar$) and the fast oscillation
condition reduces to $mc^2/\hbar\gg H$. 

{\it Remark:} To make the
link with the hydrodynamical approach, we use $|\varphi|=(\hbar/m)\sqrt{\rho}$,
$\theta=S_{\rm tot}/\hbar$ and $\omega=-E_{\rm tot}/\hbar$. Then, Eqs.
 (\ref{ext6}), (\ref{ext7}) and (\ref{ext5}) return Eqs. 
(\ref{b5}), (\ref{b6}) and (\ref{b12}), respectively.

\subsection{Spintessence}

From Eqs. (\ref{ext3}) and (\ref{ext7}) we obtain
\begin{equation}
\frac{d^2|\varphi|}{dt^2}+3H\frac{d|\varphi|}{dt}
+ \frac{m^2
c^4}{\hbar^2}|\varphi|
+2c^2\frac{dV}{d|\varphi|^2}|\varphi|-\frac{
Q^2\hbar^2c^4}{a^6|\varphi|^3}=0.
\label{suna1}
\end{equation}
This equation differs from the KG equation of a real SF by the presence of the
last term and the fact that $\varphi$ is replaced by $|\varphi|$. The last term
coming from the ``angular motion'' of the complex SF can be interpreted as a
``centrifugal force'' (see Sec. \ref{sec_spin}) whose strength depends on the
charge of the complex SF
\cite{gh}. Equation (\ref{suna1}) can be rewritten as
\begin{equation}
\frac{d^2|\varphi|}{dt^2}+3H\frac{d|\varphi|}{dt}
+ \frac{m^2
c^4}{\hbar^2}|\varphi|
+2c^2\frac{dV_{\rm eff}}{d|\varphi|^2}|\varphi|=0,
\label{suna2q}
\end{equation}
where
\begin{equation}
V_{\rm eff}(|\varphi|^2)=V(|\varphi|^2)+\frac{Q^2\hbar^2c^2}{2a^6|\varphi|^2}
\label{suna3}
\end{equation}
is an effective potential incorporating the centrifugal potential. The presence
of the centrifugal force for a complex SF is a crucial difference with the case
of a real SF (that is not charged) because the fast oscillation approximation
(\ref{ext5}) corresponds to the equilibrium between the centrifugal potential
and the total SF potential:
\begin{equation}
\frac{Q^2\hbar^2c^4}{a^6|\varphi|^4}=\frac{m^2
c^4}{\hbar^2}
+2c^2\frac{dV}{d|\varphi|^2}.
\label{suna2}
\end{equation}
This is what Boyle {\it et al.} \cite{spintessence} call
``spintessence''. Equation (\ref{suna2}) is equivalent to Eq. (\ref{b13}). Such
a
relation does not hold for a real SF. We note that $|\varphi|$ does  not
oscillate in the fast oscillation
regime when the condition (\ref{suna2}) is fulfilled. 

\subsection{EOS in the fast oscillation regime}

To establish the EOS in the fast oscillation regime, Li {\it et
al.} \cite{b36} proceed as follows (see also
\cite{turner,ford,pv,mul}). Multiplying the KG equation (\ref{h1}) by
$\varphi^*$ and averaging over a
time interval  that is much longer than the field oscillation period
$\omega^{-1}$, but much shorter than the Hubble time $H^{-1}$,  we obtain 
\begin{eqnarray}
\frac{1}{c^2}\left\langle \left |\frac{d\varphi}{dt}\right
|^2\right\rangle=\frac{m^2c^2}{\hbar^2}\langle |\varphi|^2\rangle+2\left\langle
\frac{dV}{d|\varphi|^2}|\varphi|^2\right\rangle.
\label{ext10}
\end{eqnarray}
This relation constitutes a sort of virial theorem. For a spatially
homogeneous SF, the energy density and the pressure are given
by Eqs. (\ref{h2}) and (\ref{h3}). Taking the average value of the energy
density and pressure, using Eq.
(\ref{ext10}), and making the approximation
\begin{eqnarray}
\left\langle \frac{dV}{d|\varphi|^2}|\varphi|^2\right\rangle\simeq V'(\langle
|\varphi|^2\rangle)\langle |\varphi|^2\rangle,
\label{ext11}
\end{eqnarray} 
we obtain
\begin{eqnarray}
\langle\epsilon\rangle=\frac{m^2c^2}{\hbar^2}\langle
|\varphi|^2\rangle+V'(\langle
|\varphi|^2\rangle)\langle |\varphi|^2\rangle+V(\langle
|\varphi|^2\rangle),\quad
\label{ext12}
\end{eqnarray} 
\begin{eqnarray}
\langle P \rangle=V'(\langle
|\varphi|^2\rangle)\langle |\varphi|^2\rangle-V(\langle |\varphi|^2\rangle).
\label{ext13}
\end{eqnarray} 
This returns Eqs. (\ref{b19}) and (\ref{b20}). The EOS parameter is given by
\begin{eqnarray}
w=\frac{P}{\epsilon}=\frac{V'(\langle
|\varphi|^2\rangle)\langle |\varphi|^2\rangle-V(\langle
|\varphi|^2\rangle)}{\frac{m^2c^2}{\hbar^2}\langle |\varphi|^2\rangle+V'(\langle
|\varphi|^2\rangle)\langle |\varphi|^2\rangle+V(\langle
|\varphi|^2\rangle)}.\nonumber\\
\label{ext14}
\end{eqnarray}

{\it Remark:} writing Eqs. (\ref{ext10}) and (\ref{ext11}) with
hydrodynamic variables, and ignoring the averages, we obtain
\begin{eqnarray}
\frac{\hbar^2}{8m^2c^2}\frac{1}{\rho}\left (\frac{d\rho}{dt}\right )^2+\left
(\frac{E}{2mc^2}+1\right )\frac{\rho E}{m}=V'(\rho)\rho.
\label{ext15}
\end{eqnarray}  
If we substitute this equation into Eqs. (\ref{b7}) and (\ref{b8}), we obtain
Eqs. (\ref{b19}) and (\ref{b20}) without having to neglect the term in
$\hbar^2$ in Eq. (\ref{ext15}). However, in order to be consistent with Eq.
(\ref{ext5}),
which is equivalent to Eq. (\ref{b9}), the term in
$\hbar^2$ can actually be neglected in Eq. (\ref{ext15}). 

\subsection{EOS in the slow oscillation regime: stiff matter}

For a free SF with $V=0$, Eqs. (\ref{h2}) and (\ref{h3}) reduce
to
\begin{equation}
\epsilon=\frac{1}{2c^2}\left |\frac{d\varphi}{d
t}\right|^2+\frac{m^2c^2}{2\hbar^2}|\varphi|^2,\quad P=\frac{1}{2c^2}\left
|\frac{d\varphi}{d
t}\right|^2-\frac{m^2c^2}{2\hbar^2}|\varphi|^2.
\end{equation}
For massless particles ($m=0$) or for massive particles in the slow oscillation
regime $\omega=mc^2/\hbar\ll H$, the kinetic term dominates the potential term
(kination) and we obtain the stiff EOS:
\begin{equation}
P=\epsilon.
\label{ext17}
\end{equation}
For a
self-interacting SF, we find from Eqs. (\ref{h2}) and (\ref{h3}) that the stiff
EOS (\ref{ext17}) is
valid in the slow oscillation regime $\omega\ll H$ where $\omega$ is defined by
Eq. (\ref{ext5}). In that case, the SF cannot even complete one cycle of spin
within one Hubble time so that it just rolls down the potential, without
oscillating. Therefore, the comparison of $\omega$ and $H$ determines whether
the SF oscillates or rolls. For the stiff EOS (\ref{ext17}), using the
Friedmann
equations  (\ref{h4}) and (\ref{h5}), we easily get $\epsilon\propto a^{-6}$,
$a\sim t^{1/3}$, and $\epsilon\sim c^2/24\pi Gt^2$. It is also shown in
\cite{b40} that $\rho\sim (3m^2c^4/4\pi G\hbar^2)(-\ln a)^2$ and $|\varphi|\sim
(3c^4/4\pi G)^{1/2}(-\ln a)$. We note that quantum effects (quantum potential)
give rise to a
stiff matter era but do  not prevent the initial big bang singularity since
$\epsilon\sim c^2/24\pi Gt^2$ diverges as $t\rightarrow 0$.

\section{Self-interaction constants}
\label{sec_sic}

In the main part of the paper, we have expressed all the results in terms of
the  scattering length of the bosons $a_s$. Instead of working with the
scattering length, we can work with the
dimensionless
self-interaction constant \cite{b34,cd}:
\begin{eqnarray}
\frac{\lambda}{8\pi}=\frac{a_s}{\lambda_C}=\frac{a_s mc}{\hbar},
\label{sic1}
\end{eqnarray}
where $\lambda_C=\hbar/mc$ is the Compton wavelength of the bosons. 
We can also introduce a dimensional self-interaction constant
\begin{eqnarray}
\lambda_s=\frac{4\pi a_s\hbar^2 }{m}=\frac{\lambda \hbar^3}{2m^2c}.
\label{sic3}
\end{eqnarray}
Introducing proper normalizations, we get
\begin{eqnarray}
\frac{\lambda}{8\pi}=5.07 \frac{a_s}{\rm fm}\frac{m}{\rm GeV/c^2},
\label{sic5}
\end{eqnarray}
\begin{eqnarray}
\frac{\lambda_s}{(mc^2)^2}=4.89\times 10^{-22}  \frac{a_s}{\rm
fm}\left (\frac{\rm eV/c^2}{m}\right )^3\, {\rm eV}^{-1}{\rm cm}^3.
\label{sic6}
\end{eqnarray}
In the TF regime (semiclassical approximation), the results depend on the
single parameter
\begin{eqnarray}
\frac{4\pi a_s\hbar^2 }{m^3c^4}=\frac{\lambda
\hbar^3}{2m^4c^5}=\frac{\lambda_s}{(mc^2)^2}.
\label{sic4}
\end{eqnarray}

\section{Dimensionless variables}
\label{sec_dv}

In the main part of the paper, for the sake of clarity, we have worked with
dimensional variables. However, in order to simplify the calculations and make
the
figures, it can be convenient to introduce dimensionless variables defined by
\begin{equation}
\tilde\rho=\frac{\rho}{\rho_*},\qquad \rho_*=\frac{m^3c^2}{2\pi |a_s|\hbar^2},
\end{equation}
\begin{equation}
\tilde a=\frac{a}{a_*},\qquad a_*=\left (\frac{2\pi |a_s|\hbar^2
Q}{m^2c^2}\right
)^{1/3},
\end{equation}
\begin{equation}
\tilde t=\frac{t}{t_*},\qquad t_*=\left (\frac{2\pi
|a_s|\hbar^2}{4\pi G m^3c^2}\right )^{1/2}=\frac{1}{\sqrt{4\pi G\rho_*}},
\end{equation}
\begin{equation}
\tilde\epsilon=\frac{\epsilon}{\epsilon_*},\qquad \epsilon_*=\frac{m^3c^4}{2\pi
|a_s|\hbar^2}=\rho_* c^2,
\end{equation}
\begin{equation}
\tilde P=\frac{P}{P_*},\qquad P_*=\frac{m^3c^4}{2\pi
|a_s|\hbar^2}=\epsilon_*,
\end{equation}
\begin{equation}
\tilde E=\frac{E}{E_*},\qquad E_*=mc^2,
\end{equation}
\begin{equation}
{\tilde V}_{\rm tot}=\frac{V_{\rm tot}}{\epsilon_*},\quad
{\tilde \varphi}=\frac{\varphi}{\varphi_*}, \quad \varphi_*=\left
(\frac{mc^2}{2\pi
|a_s|}\right )^{1/2}.
\end{equation}
Working with the dimensionless variables $\tilde\rho$, $\tilde a$, $\tilde t$,
$\tilde\epsilon$, $\tilde P$ and $\tilde E$ is equivalent to taking
\begin{equation}
4\pi G=c=m=Q=2\pi |a_s|\hbar^2=1
\end{equation}
in the original equations.

\section{The parameters ($m,a_s$) of the SF}
\label{sec_mas}

In order to make numerical applications, we need to specify the values of the
mass $m$ and scattering length $a_s$ of the SF. They can be obtained by the
argument developed in Appendix D of \cite{clm}. If DM is a self-gravitating 
BEC, there must
be a minimum halo radius $R$ and a minimum halo mass $M$  in the Universe
corresponding to the ground state
of the self-gravitating BEC at $T=0$. This
result is in agreement with the observations. Indeed, there is
no DM
halo with a radius less than $R\sim 1\, {\rm kpc}$ and a mass less
than $M\sim 10^8\, M_\odot$, the typical values of the radius
and mass of dwarf spheroidal galaxies (dSph). Larger halos
have a core-halo structure with a solitonic core corresponding to a pure BEC at
$T=0$ and an
``atmosphere'' made of scalar radiation that has an approximate
Navarro-Frenk-White (NFW) profile. It is the atmosphere, resulting from
gravitational cooling,  that
fixes their size. We shall consider a dwarf halo of radius $R=1\, {\rm
kpc}$ and mass $M=10^8\, M_\odot$ (Fornax).   Assuming that
this halo represents
the
ground state of a self-gravitating BEC,
we can obtain constraints on the parameters $(m,a_s)$ of the SF. In
our previous works \cite{clm,b40}, we took $M=0.39\times 10^6\,
M_\odot$ and
$R=33\, {\rm pc}$ corresponding to Willman 1 \cite{devega}. However, these
values may
not be relevant  because Willman 1 is usually not considered as a DM halo (F.
Combes, private communication).

\subsection{Noninteracting SF}

A self-gravitating BEC without self-interaction has the mass-radius
relation \cite{rb,membrado,cd}:\footnote{This relation can be understood
qualitatively by identifying the halo radius $R$ with the de Broglie wavelength
 $\lambda_{dB}=\hbar/mv$ of a boson with a  velocity  $v\sim
(GM/R)^{1/2}$ equal to the virial
velocity  of the halo.} 
\begin{equation}
MR=9.95\, \frac{\hbar^2}{G m^2}.
\label{mas0}
\end{equation}
This gives
\begin{equation}
\frac{m}{{\rm eV}/c^2}=9.22\times 10^{-17} \left (\frac{\rm pc}{R}\right
)^{1/2}\left (\frac{M_{\odot}}{M}\right )^{1/2}.
\label{mas1}
\end{equation}
Using the values of $M$ and $R$ corresponding to Fornax, we obtain a boson
mass
\begin{equation}
m=2.92\times 10^{-22}\, {\rm eV}/c^2. 
\label{mas2}
\end{equation}
We note that, inversely, the specification of $m$
does not determine the mass and the radius of the halo but only their product
$MR$.

{\it Remark:} The maximum mass of the bosonic
core (soliton) of a noninteracting SFDM halo fixed by general relativity
is $M_{\rm max}=0.633\, \hbar c/Gm$ and its minimum radius is $R_{\rm min}=9.53
\, GM_{\rm max}/c^2$ \cite{kaup}. Introducing scaled variables, we get
\begin{equation}
\frac{M_{\rm max}}{M_{\odot}}=8.48\times 10^{-11}\frac{{\rm
eV}/c^2}{m},\quad \frac{R_{\rm min}}{{\rm km}}=14.1
\frac{M_{\rm max}}{M_{\odot}}.
\label{brelat2}
\end{equation}
For $m=2.92\times
10^{-22}\, {\rm eV}/c^2$, we obtain $M_{\rm max}=2.90\times 10^{11}\, M_{\odot}$
and $R_{\rm min}=0.133\, {\rm pc}$. We note that the bosonic core of DM halos
is generally nonrelativistic ($M_c\ll M_{\rm max}$).

\subsection{Repulsive self-interaction}

A self-gravitating BEC  with a
repulsive self-interaction in the TF approximation has a unique
radius \cite{goodman,arbey,bohmer,b34}:
\begin{eqnarray}
R=\pi\left (\frac{a_s\hbar^2}{Gm^3}\right )^{1/2}
\label{mas3}
\end{eqnarray}
that is independent of its mass. This gives 
\begin{equation}
\frac{a_s}{\rm fm}\left (\frac{{\rm eV}/c^2}{m}\right
)^3=3.28\times 10^{-3} \left (\frac{R}{\rm pc}\right )^{2}.
\label{mas4}
\end{equation}
Using the value of $R$ corresponding to Fornax, we obtain
\begin{eqnarray}
\frac{a_s}{\rm fm}\left (\frac{{\rm eV}/c^2}{m}\right )^3=3.28\times 10^3.
\label{mas5}
\end{eqnarray}
This fixes the ratio $a_s/m^3$. In order to
determine the mass of the bosons, we need another relation. This relation is
provided by the constraint $\sigma/m<1.25\, {\rm cm}^2/{\rm g}$ set by the
Bullet Cluster \cite{bullet}, where $\sigma=4\pi a_s^2$ is the self-interaction
cross section. Assuming that this bound is reached, we
get $(a_s/{\rm fm})^2({\rm eV}/mc^2)=1.77\times 10^{-8}$. From this relation
and Eq. (\ref{mas5}), we obtain
\begin{equation}
m=1.10\times 10^{-3}\, {\rm eV}/c^2,\qquad  a_s=4.41\times
10^{-6}\, {\rm fm},
\label{mas6}
\end{equation}
corresponding to $\lambda/8\pi=2.46\times 10^{-17}$. This boson mass
is in
agreement with the limit $m<1.87\, {\rm eV}/c^2$ obtained from cosmological
considerations \cite{fmt}.

The TF approximation is valid when the radius given
by Eq. (\ref{mas3}) is much larger than the radius given by Eq. (\ref{mas0}).
This corresponds to $a_s\gg \hbar^2/GM^2m$ or $\lambda/8\pi\gg
\hbar c/GM^2$ (see Sec. II.G. of \cite{b34} and
Appendix A.3 of \cite{cd}). Using the value of $M$ corresponding to Fornax, we
get
\begin{equation}
\frac{a_s}{\rm fm}\frac{m}{{\rm eV}/c^2}\gg 2.36\times 10^{-84},
\label{mas6b}
\end{equation}
or, equivalently, ${\lambda}/{8\pi}\gg 1.20\times 10^{-92}$
(note the smallness of this quantity already emphasized in
Appendix A.3 of \cite{cd}).
This condition is clearly
satisfied by the parameters of Eq. (\ref{mas6}). Inversely, when  $a_s\ll
\hbar^2/GM^2m$, we can
ignore the self-interaction of the bosons. We then find that the boson mass is
given by Eqs. (\ref{mas1}) and (\ref{mas2}).  An estimate of the critical
scattering length separating the TF regime from the noninteracting regime can be
obtained by substituting Eq. (\ref{mas1}) into Eq.
(\ref{mas4}). This gives $a_c=31.4(R\hbar^2/GM^3)^{1/2}$, i.e.
\begin{equation}
\frac{a_c}{\rm fm}=2.57\times 10^{-51}\,  \left (\frac{R}{\rm pc}\right
)^{1/2}\left (\frac{M_{\odot}}{M}\right )^{3/2}.
\label{mas7ba}
\end{equation}
Using the values of $M$ and $R$ corresponding to Fornax, we obtain
\begin{equation}
m=2.92\times 10^{-22}\, {\rm eV}/c^2,\quad a_c=8.13\times 10^{-62}\, {\rm fm}.
\label{mas7b}
\end{equation}
The mass $m=2.92\times 10^{-22}\, {\rm eV}/c^2$ obtained for bosons without
self-interaction gives a lower bound on the mass of the bosonic dark matter
particle. Inversely, the mass $m=1.10\times 10^{-3}\, {\rm eV}/c^2$ obtained for
self-interacting bosons in the TF approximation gives an upper bound on the
mass of the bosonic dark matter particle. Therefore, we predict that the mass of
the bosonic particle is in the range
$2.92\times 10^{-22}\, {\rm eV}/c^2\le m\le 1.10\times 10^{-3}\, {\rm eV}/c^2$.
The TF
limit
is valid for sufficiently large scattering lengths, i.e., above $a_c$.
For $a_s<a_c$, the mass of
the bosonic
particle
is $m=2.92\times 10^{-22}\, {\rm eV}/c^2$ and for $a_c<a_s<4.41\times 10^{-6}\,
{\rm fm}$
the mass of the bosonic particle is $2.92\times 10^{-22} < m/({\rm
eV}/c^2)=6.73\times 10^{-2}\, (a_s/{\rm
fm})^{1/3}<1.10\times 10^{-3}$. We note that, inversely, the
specification of  $m$ and $a_s$ does not determine the mass of the halo but
only its radius $R$.

{\it Remark:} The maximum mass of the bosonic core (soliton) of
a self-interacting SFDM halo fixed by general relativity is $M_{\rm max}=0.307\,
\hbar c^2\sqrt{a_s}/(Gm)^{3/2}$ and its minimum radius is
$R_{\rm min}=6.25 \, GM_{\rm max}/c^2$ \cite{b45}. Introducing scaled
variables, we get
\begin{equation}
\frac{M_{\rm max}}{M_{\odot}}=1.12\, \left (\frac{a_s}{{\rm fm}}\right
)^{1/2}\left (\frac{{\rm
GeV}/c^2}{m}\right )^{3/2},
\label{brelat3}
\end{equation}
\begin{equation}
\frac{R_{\rm min}}{{\rm km}}=9.27
\frac{M_{\rm max}}{M_{\odot}}.
\label{brelat4}
\end{equation}
We note that these results do not depend 
on the specific mass $m$ and scattering length $a_s$ of the bosons, but only on
the ratio $m^3/a_s$. For $(a_s/{\rm
fm})({\rm
eV}/mc^2)^3=3.28\times 10^3$, we obtain $M_{\rm max}=2.03\times
10^{15}\, M_{\odot}$ and $R_{\rm min}=609\, {\rm pc}$. We note that the bosonic
core
of DM halos
is generally nonrelativistic ($M_c\ll M_{\rm max}$).

\subsection{Attractive self-interaction}

A self-gravitating BEC with an attractive self-interaction ($a_s<0$) is stable
only below a maximum mass $M_{\rm max}$ and above a radius $R_*$ given
by \cite{b34,cd}:
\begin{eqnarray}
\label{mas9}
M_{\rm max}=1.012\frac{\hbar}{\sqrt{Gm|a_s|}}, \qquad R_*=5.5\left
(\frac{|a_s|\hbar^2}{Gm^3}\right )^{1/2}.\nonumber\\
\end{eqnarray}
This gives
\begin{eqnarray}
\label{mas10}
\frac{M_{\rm max}}{M_{\odot}}=1.56\times 10^{-34} \left (\frac{{\rm
eV}/c^2}{m}\right )^{1/2}  \left (\frac{\rm fm}{|a_s|}\right )^{1/2}, 
\end{eqnarray}
\begin{eqnarray}
\label{mas11}
\frac{R_*}{R_{\odot}}=1.36\times 10^{9}  \left
(\frac{|a_s|}{\rm fm}\right )^{1/2} \left (\frac{{\rm
eV}/c^2}{m}\right )^{3/2}. 
\end{eqnarray}
Considering standard (QCD) axions \cite{kc} with 
\begin{eqnarray}
\label{mas12}
m=10^{-4}\,
{\rm eV}/c^2,\qquad
a_s=-5.8\times 10^{-53}\, {\rm m},
\end{eqnarray}
\begin{eqnarray}
\frac{|a_s|}{\rm fm}\left (\frac{{\rm eV}/c^2}{m}\right )^3=5.8\times 10^{-26},
\label{mas12b}
\end{eqnarray}
corresponding to $\lambda=-7.4\times 10^{-49}$, we obtain $M_{\rm
max}=6.5\times 10^{-14}\,
M_{\odot}$ and $R_*=3.3\times 10^{-4}\, R_{\odot}$. Obviously, QCD axions
cannot form DM halos of relevant
mass and size; they rather form mini axion stars \cite{bectcoll}.  DM halos
could be made of numerous mini axion
stars (mini-MACHOs) that would behave as CDM. On the other hand, ultralight
axions can
form DM halos. Assuming
that Fornax corresponds to a self-gravitating BEC with attractive
self-interaction at the limit of stability, we can use Eqs. (\ref{mas10}) and
(\ref{mas11}) to
obtain the values of $m$ and $a_s$. We get \cite{bectcoll}:
\begin{eqnarray}
\label{mas13}
m=2.19\times 10^{-22}\, {\rm eV}/c^2,\qquad  a_s=-1.11\times
10^{-62}\, {\rm fm},\nonumber\\
\end{eqnarray}
\begin{eqnarray}
\frac{|a_s|}{\rm fm}\left (\frac{{\rm eV}/c^2}{m}\right )^3=1.06\times 10^3,
\label{mas13b}
\end{eqnarray}
corresponding to $\lambda/8\pi=-1.23\times
10^{-92}$. Actually, the halo does not need to be at the
limit of stability. On the contrary, we need to impose that $M\ll M_{\rm max}$
for the halo to be robustly stable. This corresponds to $|a_s|\ll
\hbar^2/GM^2m$ or $|\lambda|/8\pi\ll
\hbar c/GM^2$
leading to the reverse of Eq. (\ref{mas6b}) with $a_s$ replaced by $|a_s|$. In
that case, we can ignore the
self-interaction of the bosons. We then find that the boson mass is given by
Eq.
(\ref{mas2}). This is valid as long as its scattering
length satisfies
\begin{eqnarray}
|a_s|\ll 1.11\times 10^{-62}\,
{\rm fm}
\label{ggg}
\end{eqnarray}
or, equivalently, $|\lambda|/8\pi\ll 1.23\times 10^{-92}$
because above this value the halo mass becomes larger than the maximum
mass and the halo undergoes gravitational collapse \cite{bectcoll}.
We note that this
value is of
the same order as the value (\ref{mas7b}) marking the transition between the
noninteracting limit and the TF limit in the case $a_s>0$. This is expected
in view of the similar scalings. In conclusion, bosons with attractive
self-interaction must have an ultralight mass and an extraordinarily small
scattering length to form stable DM halos of relevant size.

{\it Remark:} In the high resolution numerical simulations of
self-gravitating BECs performed by \cite{nature}, the self-interaction of the
bosons is not
taken into account ($\lambda=0$). Our results show that even an apparently tiny
attractive ($\lambda<0$) self-interaction with $|\lambda|\gtrsim 10^{-90}$ can
considerably change the physics of the problem. For example, the solitonic core
($M\sim 10^8\, M_{\odot}$) of the dark matter halos considered in \cite{nature}
is stable for $\lambda=a_s=0$ but becomes unstable in the case of an
attractive self-interaction with $|\lambda|/8\pi=|a_s|mc/\hbar>1.02\, \hbar
c/GM^2=1.02 (M_P/M)^2=1.23\times 10^{-92}$, or $|a_s|>1.11\times 10^{-62}\, {\rm
fm}$ for $m=2.19\times 10^{-22}\, {\rm eV/c^2}$, because in that case $M>M_{\rm
max}$ \cite{b34,bectcoll}. Therefore $\lambda/8\pi=-1.23\times 10^{-92}$ is
very different from $\lambda=0$ (!). It would be therefore extremely interesting
to perform numerical simulations of the GPP and KGE equations for
self-interacting bosons.

\subsection{Cosmological constraints}

Li {\it et al.} \cite{b36} have obtained stringent bounds on the
values of $m$ and $a_s$ (assuming $a_s\ge 0$) by
using cosmological constraints coming from the CMB and from the abundances of
the light elements produced by the BBN.
First of all, their bounds exclude
the possibility that the bosons are noninteracting ($a_s\neq 0$). On the other
hand, by combining all their constraints they obtain a fiducial model:
\begin{equation}
m=3\times 10^{-21}\, {\rm eV}/c^2, \qquad a_s=1.11\times
10^{-58}\, {\rm fm},
\label{mas7}
\end{equation}
\begin{eqnarray}
\frac{a_s}{\rm fm}\left (\frac{{\rm eV}/c^2}{m}\right )^3=4.10\times 10^3,
\label{mas8}
\end{eqnarray}
corresponding to  $\lambda/8\pi=1.69\times 10^{-87}$. We note that the ratio 
(\ref{mas8}) obtained by Li {\it et al.} \cite{b36} from
cosmological (large scales) arguments is of the same order as the ratio
(\ref{mas5}) obtained from astrophysical (small scales)
arguments (it corresponds to a minimum halo radius $R=1.12\, {\rm
kpc}$). This
agreement is very satisfactory. On the other hand, the value of the boson mass
(\ref{mas7}) obtained by Li {\it et al.} \cite{b36} is relatively close to the
mass (\ref{mas2}) of a noninteracting boson. This is because their fiducial
model is relatively close to the transition between the noninteracting limit and
the TF limit [compare Eq. (\ref{mas7}) with Eq. (\ref{mas7b})]. However, the
fact that their mass (\ref{mas7}) is substantially larger (by one order of
magnitude) than the mass (\ref{mas2}) reflects
the fact that the bosons are self-interacting (their fiducial model is at the
begining of the TF regime). Actually the mass (\ref{mas2}) of a noninteracting
SF is excluded by their bound $m\ge 2.4\times 10^{-21}\, {\rm eV}/c^2$
\cite{b36}. Note
that their fiducial model uses a mass close to the minimum allowed mass.
However, the mass of the SF could be much larger than this bound like the mass
of Eq. (\ref{mas6}) which is deeper in the TF regime. 

{\it Remark:} In a very recent paper, Hui {\it et al.}
\cite{hotw} have given further support to the BECDM/SFDM
model. They focused on the
noninteracting case ($a_s=0$), considering an ultralight axion of mass $m\sim
1-10\times 10^{-22}\, {\rm eV}/c^2$ (consistent with Eq. (\ref{mas2})). While
mentioning several virtues of this model, they noted that this mass is in
tension with observations of the Lyman-$\alpha$ forest, which favor masses
$10-20\times 10^{-22}\,
{\rm eV}/c^2$ or higher. A similar conclusion was reached by
Menci {\it et al.} \cite{menci} based on the measured abundance of ultra-faint
lensed galaxies at redshift $z\simeq 6$ in the Hubble Frontier Fields (HFF). We
note that such larger masses are in agreement with
Eq. (\ref{mas7}). Therefore, large-scale observations
could reflect the fact that axions are
self-interacting. In that case, all the known observational constraints
seem to
be satisfied.

\subsection{Fermions}

In this paper, we have assumed that DM halos are made of bosons.
If they are made of fermions, their mass-radius relation is
$MR^3=1.49\times 10^{-3}\, h^6/(G^3 m^8)$ \cite{chandra}. This gives
\begin{equation}
\frac{m}{{\rm eV}/c^2}=2.27\times 10^4 \left (\frac{\rm pc}{R}\right
)^{3/8}\left (\frac{M_{\odot}}{M}\right )^{1/8}.
\label{fb1}
\end{equation}
Using the values of
$M$ and $R$ corresponding to Fornax, we find a fermion mass $m=170 \, {\rm
eV}/c^2$. We note
that, inversely, the
specification of  $m$ does not determine the mass and the radius of the halo
but only the product $MR^3$.

{\it Remark:} The maximum mass of the fermionic core (fermion
ball) of a DM halo fixed by general relativity is  $M_{\rm max}=0.376\, (\hbar
c/G)^{3/2}/m^2$ and its minimum radius is $R_{\rm min}=9.36 \,
GM_{\rm max}/c^2$ \cite{ov}. Introducing scaled variables, we get
\begin{equation}
\frac{M_{\rm max}}{M_{\odot}}=6.13\times 10^{17}\left (\frac{{\rm
eV}/c^2}{m}\right
)^2,\quad \frac{R_{\rm min}}{{\rm km}}=13.8 \frac{M_{\rm max}}{M_{\odot}}.
\label{relat1}
\end{equation}
For $m=170\, {\rm eV}/c^2$, we obtain $M_{\rm max}=2.12\times 10^{13}\,
M_{\odot}$
and $R_{\rm min}=9.49\, {\rm pc}$. We note that the fermionic core
of DM halos is generally nonrelativistic ($M_c\ll M_{\rm max}$).

\section{Comparison between the standard model and the SF model}
\label{sec_mr}

In this Appendix, we  compare the standard model and the SF model with
$a_s\ge 0$. In the first two subsections, for the clarity of the presentation,
we do not take the baryonic matter and the DE (or cosmological constant) into
account. The complete model is discussed in the third
subsection. For the numerical applications, we adopt the values of the
cosmological parameters given by Li {\it et al.} \cite{b36}. They are
listed in the fourth subsection. 

\subsection{The standard model}
\label{sec_mrs}

In the standard model, DM (which corresponds to WIMPs) and
radiation (which accounts for the photons and neutrinos of the CMB) are two
different species described by the EOSs $P_{\rm
dm}=0$ and $P_{\rm r}=\epsilon_{\rm r}/3$ respectively. Solving the 
continuity equation (\ref{h4}) for each species, we
obtain
\begin{eqnarray}
\frac{\epsilon_{\rm dm}}{\epsilon_0}=\frac{\Omega_{\rm dm,0}}{a^3},\qquad
\frac{\epsilon_{\rm r}}{\epsilon_0}=\frac{\Omega_{\rm r,0}}{a^4},
\label{mrs1}
\end{eqnarray}
where $\Omega_{\rm dm,0}$ is the present fraction of dark matter and 
$\Omega_{\rm r,0}$
is the present fraction of radiation. We have taken $a_0=1$. The total energy
density of these two species is
\begin{eqnarray}
\frac{\epsilon}{\epsilon_0}=\frac{\Omega_{\rm
r,0}}{a^4}+\frac{\Omega_{\rm dm,0}}{a^3} .
\label{mrs2}
\end{eqnarray}
From the Friedmann equation (\ref{h5}) we obtain  the differential equation
\begin{eqnarray}
\left
(\frac{\dot
a}{a}\right
)^2=\frac{8\pi
G\epsilon_0}{3c^2}\left (\frac{\Omega_{\rm r,0}}{a^4}+\frac{\Omega_{\rm dm, 0
} } { a^3 }\right )
\label{mrs3}
\end{eqnarray}
that determined the evolution of the scale factor
$a$. This equation  can be integrated exactly, giving
\cite{stiff}:
\begin{equation}
H_0 t=-\frac{2}{3}\frac{1}{\Omega_{\rm dm,0}^{1/2}}\left
(\frac{2\Omega_{\rm r,0}}{\Omega_{\rm dm,0}}-a\right
)\sqrt{\frac{\Omega_{\rm r,0}}{\Omega_{\rm dm,0}}+a}
+\frac{4\Omega_{\rm r,0}^{3/2}}{3\Omega_{\rm dm,0}^2}.
\label{radd1}
\end{equation}
Equation (\ref{radd1}) can also be written as
\begin{equation}
a^3-3\frac{\Omega_{\rm r,0}}{\Omega_{\rm dm,0}}a^2=\frac{9}{4}\Omega_{\rm
dm,0}H_0^2t^2-6\frac{\Omega_{\rm r,0}^{3/2}}{\Omega_{\rm dm,0}}H_0t.
\label{radd1cub}
\end{equation}
This is a cubic equation
for $a$ of the form $a^3+Aa^2+Ba+C=0$ which can be solved by standard methods.
Using Cardano's formula, the real root is given by
\begin{equation}
a=-\frac{A}{3}+\left (-\frac{q}{2}+\sqrt{R}\right )^{1/3}+\left
(-\frac{q}{2}-\sqrt{R}\right )^{1/3}
\label{radd1cub2}
\end{equation}
with $R=(p/3)^3+(q/2)^2$, $p=B-A^2/3$, and $q=C-AB/3+2A^3/27$ (in our
case $R>0$). However, to
obtain $a(t)$, it is easier to use Eq. (\ref{radd1}) that gives $t(a)$, and plot
the inverse function.
For $a\rightarrow 0$ (radiation era):
\begin{eqnarray}
\frac{\epsilon}{\epsilon_0}\sim \frac{\Omega_{\rm r,0}}{a^4},\qquad a\sim \left
(\frac{32\pi G\epsilon_0\Omega_{\rm r,0}}{3c^2}\right )^{1/4}t^{1/2}.
\label{mrs4}
\end{eqnarray}
For $a\rightarrow +\infty$ (matter era, EdS  solution):
\begin{eqnarray}
\frac{\epsilon}{\epsilon_0}\sim \frac{\Omega_{\rm dm,0}}{a^3},\qquad a\sim \left
(\frac{6\pi G\epsilon_0\Omega_{\rm dm,0}}{c^2}\right )^{1/3}t^{2/3}.
\label{mrs5}
\end{eqnarray}
The epoch of matter-radiation equality ($\epsilon_{\rm m}=\epsilon_{\rm r}$)
corresponds
to\footnote{Here, we take  the contribution of baryonic matter into account
noting that baryonic matter behaves as dark matter, i.e., it is described
by an EOS $P_{\rm b}=0$.}
\begin{eqnarray}
a_{\rm eq}=\frac{\Omega_{\rm r,0}}{\Omega_{\rm m,0}}=\frac{\Omega_{\rm
r,0}}{\Omega_{\rm dm,0}+\Omega_{\rm b,0}}.
\label{mrs6}
\end{eqnarray}
Numerically, $a_{\rm eq}=2.95\times 10^{-4}$.
For $a<a_{\rm eq}$, we are in the
radiation-dominated regime and for  $a>a_{\rm eq}$, we are in the
matter-dominated regime.

The energy density of radiation can be written as
\begin{eqnarray}
\epsilon_{\rm r}=\kappa\sigma T^4=\frac{\kappa\pi^2}{15c^3\hbar^3}(k_B T)^4,
\label{mrs7}
\end{eqnarray}
where $T$ is the temperature, $\sigma$ is the Stefan-Boltzmann constant, and
$\kappa=\kappa_\gamma+\kappa_\nu=1+3.046(7/8)(4/11)^{4/3}\simeq 1.692$ is a
constant that accounts for the fact that radiation comes from photons and
neutrinos in thermal equilibrium \cite{planck}. According to
Eqs. (\ref{mrs1}) and (\ref{mrs7}), the relation between the temperature and the
scale factor is
given by
\begin{eqnarray}
k_B T=\left (\frac{15c^3\hbar^3\Omega_{\rm r,0}\epsilon_0}{\kappa\pi^2}\right
)^{1/4}\frac{1}{a}=\frac{k_B T_0}{a},
\label{mrs8}
\end{eqnarray}
where $T_0$ is the present temperature of radiation. Numerically, $T_0=2.7255\,
{\rm K}$. 

\subsection{The SF model}
\label{sec_mrf}

In the standard model, the Universe undergoes a radiation era followed by a
pressureless DM era. Similarly, a SF with $a_s>0$ undergoes a
radiationlike era followed by a pressureless dark matter era (see
Sec. \ref{sec_erapos}). However, the
two models are physically different. In the standard model, radiation
and dark matter correspond to different species that exist in
permanence. For $a<a_{\rm eq}$ radiation dominates over matter and for $a>a_{\rm
eq}$ matter dominates over radiation. In the SF model, there is just one
species. The radiation and the matter are two manifestations of the  same
entity. For $a<a_t$, the SF behaves as radiation and for $a>a_t$  it behaves as
pressureless matter. The relation between the energy
density and the scale factor is different in the two models [see Eq.
(\ref{mrs2}) for the standard model and Eqs. (\ref{posba1}) and (\ref{posba3})
for the SF model]. However, their asymptotic behaviors for $a\rightarrow 0$ and
$a\rightarrow +\infty$ are similar.

For $a\rightarrow +\infty$, the SF  behaves as pressureless matter. Since
the SF is expected to describe DM, we can identify the matterlike era
of the SF with the DM era of the standard model. Comparing Eqs. (\ref{posea4})
and (\ref{mrs5}) valid for $a\rightarrow +\infty$, we find that the charge of
the SF is given by
\begin{eqnarray}
Q=\frac{\Omega_{\rm dm,0}\epsilon_0}{m c^2}.
\label{mrf2}
\end{eqnarray}
This relation is also valid for the SF model with $a_s<0$ on the normal branch
since it also behaves as pressureless matter for $a\rightarrow +\infty$.

For $a\rightarrow 0$, the SF behaves as radiation which adds
to the standard radiation (photons, neutrinos...). We define
the initial ratio between the radiation of the SF and the
standard radiation by 
\begin{eqnarray}
\mu=\lim_{a\rightarrow 0}\frac{\epsilon_{\rm SF}}{\epsilon_{\rm r}},
\label{mrf3}
\end{eqnarray}
where $\epsilon_{\rm SF}$ is the energy density of the SF and $\epsilon_{\rm r}$
is the energy density of the standard radiation. Using
Eqs. (\ref{posea3})
and (\ref{mrs4}) valid for $a\rightarrow 0$, we obtain 
\begin{eqnarray}
\mu=\left (\frac{27\pi Q^4 m a_s \hbar^2
c^4}{8\Omega_{\rm r,0}^3\epsilon_0^3}\right )^{1/3}.
\label{mrf4}
\end{eqnarray}
Substituting the expression of the charge from Eq. (\ref{mrf2}) into Eq.
(\ref{mrf4}), we get
\begin{eqnarray}
\mu=\left (\frac{27\pi  a_s \hbar^2
\Omega_{\rm dm,0}^4\epsilon_0}{8\Omega_{\rm r,0}^3 m^3 c^4}\right )^{1/3}.
\label{mrf5}
\end{eqnarray}
We see that $\mu$ is determined by the ratio $a_s/m^3$. Inversely, if we know
the value of $\mu$, Eq. (\ref{mrf5}) determines the ratio $a_s/m^3$ through the
relation
\begin{eqnarray}
\frac{a_s}{m^3}=\frac{8\mu^3\Omega_{\rm r,0}^3c^4}{27\pi\Omega_{\rm dm,0}
^4\epsilon_0\hbar^2 }.
\label{mrf6}
\end{eqnarray}
Introducing proper normalizations, we get
\begin{eqnarray}
\frac{a_s}{\rm fm}\left (\frac{{\rm eV}/c^2}{m}\right )^3= 8.18\times
10^7\, \mu^3.
\label{mrf6b}
\end{eqnarray}
On the other hand, in the radiationlike era of the SF valid for $a\rightarrow
0$, we can write
\begin{eqnarray}
\epsilon_{\rm SF}=\kappa\sigma T_{\rm
eff}^4=\frac{\kappa\pi^2}{15c^3\hbar^3}(k_B T_{\rm
eff})^4,
\label{mrf7}
\end{eqnarray}
where $T_{\rm eff}$ is an effective temperature of SF radiation. Using
Eq. (\ref{posea3}), we obtain
\begin{eqnarray}
k_B T_{\rm eff}=\left (\frac{91125
c^{13}\hbar^{11}Q^4 m a_s}{8\pi^5\kappa^3}\right
)^{1/12}\frac{1}{a}.
\label{mrf8}
\end{eqnarray}
Although the SF is at $T=0$, we can define an effective temperature of
radiation for the SF that
depends on its charge $Q$ and on the self-interaction strength $a_s$. Using
Eq. (\ref{mrf2}) to evaluate $Q$, we get
\begin{eqnarray}
k_B T_{\rm eff}=\left (\frac{91125
c^{5}\hbar^{11}a_s \Omega_{\rm dm,0}^4\epsilon_0^4}{8\pi^5m^3\kappa^3}\right
)^{1/12}\frac{1}{a}.
\label{mrf9}
\end{eqnarray}
Using Eq. (\ref{mrf6}) to evaluate $a_s/m^3$, we obtain
\begin{eqnarray}
k_B T_{\rm eff}=\left (\frac{15\mu
c^{3}\hbar^{3}\epsilon_0\Omega_{r,0}}{\pi^2\kappa}\right
)^{1/4}\frac{1}{a}.
\label{mrf10}
\end{eqnarray}
Comparing Eqs. (\ref{mrs8}) and (\ref{mrf10}), we find that
\begin{eqnarray}
\frac{T_{\rm eff}}{T}=\mu^{1/4}
\label{mrf11}
\end{eqnarray}
in the radiative regime of the SF. 

{\it Remark:}  In the standard model we can calculate the present temperature of
radiation $T_0$. We cannot define $(T_{\rm eff})_0$ for the SF because $T_{\rm
eff}$ is only defined in the radiationlike regime so this effective temperature
has no
meaning in the present Universe.

\subsection{The complete model}
\label{sec_compl}

Since the SF is expected to represent DM (replacing the WIMP hypothesis), the
complete
model incorporating standard radiation and SFDM is
obtained by replacing the second term in Eq. (\ref{mrs2}) by the energy density
of the SF. To be
even more complete, we must also include baryonic matter and DE
(cosmological constant).  Therefore, the total energy
reads 
\begin{eqnarray}
\frac{\epsilon}{\epsilon_0}=\frac{\Omega_{\rm
r,0}}{a^4}+\frac{\epsilon_{\rm SF}(a)}{\epsilon_0}+\frac{\Omega_{\rm
b,0}}{a^3}+\Omega_{\Lambda,0}.
\label{compl1}
\end{eqnarray}
This complete model has been studied by Li {\it et al.} \cite{b36}. In the fast
oscillation regime, introducing the dimensionless variables of Appendix
\ref{sec_dv}, the energy density of the SF is given by
\begin{eqnarray}
\frac{\epsilon_{\rm
SF}}{\epsilon_0}=\frac{27}{16}\frac{\Omega_{{\rm
dm},0}^4}{\Omega_{\rm
r,0}^3}\frac{1}{\mu^3}{\tilde\epsilon}_{\rm SF}(\tilde a) 
\label{compl2}
\end{eqnarray}
with
\begin{eqnarray}
a=\left (\frac{16}{27}\right )^{1/3}\frac{\Omega_{{\rm
r},0}}{\Omega_{\rm
dm,0}}\mu{\tilde a},
\label{compl2b}
\end{eqnarray}
where the function ${\tilde\epsilon}_{\rm SF}(\tilde a)$ is given in parametric
form by Eqs. (\ref{posba1}) and (\ref{posba3}). This fast oscillation regime
has been investigated in detail in Sec. \ref{sec_pos}.

\subsection{Values of the cosmological parameters}
\label{sec_vcp}

For the values of the  cosmological parameters, following Li {\it et al.}
\cite{b36}, we take $\Omega_{\rm r,0}=9.23765\times 10^{-5}$, $\Omega_{\rm
dm,0}=0.2645$, $\Omega_{\rm b,0}=0.0487273$, $\Omega_{\rm m,0}=0.313228$,
$\Omega_{\Lambda,0}=0.687$, $H_0=2.18\times 10^{-18}\, {\rm s}^{-1}$, and
$\epsilon_0=7.64\times 10^{-7} {\rm g}\, {\rm m}^{-1}\,{\rm s}^{-2}$

\section{Match asymptotics}
\label{sec_stiff}

In the stiff matter era, corresponding to the slow oscillation regime, the
pseudo rest-mass density and the energy density evolve with the scale factor as
\cite{b40}:
\begin{equation}
\rho_{\rm stiff}\sim \frac{3m^2c^4}{4\pi G\hbar^2}(-\ln a)^2,\qquad
\epsilon_{\rm stiff}\sim
\frac{K}{a^6},
\label{stiff1}
\end{equation}
where $K$ is a constant. We note that the pseudo rest-mass density changes very
slowly with the scale factor as compared to the energy density. We want to see
when it is possible to connect the  slow oscillation regime $\rho_{\rm
stiff}(a)$ to the fast oscillation regime $\rho(a)$.
Using the dimensionless variables introduced in Appendix
\ref{sec_dv}, the condition $\rho(a)\sim \rho_{\rm stiff}(a)$ corresponds to
\begin{eqnarray}
\tilde\rho(\tilde a)\sim \tilde\rho_{\rm stiff}(\tilde a)=2\sigma (-\ln a)^2,
\label{stiff1b}
\end{eqnarray}
where $\sigma$ is defined by Eq. (\ref{valp3}).
Therefore, the matching point corresponds to the intersection of the curve
$\tilde\rho(\tilde a)$ drawn in Figs. \ref{arhopos} and \ref{arhoneg}
with the curve $\rho_{\rm stiff}(\tilde a)=2\sigma (-\ln a)^2$ that is almost a
straight line due to the slow variation of the logarithmic term.

We first consider a SF with $a_s\ge 0$. Matching the pseudo rest-mass density of
the slow oscillation regime [see Eq. (\ref{stiff1})] with the pseudo rest-mass
density of the fast oscillation regime [see Eq. (\ref{posba1})], we obtain
\begin{eqnarray}
a_v^*= \left (\frac{2\pi a_s\hbar^2 Q}{m^2c^2}\right
)^{1/3}\, f_*\left (\frac{3 a_s c^2}{4 Gm}\right ),
\label{stiff2}
\end{eqnarray}
where the function $f_*(\sigma)$ is defined by
\begin{eqnarray}
f_*(\sigma)=\frac{1}{r^{1/3}(1+4r)^{1/6}}
\label{stiff3}
\end{eqnarray}
with
\begin{eqnarray}
r=2\sigma(-\ln a_v^*)^2.
\label{stiff4}
\end{eqnarray}
It is easy to see that, up to logarithmic corrections, $a_v^*$ is of  the same
order of magnitude as the scale $a_v$ marking the transition between the slow
and fast oscillation regimes given by Eq.
(\ref{valp5}). Mathematically, this is
because the function $r$ defined by Eq. (\ref{valp7}) behaves as $r\sim
4\sigma/3$
for $\sigma\rightarrow +\infty$ and as $r\sim \sigma$ for $\sigma\rightarrow
0$
which is the same scaling as the function $r\propto \sigma$ defined by Eq.
(\ref{stiff4}). This is most easily seen by considering the asymptotic limits.
Matching the pseudo rest-mass density of the stiff matter era [see Eq.
(\ref{stiff1})]
with the pseudo rest-mass density of the radiationlike era [see Eq.
(\ref{posea1})], we obtain a transition scale of the same order as Eq.
(\ref{valp10}). Matching the pseudo rest-mass density of the stiff matter era
[see Eq. (\ref{stiff1})] with the pseudo rest-mass density of the matterlike
era [see Eq. (\ref{posea2})], we obtain a transition scale of the same order as
Eq. (\ref{valp11}). Of course, these qualitative  agreements are to be
expected. They just provide a consistency check of our approximations.

We can also use match asymptotics to estimate the constant $K$ in Eq.
(\ref{stiff1}). Matching the energy density of the slow oscillation regime [see
Eq. (\ref{stiff1})] with the energy density of the fast oscillation regime [see
Eq. (\ref{posba3})] at the transition scale $a_v$ determined by Eq.
(\ref{valp5}), we obtain
\begin{eqnarray}
K=\frac{2\pi a_s\hbar^2 Q^2}{m}\, h\left (\frac{3 a_s
c^2}{4 Gm}\right ),
\label{stiff5}
\end{eqnarray}
where the function $h(\sigma)$ is defined by
\begin{eqnarray}
h(\sigma)=\frac{1+3r}{(1+4r)r}
\label{stiff6}
\end{eqnarray}
with
\begin{eqnarray}
r=\frac{4\sigma-1+\sqrt{(4\sigma-1)^2+12\sigma}}{6}.
\label{stiff7}
\end{eqnarray}
For $\sigma\rightarrow +\infty$, we obtain
\begin{eqnarray}
K=\frac{3\pi G\hbar^2 Q^2}{2c^2}
\label{stiff8}
\end{eqnarray}
which can also be obtained by  matching  the energy density of
the stiff matter era  [see
Eq. (\ref{stiff1})] with the energy density of the radiationlike
era [see Eq. (\ref{posea3})] at the transition scale (\ref{valp10}). For
$\sigma\rightarrow 0$, we obtain
\begin{eqnarray}
K=\frac{8\pi G\hbar^2 Q^2}{3c^2}
\label{stiff9}
\end{eqnarray}
which can also be obtained by  matching  the energy density of
the stiff matter era [see
Eq. (\ref{stiff1})]  with the energy density of the matterlike
era [see Eq. (\ref{posea4})] at the transition scale (\ref{valp11}). We note
that the value of $K$ does not sensibly depend on $a_s$. Now that the constant
$K$ is known, the Friedmann equation (\ref{h5}) can be integrated with the stiff
energy density given by Eq. (\ref{stiff1}) yielding  
\begin{eqnarray}
a=\left (\frac{24\pi GK}{c^2}\right )^{1/6}t^{1/3}.
\end{eqnarray}
Using the foregoing results, we can determine the transition
between the stiff matter era of the SF and the standard radiation era.
Equating Eqs. (\ref{mrs1}) and (\ref{stiff1}), and using Eq. (\ref{stiff8})
valid for $\sigma\gg 1$ (the most relevant case), we
obtain
\begin{eqnarray}
a_{\rm sr}=\left (\frac{3\pi G\hbar^2\Omega_{\rm
dm,0}^2\epsilon_0}{2m^2c^6\Omega_{r,0}}\right )^{1/2}.
\end{eqnarray}
For a SF with $m=3\times 10^{-21}\, {\rm
eV}/c^2$ and $a_s=1.11\times 10^{-58}\, {\rm fm}$, corresponding to the fiducial
model of Li {\it et al.} \cite{b36}, we obtain $a_{\rm sr}=9.87\times 10^{-12}$.
This analytical result is in qualitative agreement with their numerical result
(see
their Fig. 3).

We now consider a SF with $a_s<0$. We first consider the normal branch. 
If $\sigma\gg 1$, it is not possible to match the pseudo rest-mass density of
the slow oscillation regime [see Eq. (\ref{stiff1})] with the pseudo rest-mass
density of the fast oscillation regime [see Eq. (\ref{negba1})]. This suggests
that there is no stiff matter era when  $\sigma\gg 1$, or that it cannot be
smoothly connected to the matterlike era. When $\sigma\ll 1$, we
obtain
\begin{eqnarray}
(a'_v)^*= \left (\frac{2\pi |a_s|\hbar^2 Q}{m^2c^2}\right
)^{1/3}\, g_*\left (\frac{3 |a_s| c^2}{4 Gm}\right ),
\label{stiff10}
\end{eqnarray}
where the function $g_*(\sigma)$ is defined by
\begin{eqnarray}
g_*(\sigma)=\frac{1}{r^{1/3}(1-4r)^{1/6}}
\label{stiff11}
\end{eqnarray}
with
\begin{eqnarray}
r=2\sigma(-\ln a_v^*)^2.
\label{stiff12}
\end{eqnarray}
It is easy to see
that, up to logarithmic corrections, $(a'_v)^*$ is of  the same
order of magnitude as the scale $a'_v$ marking the transition between the slow
and fast oscillation regimes given by Eq.
(\ref{vip2}). Mathematically, this is
because the function $r$ defined by Eq. (\ref{vip4}) behaves as $r\sim
\sigma$ for $\sigma\rightarrow 0$
which is the same scaling as the function $r\propto \sigma$ defined by Eq.
(\ref{stiff12}). This is most easily seen by
considering asymptotic limits.
Matching the pseudo rest-mass density of the stiff matter era
[see Eq. (\ref{stiff1})] with the pseudo rest-mass density of the matterlike
era [see Eq. (\ref{negea4})], we obtain a transition scale of the same
order as
Eq. (\ref{valp11}).

We now consider the peculiar branch. In that case, we find that it is
not possible to match the pseudo rest-mass density of
the slow oscillation regime with the pseudo rest-mass
density of the fast oscillation regime, except when $\sigma\sim 1$. This
suggests that the  peculiar branch cannot be connected to a stiff matter era
when $\sigma\neq 1$. On the other hand, when $\sigma\sim 1$ the domain of
validity of the fast oscillation regime is very small so this case is not
very relevant.

\section{Mass and length scales}
\label{sec_c}

In Sec. \ref{sec_effl} we have obtained a relation  between the
effective cosmological constant $\Lambda$ associated with the energy
$\epsilon_\Lambda$ and the mass $m$ and the scattering length $a_s<0$ of the
SF [see Eqs. (\ref{pb1}) and (\ref{pb3})]. The energy scale
$\epsilon_\Lambda$ corresponds to the maximum of the total potential $V_{\rm
tot}$ [see Eq.
(\ref{tp4b})]. If we assume that $|a_s|\sim r_S=2Gm/c^2$ (effective
Schwarzschild radius),
and substitute this
relation into Eqs.
(\ref{pb1}) and (\ref{pb3}), we obtain a mass scale
\begin{eqnarray}
m_\Lambda=\frac{\hbar}{c^2}\sqrt{\frac{\Lambda}{8\pi}}=\frac{\hbar}{c^2}
\sqrt{G\rho_{\Lambda}}
\label{c1}
\end{eqnarray}
and a length scale
\begin{eqnarray}
r_\Lambda=\frac{G\hbar}{c^4}\sqrt{\frac{\Lambda}{8\pi}}=\frac{G\hbar}{c^4}\sqrt{
G\rho_\Lambda}.
\label{c2}
\end{eqnarray}
Of course, $|a_s|$ can be different from $r_S$ but
the scales (\ref{c1}) and (\ref{c2}) can be introduced on a dimensional basis,
just like the Planck scales. Actually, they can be written as
\begin{eqnarray}
m_\Lambda=\left (\frac{\rho_{\Lambda}}{\rho_P}\right
)^{1/2}M_P,\qquad r_\Lambda=\left
(\frac{\rho_{\Lambda}}{\rho_P}\right )^{1/2}l_P,
\label{c3}
\end{eqnarray}
so they reduce to
the Planck mass $M_P=(\hbar
c/G)^{1/2}=1.22\times 10^{19}\, {\rm GeV}/c^2$ and Planck length $l_P=(\hbar
G/c^3)^{1/2}=1.62\times 10^{-35}\, {\rm m}$ if we identify
$\rho_\Lambda$ to the Planck
density $\rho_P=c^5/\hbar G^2=5.16\times 10^{99}\, {\rm g}\, {\rm
m}^{-3}$. However, we shall consider here that $\rho_\Lambda$ represents
the cosmological density. Writing
$\rho_{\Lambda}=\Omega_{\Lambda,0}\epsilon_0/c^2$ and using the Friedmann
equation (\ref{h5}), we get
\begin{eqnarray}
m_\Lambda=\sqrt{\frac{3\Omega_{\Lambda,0}}{8\pi}}\frac{H_0\hbar}{
c^2
},\qquad r_\Lambda=\sqrt{\frac{3\Omega_{\Lambda,0}}{8\pi}}\frac{GH_0\hbar}{c^4}.
\label{c4}
\end{eqnarray}
Numerically (see Appendix \ref{sec_vcp}),
\begin{equation}
m_\Lambda=4.11\times 10^{-34}\, {\rm eV}/c^2,\quad r_\Lambda=5.44\times
10^{-82}\, {\rm fm}.
\label{c5}
\end{equation}

Other mass and length scales can be introduced similarly. If we assume that
$|a_s|\sim \lambda_C=\hbar/mc$ (Compton wavelength), and substitute this
relation
into Eq. (\ref{pb1}), we obtain a mass scale and a length scale
\begin{eqnarray}
m^*_\Lambda=\left (\frac{\rho_\Lambda \hbar^3}{c^3}\right )^{1/4},\qquad
r^*_\Lambda=\left (\frac{\hbar}{\rho_\Lambda c}\right )^{1/4}.
\label{c6}
\end{eqnarray}
Numerically (see Appendix \ref{sec_vcp}),
\begin{equation}
m^*_\Lambda=2.24\times 10^{-3}\, {\rm eV}/c^2,\quad r^*_\Lambda=8.81\times
10^{10}\, {\rm fm}.
\label{c7}
\end{equation}

The Compton wavelength of a particle of mass $M_P$ is the Planck length
$l_P$. It is also equal to the effective semi-Schwarzschild radius $r_S/2$. The
Compton
wavelength
of a
particle of mass $m_\Lambda$ is the cosmological length
$l_\Lambda=c(8\pi/\Lambda)^{1/2}=4.80\times 10^{26}\, {\rm m}$, i.e., the
typical size of
the visible Universe (since $\Lambda\sim G\rho_\Lambda\sim H_0^2$ implies
$l_\Lambda\sim
c/\sqrt{\Lambda}\sim c/H_0$).

\section{The initial condition for the SF}
\label{sec_ic}

The KG equation is a second order differential equation in time. To
solve this equation, we need to specify the values of $\varphi$
and $\dot\varphi$ at $t=0$. In this Appendix, we show how they are related to
the hydrodynamic variables used in Secs. \ref{sec_post} and \ref{sec_negt}.

If we restrict ourselves to a spatially homogeneous SF, we have
\begin{eqnarray}
\varphi(t)=\frac{\hbar}{m}\sqrt{\rho(t)} e^ { i S_{\rm
tot}(t)/\hbar}.
\label{ic1}
\end{eqnarray}
Taking the time derivative of Eq. (\ref{ic1}), we get
\begin{eqnarray}
\dot\varphi=\frac{\hbar}{2m\sqrt{\rho}}
\left ( \dot\rho-i\frac{2\rho}{\hbar}E_{\rm tot}\right ) e^ { i S_{\rm
tot}/\hbar}.
\label{ic2}
\end{eqnarray}
Substituting the results of Secs. \ref{sec_post} and \ref{sec_negt} into Eq.
(\ref{ic2}), we obtain the asymptotic behaviors of $\varphi$ and
$\dot\varphi$ for $t\rightarrow 0$. Considering the modulus of the SF, we find
\begin{eqnarray}
|\varphi(t)|=\frac{\hbar}{m}\sqrt{\rho(t)}, \qquad
\dot{|\varphi|}=\frac{\hbar}{2m\sqrt{\rho}}\dot\rho.
\label{ic3}
\end{eqnarray}
For a SF with $a_s>0$, using the results of Sec. \ref{sec_post}, we get
for $t\rightarrow 0$:
\begin{eqnarray}
|\varphi|\propto t^{-1/2},\qquad
\dot{|\varphi|}\propto -t^{-3/2}.
\label{ic4}
\end{eqnarray}
Therefore, $|\varphi|\rightarrow +\infty$ and $\dot{|\varphi|}\rightarrow
-\infty$ for $t\rightarrow 0$. For a SF with $a_s<0$, using the results of Sec. 
\ref{sec_negt}, we get for $t\rightarrow 0$:
\begin{eqnarray}
|\varphi|\rightarrow |\varphi_i|,\qquad
\dot{|\varphi|}\propto \pm t^{-1/2}.
\label{ic5}
\end{eqnarray}
In that case, the initial value of  $|\varphi|$ is finite while
$\dot{|\varphi|}\rightarrow -\infty$ on the normal branch and 
$\dot{|\varphi|}\rightarrow +\infty$ on the peculiar branch. This is a very
singular initial condition. The choice of the branch is selected by the initial
condition, i.e., by the sign of $\dot{|\varphi|}$.

\section{The case of power-law SF potentials}
\label{sec_plsf}

In this Appendix, we briefly discuss the
evolution of a homogeneous SF with a general power-law potential.
For a power-law SF potential of the form (see Appendix C.5.2 of
\cite{logotrope}):
\begin{eqnarray}
V(|\varphi|^2)=\frac{K}{\gamma-1}\left (\frac{m}{\hbar}\right
)^{2\gamma}|\varphi|^{2\gamma},
\label{plsf1}
\end{eqnarray}
we obtain
\begin{eqnarray}
V(\rho)=\frac{K}{\gamma-1}\rho^{\gamma}, \qquad
h(\rho)=V'(\rho)=\frac{K\gamma}{\gamma-1}\rho^{\gamma-1},
\label{plsf2}
\end{eqnarray}
\begin{eqnarray}
P(\rho)=K\rho^{\gamma},\qquad 
c_s^2=K\gamma\rho^{\gamma-1}.
\label{plsf3}
\end{eqnarray}
The pressure law $P(\rho)$ is
that of a polytrope of index $\gamma$ ($h$ is the enthalpy).
For a quartic potential, we recover the polytropic EOS
(\ref{kge31}) with the exponent $\gamma=2$. For a $ |\varphi|^6$ potential,
which is the next order term in an expansion of the SF potential
$V(|\varphi|^2)$ in powers of $|\varphi|^2$, we get a polytropic EOS
$P=K\rho^3$ with the exponent $\gamma=3$.

The equations of the problem are 
\begin{eqnarray}
\rho
\sqrt{1+\frac{2}{c^2}\frac{K\gamma}{\gamma-1}\rho^{\gamma-1}}=\frac{Qm}{a^3},
\label{plsf4}
\end{eqnarray}
\begin{eqnarray}
\frac{3H^2}{8\pi G}=\rho+\frac{\gamma+1}{\gamma-1}\frac{K}{c^2}\rho^{\gamma},
\label{plsf5}
\end{eqnarray}
\begin{eqnarray}
\epsilon=\rho c^2+\frac{\gamma+1}{\gamma-1}K\rho^{\gamma},
\label{plsf6}
\end{eqnarray}
\begin{eqnarray}
P=K\rho^{\gamma},
\label{plsf7}
\end{eqnarray}
\begin{eqnarray}
w=\frac{\frac{K}{c^2}\rho^{\gamma-1}}{1+\frac{\gamma+1}{\gamma-1}\frac{K}{c^2}
\rho^{\gamma-1}},
\label{plsf8}
\end{eqnarray}
\begin{eqnarray}
E_{\rm tot}=mc^2
\sqrt{1+\frac{2}{c^2}\frac{K\gamma}{\gamma-1}\rho^{\gamma-1}}.
\label{plsf9}
\end{eqnarray}
From Eqs.
(\ref{plsf6}) and (\ref{plsf7}), we obtain
\begin{eqnarray}
\epsilon=\left (\frac{P}{K}\right )^{1/\gamma}c^2+\frac{\gamma+1}{\gamma-1}P,
\label{plsf10}
\end{eqnarray}
which determines the EOS $P(\epsilon)$ of the SF under the inverse
form $\epsilon(P)$.  The differential equation governing the temporal
evolution of
the pseudo rest-mass density is 
\begin{eqnarray}
\frac{c^2}{24\pi G}\left
(\frac{\dot\rho}{\rho}\right )^2=\frac{\rho
c^2+\frac{K(\gamma+1)}{\gamma-1}\rho^{\gamma}}{\left\lbrack
1+\frac{K\gamma\rho^{\gamma-1}}{c^2+\frac{2K\gamma}{\gamma-1}
\rho^{\gamma-1}}\right\rbrack^2 }.
\label{plsf11}
\end{eqnarray}
These equations can be used to determine the cosmological evolution of a
homogeneous SF for any value of $K$ and $\gamma$. This general study will be
considered in a future work. 

To be more specific, we now assume
$\gamma>1$ and $K>0$. For $a\rightarrow +\infty$, the SF
experiences a pressureless matterlike era. For  $a\rightarrow
0$, we get
\begin{eqnarray}
\rho\sim \left\lbrack
\frac{(\gamma-1)Q^2m^2c^2}{2K\gamma}\right\rbrack^{1/(\gamma+1)}
\frac{1}{a^{6/(\gamma+1)}},
\label{plsf12}
\end{eqnarray}
\begin{eqnarray}
\epsilon\sim \frac{\gamma+1}{\gamma-1}K\rho^{\gamma}\propto
\frac{1}{a^{6\gamma/(\gamma+1)}},
\label{plsf13}
\end{eqnarray}
\begin{eqnarray}
P\sim \frac{\gamma-1}{\gamma+1}\epsilon\propto
\frac{1}{a^{6\gamma/(\gamma+1)}},
\label{plsf14}
\end{eqnarray}
\begin{eqnarray}
w_i=\frac{\gamma-1}{\gamma+1},
\label{plsf15}
\end{eqnarray}
\begin{eqnarray}
\frac{E_{\rm tot}}{mc^2}\sim \left (\frac{\gamma}{\gamma-1}\frac{2K}{c^2}\right
)^{1/2}\rho^{(\gamma-1)/2}\propto \frac{1}{a^{3(\gamma-1)/(\gamma+1)}}.\qquad
\label{plsf16}
\end{eqnarray}
In that limit, the SF behaves as a fluid with a linear EOS
$P=\alpha  \epsilon$ where $\alpha=(\gamma-1)/(\gamma+1)$. For a quartic
potential ($\gamma=2$), we recover the EOS of radiation  $P=
 \epsilon/3$. For a $ |\varphi|^6$ 
potential ($\gamma=3$), we get $\alpha=1/2$. More generally, for $\gamma\ge 1$,
the exponent $\alpha$ goes from $0$ to $1$. These results coincide with those
obtained in Refs. \cite{turner,ford} for a real SF. 

The case $\gamma<1$ and $K<0$ is interesting because it leads to a model of
Universe that behaves as pressureless DM for $a\rightarrow 0$ and as DE for
$a\rightarrow +\infty$. Therefore, it provides a unification of DM and DE. For
$\gamma=0$ we recover the $\Lambda$CDM model and for $\gamma=-1$ we recover the
Chaplygin gas model \cite{chaplygin}. It is interesting that these two
famous models are selected by our approach among the infinite family of
polytropic models described by an EOS of the form $P=K\epsilon^{\gamma}$
\cite{cosmopoly1,cosmopoly2}. For $\gamma<1$ and $K>0$ we obtain models of
Universe that oscillate (phoenix Universes). For $\gamma=0$ we recover the
anti-$\Lambda$CDM model and for $\gamma=-1$ we recover the anti-Chaplygin gas
model (see, e.g., \cite{cosmopoly1,cosmopoly2} for more details).

{\it Remark:} Some of these results were previously obtained by
Bilic {\it et al.} \cite{bilic} by considering the inverse problem (we are
grateful to the referee for this remark). Assuming an EOS of the form
$P=-A/\epsilon$ (Chaplygin gas) and
using Eq. (\ref{bi1}), we easily obtain $\epsilon=\rho
c^2$ and 
\begin{eqnarray}
V_{\rm tot}(\rho)=\frac{1}{2}\left (\rho c^2+\frac{A}{\rho c^2}\right ).
\label{plsf17}
\end{eqnarray}
This corresponds to Eq. (13) of \cite{bilic} if we recall Eq. (\ref{kge7}). This
is also a particular case of Eq. (\ref{plsf2}) corresponding to $\gamma=-1$ and
$K=-A/c^2$.
If we consider a constant EOS of the form $P=-\rho_{\Lambda}c^2$ ($\Lambda$CDM
model \cite{cosmopoly2}), we
obtain $\epsilon=\rho c^2+\rho_{\Lambda}c^2$ and  
\begin{eqnarray}
V_{\rm tot}(\rho)=\frac{1}{2}\rho c^2+\rho_{\Lambda}c^2.
\label{plsf18}
\end{eqnarray}
This is a particular case of Eq. (\ref{plsf2}) corresponding to $\gamma=0$ and
$K=-\rho_{\Lambda}c^2$.

\end{document}